\documentclass[fleqn,usenatbib]{mnras}

\usepackage{newtxtext,newtxmath}

\usepackage[T1]{fontenc}

\DeclareRobustCommand{\VAN}[3]{#2}
\let\VANthebibliography\thebibliography
\def\thebibliography{\DeclareRobustCommand{\VAN}[3]{##3}\VANthebibliography}

\usepackage{graphicx}	
\usepackage{amsmath}	
\usepackage{float}
\usepackage{url}
\usepackage{caption}
\usepackage{subcaption} 

\defcitealias{Ma_2023}{M23}
\graphicspath{{./}{figures/}}
\usepackage{xcolor}

\title[EoR 21-cm signal for different cosmologies]{Exploring the effect of different cosmologies on the Epoch of Reionization 21-cm signal with \textsc{Polar}}

\author[A. Acharya et al.]{Anshuman Acharya$^{1}$\thanks{E-mail: anshuman@mpa-garching.mpg.de},
Qing-bo Ma$^{2,3}$,
Sambit K. Giri$^{4}$,
Benedetta Ciardi$^{1}$, 
Raghunath Ghara$^{5}$,
\newauthor Garrelt Mellema$^{6}$, 
Saleem Zaroubi$^{7,8}$, 
Ian Hothi$^{9,10}$, 
Ilian T. Iliev$^{11}$, 
L\'eon V. E. Koopmans$^{7}$, 
\newauthor Michele Bianco$^{12}$
\\
$^{1}$Max-Planck-Institut für Astrophysik, Garching 85748, Germany \\
$^{2}$School of Physics and Electronic Science, Guizhou Normal University, Guiyang 550001, PR China \\
$^{3}$Guizhou Provincial Key Laboratory of Radio Astronomy and Data Processing, Guizhou Normal University, Guiyang 550001, PR China \\
$^{4}$Nordita, KTH Royal Institute of Technology and Stockholm University, Hannes Alfv\'ens v\"ag 12, SE-106 91 Stockholm, Sweden \\
$^{5}$Department of Physical Sciences, Indian Institute of Science Education and Research Kolkata, Mohanpur, WB 741 246, India \\
$^{6}$The Oskar Klein Centre, Department of Astronomy, Stockholm University, AlbaNova, SE-10691 Stockholm, Sweden \\
$^{7}$Kapteyn Astronomical Institute, University of Groningen, P.O. Box 800, 9700AV Groningen, The Netherlands \\
$^{8}$Astrophysics Research Centre of the Open University of Israel, Ra’anana 4353701, Israel \\
$^{9}$LERMA, Observatoire de Paris, PSL Research University, CNRS, Sorbonne Universit\'{e}, F-75014 Paris, France \\
$^{10}$Laboratoire de Physique de l’ENS, ENS, Universit\'{e} PSL, CNRS, Sorbonne Universit\'{e}, Universit\'{e}e Paris Cit\'{e}, 75005, Paris, France \\
$^{11}$Astronomy Centre, Department of Physics \& Astronomy, Pevensey III Building, University of Sussex, Falmer, Brighton, BN1 9QH, United Kingdom \\
$^{12}$Institute for Particle Physics and Astrophysics, ETH Zurich, Wolfgang-Pauli-Str 27, 8093 Zurich, Switzerland \\
}

\date{Accepted XXX. Received YYY; in original form ZZZ. NORDITA-2024-035}

\pubyear{2024}

\begin{document}
\label{firstpage}
\pagerange{\pageref{firstpage}--\pageref{lastpage}}
\maketitle

\begin{abstract}
A detection of the 21-cm signal power spectrum from the Epoch of Reionization is imminent, thanks to consistent advancements from telescopes such as LOFAR, MWA, and HERA, along with the development of SKA. In light of this progress, it is crucial to expand the parameter space of simulations used to infer astrophysical properties from this signal. In this work, we explore the role of cosmological parameters such as the Hubble constant $H_0$ and the matter clustering amplitude $\sigma_8$, whose values as provided by measurements at different redshifts are in tension. We run \textit{N}-body simulations using \textsc{Gadget-4}, and post-process them with the reionization simulation code \textsc{Polar}, that uses \textsc{L-Galaxies} to include galaxy formation and evolution properties and \textsc{Grizzly} to execute 1-D radiative transfer of ionizing photons in the intergalactic medium (IGM). We compare our results with the latest JWST observations and explore which astrophysical properties for different cosmologies are necessary to match the observed UV luminosity functions at redshifts $z = 10$ and 9. Additionally, we explore the impact of these parameters on the observed 21-cm signal power spectrum upper limits, focusing on the redshifts within the range of LOFAR 21-cm signal observations ($z \approx 8.5-10$). Despite differences in cosmological and astrophysical parameters, our models cannot be ruled out by the current upper limits. This suggests the need for broader physical parameter spaces for inference modeling to account for all models that agree with observations. However, we also propose stronger constraining power by using a combination of galactic and IGM observables.
\end{abstract}



\section{Introduction}\label{sec:intro}

The Epoch of Reionization (EoR) refers to the period of the Universe when sufficient astrophysical objects formed and emitted UV photons to (re)ionize the neutral hydrogen of the intergalactic medium (IGM). One of the most important probes for studying this epoch is the brightness temperature fluctuations of the 21-cm line of neutral hydrogen (HI) as observed in emission or absorption against the Cosmic Microwave Background radiation (CMB; \citealt{Field_1959,Hogan_1979,Madau_1997,Shaver_1999,Tozzi_2000,Ciardi_2003,Zaroubi_2013}). The 21-cm signal is produced by the forbidden hyperfine spin-flip transition of the ground state of neutral hydrogen. As the excited state has an extremely long lifetime and neutral hydrogen is abundant in the IGM at the start of the EoR, this signal can be used as a tracer of the IGM ionization state. A statistical detection of the strength of the fluctuations of its brightness temperature allows us to constrain models of the Universe during this early formation stage.

For this purpose, multiple interferometric low-frequency radio telescopes have been designed, such as PAPER\footnote{Precision Array to Probe EoR, \url{http://eor.berkeley.edu}}, MWA\footnote{Murchison Widefield Array, \url{http://www.mwatelescope.org}}, HERA\footnote{Hydrogen Epoch of Reionization Array, \url{https://reionization.org/}}, LOFAR\footnote{Low-Frequency Array, \url{http://www.lofar.org}}, and the upcoming SKA\footnote{Square Kilometre Array, \url{https://www.skao.int/en}}. While the signal is yet to be detected, upper limits of the 21-cm power spectrum have been obtained and are becoming increasingly tighter \citep[e.g.,][]{Mertens2020, Trott_2020, HERA_2023, Acharya_2024gprletter}. 
These have allowed us to rule out some extreme astrophysical models through comparisons with simulations \citep[e.g.,][]{Ghara2020, Mondal_2020, Greig_2021, Greig_2021lofar, Abdurashidova_2022}. New methods that utilise multi-redshift power spectrum observations \citep{Ghara_2024, Choudhury_2024}, wavelet statistics \citep{Hothi_2024}, and simulation-based inference \citep{Saxena_2023, Greig_2024} have also been developed to improve the understanding of parameters governing the properties of the IGM.

At the same time, galaxies that formed during the EoR that are responsible for HI ionization have also been studied using hydrodynamical and/or radiative transfer simulations with different prescriptions, such as {\sc CROC} \citep{Gnedin_2014,Gnedin_2014b, Esmerian_2021, Esmerian_2022, Esmerian_2024},  {\sc FIRE} \citep{Ma_2018fire}, {\sc Technicolor Dawn} \citep{Finlator_2018}, {\sc Sphinx} \citep{Rosdahl_2018}, {\sc CRASH} simulations  \citep{Eide_2018,Eide_2020,Ma_2021, Kostyuk_2023, Basu_2024}, {\sc C$^2$Ray} simulations \citep{Mellema_2006,Dixon_2016,Giri_2021,Hirling_2024,Giri_2024}, {\sc CoDa} I,II, III \citep{Ocvirk_2016,Ocvirk_2020,Lewis_2022}, {\sc Astrid} \citep{Bird_2022}, {\sc ColdSIM} \citep{Maio_2022, Maio_2023, Casavecchia_2024}, {\sc Thesan} \citep{Kannan_2022,Garaldi_2022,Smith_2022,Garaldi_2024}, and {\sc SPICE} \citep{Bhagwat_2024}. Although these simulations are ideal tools to investigate the IGM and galactic properties, they are computationally expensive and thus limited in box size and/or in the number of simulations that can be run. Thus, they cannot be used for building a wide range of models for comparison with the 21-cm power spectrum or their upper limits. 

To this aim, semi-numerical approaches such as {\sc SimFast21} \citep{Santos_2010code, Santos_2010}, {\sc Bears} \citep{Thomas_2009}, {\sc 21cmFAST} \citep{Mesinger_2007,Mesinger_2011}, {\sc Grizzly} \citep{Ghara_2015,Ghara_2018}, {\sc ReionYuga} \citep{Mondal_2017}, {\sc Artist} \citep{Molaro_2019}, {\sc AMBER} \citep{Trac_2022} and {\sc BEORN} \citep{Schaeffer_2023} are typically employed. To incorporate a more realistic modeling of galactic properties into these methods, codes such as {\sc ASTRAEUS} \citep{Hutter_2021, Hutter_2024}, {\sc MERAXES} \citep{Mutch_2016,Balu_2023} and {\sc Polar} \citep[][hereafter M23]{Ma_2023} have been developed.

The LOFAR EoR Key Science Project (KSP) team has designed {\sc Polar}, a semi-numeric approach that strikes a balance between speed and complexity of relevant physical processes, by post-processing $N$-body Dark Matter (DM) simulations with Semi-Analytic Models (SAMs) of galaxy formation and evolution, and subsequently applying the 1D radiative transfer (RT) code  {\sc Grizzly} \citep{Ghara_2015, Ghara_2018}, an updated version of {\sc Bears} \citep{Thomas_2009}. This allows a fast modelling of the 21-cm signal power spectrum, while at the same time accounting for important galactic properties and RT effects. The 21-cm signal can be used to derive physical properties of the Universe \citep{McQuinn_2006}, but as it primarily comes from the IGM, it provides only a partial picture. Thus, instead, we combine observations of the 21-cm line with data obtained in other frequency bands (e.g. from JWST) to jointly constrain IGM and galactic properties. This has been explored in the past, for example, by using cross-power spectra of galaxy clustering observations \citep{Lidz_2009, Vrbanec_2016, Berti_2024}, or intensity mapping \citep{Mao_2008, LaPlante_2023, Berti_2024} at different redshifts. Eventually, the goal of such studies would be to jointly constrain the physical properties of the Universe from multiple observables of galaxies and the IGM.

In this work, we investigate the impact of varying cosmological parameters on the 21-cm signal at $z \approx 10.11, 9.16$, and 8.3, i.e. the three redshifts targeted by LOFAR \citep{Mertens_2025}. For this, we develop models with the most extreme choices of cosmological parameters observed so far. Thus, we focus on those parameters whose low and high redshift measurements show maximum tension (here we define ``tension'' as having more than 2$\sigma$ difference in the measurements of the parameters), i.e. the Hubble constant $H_0 = 100 h~\rm km~s^{-1}~Mpc^{-1}$ (where $h$ is a dimensionless parameterisation of the Hubble constant), and the matter clustering amplitude $\sigma_8$
\footnote{While the $\rm S_8$ parameter is also occasionally used, it is a derived quantity given by $S_8 = \sigma_8 \sqrt{\Omega_{\rm m}/0.3}$. Thus, when a tension is present, it is driven by the underlying tensions of $\sigma_8$ and/or $\Omega_{\rm m}$. For example, the $3.4 \sigma$ tension between the \citet{Troster_2020} and \citet{Planck_2020} measurements is completely driven by the $3.7\sigma$ tension between the $\sigma_8$ parameters.
We note that in some cases $S_8$ is measured directly (and not derived from $\sigma_8$ and $\Omega_{\rm m}$), and a tension up to $\approx 2 \sigma$ can be  found {\citep[for example][]{Amon_2023, Longley_2023, DES_2023}}. Most simulations, though, adopt $\sigma_8$ and $\Omega_{\rm m}$ as independent parameters, and we follow this typical convention.
}.
In future work, though, we plan to run models with various combinations of observed cosmological parameters to build a suite of $N$-body simulations that span a large parameter space. Specifically, we will build a range of models whose corresponding astrophysical parameters are derived by fitting multi-wavelength observables with an MCMC-based algorithm, which in turn can be used for parameter forecasting
(see \citetalias{Ma_2023} and \citealt{Ma_2025} for more details).

Our fiducial model is the one of \citet{Planck_2020}, with $h = 0.6766$ and $\sigma_8 = 0.8102$, while a higher $h$ value is adopted to explore the low redshift measurement of $h = 0.733$ from studies of Cepheid variables in the host galaxies of 42 Type Ia supernovae \citep{Riess_2022}. Similarly, a lower $\sigma_8$ value of 0.702 is adopted from the low-redshift measurement of anisotropic galaxy clustering measurement analysis \citep{Troster_2020}. Lastly, we also adopt a $\sigma_8=0.88$ from recent eROSITA results \citep{Ghirardini_2024} to consider both extremes of $\sigma_8$ measurements. Motivated by results obtained in the context of the 21-cm signal shown by \citet{Giri_2023} and \citet{Acharya_2024}, we additionally implement the Fixed \& Paired (F\&P) approach \citep{Angulo_2016} to boost effective volumes of the simulations and suppress cosmic variance.

The paper is structured as follows. 
In Section~\ref{sec:methodology}, we discuss the setup of the {\sc Polar} simulations with different cosmologies. In Section~\ref{sec:results} we present the resulting galactic and IGM  properties, and in Section~\ref{sec:discuss} we discuss the implications of varying astrophysical and cosmological parameters for inference modeling from 21-cm signal observations, jointly constrained with other observables of the EoR. Finally, we summarize our results in Section~\ref{sec:summary}.

\section{Methodology}\label{sec:methodology}

Building on \citetalias{Ma_2023}, we utilise a similar setup by running $N$-body DM simulations, and then post-processing them with the {\sc L-Galaxies} SAM \citep{Barrera_2023,Henriques_2015,Henriques_2020} to model the formation and evolution of galaxies, and with the 1D radiative transfer code {\sc Grizzly} \citep{Ghara_2015,Ghara_2018} to model the gas ionization and the 21-cm signal from neutral hydrogen. In Sections~\ref{sec:dmsims} and~\ref{sec:polar}, we highlight the key parameters used for setting up the simulations and  the analysis done in this work. Further, in Section~\ref{sec:compareobs} we propose two cases: one where we keep the astrophysical parameters the same for all four cosmological models, and one in which we tune them to match UV luminosity functions (UVLFs) observed with JWST and HST at $z = 10$ and 9. We refer to the first and second case as ``unconstrained'' and ``constrained'', respectively.

\subsection{Dark-matter simulations}\label{sec:dmsims}

We use the {\sc Gadget-4} code \citep{Springel_2021} for running $N$-body DM simulations with a volume of $(150 h^{-1}~\rm cMpc)^3$ and $2048^3$ DM particles. This translates into a DM particle mass of $5 \times 10^7~\rm M_{\odot}$, matching the particle resolution of larger simulations used by the LOFAR EoR KSP team \citep[see for example][]{Giri_2019, Giri_2019b}. The box size is chosen to probe wave-modes of $k \leq 1.0 h~\rm cMpc^{-1}$, where the best results from observations with LOFAR, HERA, MWA (and eventually SKA) are expected \citep{Koopmans_2015}. This condition requires simulations with box sizes $> 100 h^{-1}~\rm cMpc$  \citep{Iliev_2014}, although \citet{Kaur_2020} suggests box sizes  $> 175 h^{-1}$ cMpc being necessary when accounting for X-ray heating of the IGM. To sidestep this issue, we assume that at our redshifts of interest, the IGM has already been heated above the CMB temperature (see Section~\ref{sec:grizzly}). We additionally employ the F\&P approach \citep{Angulo_2016} to mitigate sample variance, and thus run two realisations of each DM simulation. Each pair of realisations has the mode amplitudes fixed to the square root of the initial matter power spectrum, and the phases of the second realisations (B-series) are obtained by mirroring those of the first realisations (A-series). This is in contrast to the traditional method of generating initial conditions, where the mode amplitudes are sampled from a Gaussian distribution centred on the expectation value, which is given by the initial matter power spectrum. Taking the F\&P approach suppresses the variance-induced fluctuations (which in traditional initial conditions mainly affect the large-scale modes, where the sampling of the power spectrum amplitude is scarce and therefore more susceptible to deviations from its expectation value) up to the fourth perturbative order \citep{Angulo_2016}. This means that while small-scale properties, such as individual halos, may not show any noticeable difference, averaged large-scale statistics of an F\&P pair reproduce very closely the ensemble averages of traditional simulations. The F\&P approach of averaging observables of these two realisations has been shown to boost the statistical precision of the matter power spectrum, bispectrum and halo mass function \citep{Angulo_2016, Chartier_2021,Maion_2022, Villaescusa-Navarro_2018}, Lyman-$\alpha$ power spectra \citep{Anderson_2019}, 21-cm signal power spectrum and bispectrum \citep{Giri_2023,Acharya_2024} and several other quantities derived from simulations \citep{Villaescusa-Navarro_2018,Klypin_2020}. Beyond this, we use the same random seed to minimise the effect of randomized initial conditions for both the A and B-series of simulations.

\begin{table}
\centering
\caption{The four $N$-body dark matter simulation models considered in this work. From left to right, the model name, the adopted value of $h$ and $\sigma_8$, and the reference for the values. All other cosmological parameters are fixed to the \citet{Planck_2020} values. 
}
\begin{tabular}{llll}
\hline 
Model & $h$ & $\sigma_8$ & Reference \\
\hline 
fiducial & 0.6766 & 0.8102 & \citet{Planck_2020} \\
$h$ high & 0.7330 & 0.8102 & \citet{Riess_2022} \\
$\sigma_8$ low & 0.6766 & 0.7020 & \citet{Troster_2020} \\
$\sigma_8$ high & 0.6766 & 0.8800 & \citet{Ghirardini_2024} \\
\hline
\end{tabular}
\label{table:dmsims}
\end{table}

For our reference simulation, we assume a ``fiducial'' $\Lambda$-Cold Dark Matter ($\Lambda$CDM) cosmological model based on \citet[][specifically the TT,TE,EE+lowE+lensing+BAO case]{Planck_2020}, setting the dark energy density parameter as $\Omega_{\Lambda} = 0.6889$, the total matter density parameter, including both baryons and dark matter as $\Omega_{\rm m} = \Omega_{\rm b} + \Omega_{\rm dm} = 0.3111$, the baryonic matter density parameter as $\Omega_{\rm b} = 0.04897$, the Hubble constant as $H_0 = 100 h~\rm km~s^{-1}~Mpc^{-1}$ with $h = 0.6766$, the amplitude of matter density fluctuations on scales of 8$h^{-1}$ Mpc as $\sigma_8 = 0.8102$, and the tilt of the primordial power spectrum of scalar perturbations as $n_s = 0.9665$. Further, we consider three additional cosmologies, where we vary $h$ and $\sigma_8$ while keeping the other parameters fixed to the above-mentioned values. This is necessary to consider the maximum impact of changing cosmological parameters across the range of observed values. First, we adopt $h = 0.7330$ \citep[from][]{Riess_2022}, and refer to this as the ``$h$ high'' model. In addition, we explore two extreme values of $\sigma_8$, namely the ``$\sigma_8$ low'' case with $\sigma_8 = 0.702$ \citep{Troster_2020}, and the ``$\sigma_8$ high'' case with $\sigma_8 = 0.88$ \citep{Ghirardini_2024}. The details of the four simulation models are listed in Table~\ref{table:dmsims}. As the $\sigma_8$ parameter represents the amplitude of matter density fluctuations, a higher (lower) $\sigma_8$ value leads to larger (smaller) dark matter halos, which in turn can lead to more (less) numerous and more (less) massive galaxies as compared to the fiducial case.
If all astrophysical parameters are kept the same, this would result in a faster (slower) reionization.
Thus, if we want, e.g. to produce a similar overall star formation history, the astrophysical parameters which affect the amount of stars formed would have to be reduced (boosted) in their impact. A higher (lower) Hubble constant leads to faster (slower) expansion of spacetime, i.e. to larger (smaller) voids. This translates into more (less) matter being pushed together, which leads to the formation of larger (smaller) dark matter halos. However, as the Hubble constant does not directly control the matter density, halo sizes are less sensitive to changes in $h$ than in $\sigma_8$. As a consequence, the impact on galaxy formation and evolution of the former is expected to be similar to the latter, albeit smaller. As the interplay between $\sigma_8$ and/or $h$ and the astrophysical parameters governing galaxy formation and evolution is complex, understanding their impact on the reionization process is not straightforward and should be explored carefully.

\begin{figure}
\centering
\includegraphics[width=\columnwidth,keepaspectratio]{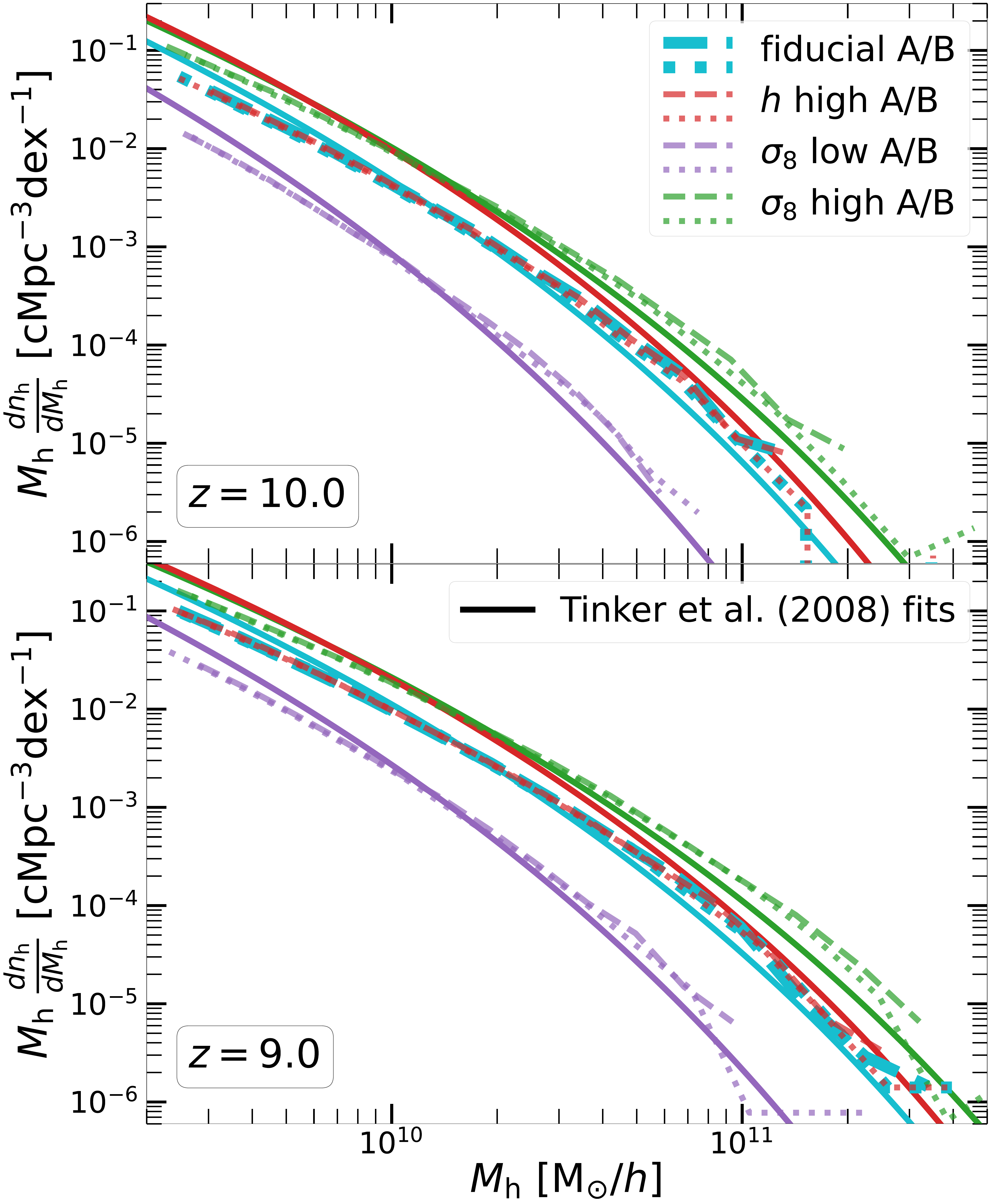}
\caption{Halo mass functions (HMFs) for our four models: fiducial (cyan), $h$ high (red), $\sigma_8$ low (purple) and $\sigma_8$ high (green) at $z = 10$ (top panel) and 9 (bottom). The A- and B-series are shown with dashed and dotted lines, respectively. The corresponding fit for each simulation using the \citet{Tinker_2008} model is shown with solid lines of the corresponding colours.}

\label{fig_hmf}
\end{figure}

The initial conditions for all simulations are generated at $z = 199$ with a second-order Lagrangian perturbation theory based on the {\sc NgenIC} algorithm implemented into {\sc Gadget-4}, using the same linear theory power spectrum as \citet{Hernandez_2023}. We generate 90 snapshots between $z = 20$ and $z = 5$, for which we save the full snapshot information, i.e., the positions and velocities of all the dark matter particles. However, in order to ensure a maximum step size of 10 Myrs between snapshots used to build halo merger trees with the Friend-Of-Friends (FOF) group finding algorithm \citep{Springel_2001}, we perform a finer gridding to generate a total of 156 output time steps for FOF groups. This is necessary because the dynamical time of a typical galactic disc at high redshift is of the order of tens of Myrs \citep{Poole_2016, Balu_2023}, which also is the lifetime of massive stars \citep{Barkana_2001, Schaerer_2002, Heger_2002, Bromm_2004}. Thus, a longer time step would miss how reionization progresses when post-processed with SAMs like {\sc L-Galaxies} and radiative transfer codes like {\sc Grizzly}. We use a standard linking length of 0.2 times the mean particle spacing and a minimum group size of 64 DM particles (corresponding to a minimum halo size of $\approx 3 \times 10^9~\rm M_{\odot}$). Substructures within halos are identified using the {\sc Subfind-HBT} algorithm \citep{Han_2018, Springel_2021}. Finally, the gravitational softening length is set to 0.025 of the mean particle spacing, i.e. $\approx 1.83 h^{-1}~\rm ckpc$.

To evaluate the performance of the simulations, in Figure~\ref{fig_hmf} we present the halo mass function (HMF) as  $M_{\rm h}~\frac{d n_{\rm h}}{d M_{\rm h}}$, where $M_{\rm h}$ and $n_{\rm h}$ are the halo mass and number density respectively, and compare them with fits by \citet{Tinker_2008}, computed using the Python package {\sc hmf} \citep{Murray_2014} at $z = 9~\rm and~10$. As in this study, \citet{Tinker_2008} takes into account the non-universality of the HMF by considering the impact of varying the cosmological parameters in the $\Lambda$CDM model. We note that the fiducial and $h$ high models at these redshifts produce very similar HMFs, despite some noticeable differences in their corresponding fits. A deeper exploration indicates that our simulation setup is less sensitive to differences in the Hubble parameter, given the choice of the starting redshift, and that the initial conditions were generated using second-order Lagrangian perturbation theory. A significantly higher starting redshift or third/fourth-order Lagrangian perturbation theory may thus be required to better match the fits. Nevertheless, some differences do exist, especially at the extremities of the HMF, which lead to the differences noted in the subsequent sections.

Further, we also find a mild under-prediction of the HMF at the low mass end and an over-prediction at the high mass end. This may be due to the choice of the linking length of 0.2, which at significantly high redshifts may lead to overlinking. A lower value of 0.17, as suggested by \citet{Watson_2013}, may resolve this issue. However, we still use 0.2 for minimizing differences between the setups used for past applications of {\sc L-Galaxies} \citep[e.g.,][and \citetalias{Ma_2023}]{Henriques_2015, Henriques_2020, Barrera_2023} and our setup. Lastly, we note that the B-series has slightly more massive halos in all models. This is purely because of the choice of the random seed; the same seed that leads to voids in the A-series produces regions of massive halo formation in the B-series due to the F\&P method described above.

\subsection{Galactic and IGM properties with {\sc Polar}}\label{sec:polar}

{\sc Polar} is a semi-numerical model designed to obtain high-$z$ galaxy properties and the 21-cm signal from the IGM in a fast and robust manner. It combines the semi-analytical galaxy formation and evolution model of {\sc L-Galaxies} \citep{Henriques_2015, Henriques_2020, Barrera_2023} with the one-dimensional radiative transfer code {\sc Grizzly} \citep{Ghara_2015, Ghara_2018}.

\subsubsection{Semi-Analytic modeling of galaxies}\label{sec:lgalaxies}

We use {\sc L-Galaxies} as described by \citet{Barrera_2023} as a post-processing module of {\sc Gadget-4}. This is an updated version of the publicly available {\sc L-Galaxies 2020} that was used by \citetalias{Ma_2023}. {\sc L-Galaxies} implements most major physical processes of gas cooling, star formation, galaxy mergers, supernovae feedback, black hole growth, AGN feedback, and dust attenuation. While \citet{Barrera_2023} largely builds on \citet{Henriques_2015}, we also consider the parameters used by \citet{Henriques_2020} before adapting them for our purposes. In particular, we focus on those that scale the star formation efficiency ($\alpha_{\rm SF}$), star formation efficiency during galaxy mergers ($\alpha_{\rm SF, burst}$), AGN accretion rate ($k_{\rm AGN}$), reheating of cold gas by star formation ($\epsilon_{\rm reheat}$, $V_{\rm reheat}$), and the energy released by each supernova ($E_{\rm SN}$). {\sc L-Galaxies} assumes that stars form from the cold gas within the disk of each galaxy. Thus, these parameters control the amount of cold gas available in the disk and the efficiency of converting it into stars. For more details, refer to the supplementary material of \citet{Henriques_2015}.

We tune the parameters to match photometric observations of the UVLFs from JWST and HST at $z = 10$ and 9, as these redshifts are observationally relevant for the LOFAR EoR KSP. We note that {\sc L-Galaxies} assumes a 100\% escape fraction for the UV photons, and thus we also implement the dust attenuation approach of \citet{Henriques_2015}. However, this is a simplified model and more complex ones may lead to greater suppression of the UV luminosity function. In Table~\ref{table:LGal_all} we list all the possible parameters available in {\sc L-Galaxies}, and their values set to match UVLF observations as discussed in Section~\ref{sec:compareobs}.

\subsubsection{Radiative transfer and the 21-cm signal}\label{sec:grizzly}

Modeling the EoR requires the inclusion of radiative transfer to describe the hydrogen ionization and heating. For this, we take the results of the $N$-body simulations from Section~\ref{sec:dmsims} and the semi-analytic modeling of galaxies from Section~\ref{sec:lgalaxies}, and post-process them with the 1D radiative transfer code {\sc Grizzly}, as done by \citetalias{Ma_2023}. {\sc Grizzly} uses pre-computed ionization and temperature profiles of gas for different source and density properties at various redshifts to model the ionization
and heating processes and the differential brightness temperature of the 21-cm signal ($\delta T_{\rm b}$).

More specifically, {\sc Grizzly} requires as input the gridded matter density field and dark matter halo masses from the $N$-body simulation, as well as the corresponding galactic stellar masses and stellar ages obtained with {\sc L-Galaxies}. Our reference simulation has a grid of 256$^3$ cells, resulting in a cell size of $\approx 600~h^{-1}~\rm ckpc$\footnote{The grid size is chosen after checking the convergence of the volume-averaged neutral hydrogen fraction and the 21-cm power spectrum in the fiducial model for a range of grid sizes (from 64$^3$ to 512$^3$ cells in steps of 2$^3$) across redshifts $12>z>5$. We note that the values are within 5\%, and thus opt for 256$^3$ to ensure we do not miss out on the impact of low-density regions in any redshift regime.}. Next, we assume that the gas density scales by a factor of $\Omega_{\rm b}/\Omega_{\rm dm}$ with the dark matter density. Lastly, as done in \citetalias{Ma_2023}, we use the Binary Population and Spectral Synthesis \citep[BPASS;][]{Stanway_2018} code to model the Spectral Energy Distributions (SEDs) of stellar sources. In future work, we will also explore the impact of other source types such as X-ray binaries, shock-heated interstellar medium, and accreting black holes \citep[as done in][]{Eide_2018, Eide_2020, Ma_2021}.

{\sc Grizzly} computes $\delta T_{\rm b}$ as follows  \citep[see][]{Furlanetto_2006eq}: \begin{multline}\label{eq:dtbfull} \delta T_{ \rm b} = 27 x_{\rm HI} (1 + \delta_{\rm B}) \left(1 - \frac{T_{\rm CMB}}{T_{\rm S}} \right) \\
        \times\biggl[ \left( \frac{\Omega_{\rm b}h^2}{0.023} \right) \left( \frac{0.15}{\Omega_{\rm m} h^2} \frac{1+z}{10} \right)^{1/2} \biggr]~\rm mK,
\end{multline} 
where $x_{\rm HI}$ is the fraction of neutral hydrogen, $\delta_{\rm B}$ is the fractional overdensity of baryons, $T_{\rm S}$ is the hydrogen spin temperature, $T_{\rm CMB}$ is the temperature of the CMB photons at redshift $z$, $h$ is the dimensionless Hubble constant, and $\Omega_{\rm b}$ and $\Omega_{\rm m}$ are the baryonic matter density and total matter density parameters respectively. As done by \citetalias{Ma_2023}, we assume $T_{\rm S} \gg T_{\rm CMB}$ which is valid when the IGM has been sufficiently heated by X-ray sources and expected to be the case in the range of redshift of interest here. Additionally, we also ignore the impact of redshift space distortions. Thus Equation~\ref{eq:dtbfull} is reduced to\begin{equation}\label{eq:dtb}
   \delta T_{ \rm b} = 27 x_{\rm HI} (1 + \delta_{\rm B}) \biggl[ \left( \frac{\Omega_{\rm b}h^2}{0.023} \right) \left( \frac{0.15}{\Omega_{\rm m} h^2} \frac{1+z}{10} \right)^{1/2} \biggr]~\rm mK.
\end{equation}

We define the power spectrum of $\delta T_{\rm b}$ as \citep{Mertens2020, Metha_2024}: \begin{equation}\label{eq:ps}
P_{\rm 21cm} (\mathbf{k}) = \delta_D(\mathbf{k}+\mathbf{k}') \langle \delta T_{\rm b}(\mathbf{k}) \delta T_{\rm b}(\mathbf{k}') \rangle,
\end{equation} where $\delta_D$ is the Dirac delta function, $\delta T_{\rm b}(\mathbf{k})$ is the differential brightness temperature in Fourier space, and $\langle ... \rangle$ is the ensemble average. In the following, we will report our results in terms of the normalized form of the power spectrum, given by \citep[see][for details]{Peacock_1999}:\begin{equation}\label{eq:psnorm}
    \Delta^{2}_{\rm 21cm} (k) = \frac{k^3}{2\pi^2} \times P_{\rm 21cm} (k).
\end{equation} 

\begin{table*}
\centering
\caption{List of astrophysical parameters in {\sc L-Galaxies} \citep[see][for more details]{Henriques_2015,Henriques_2020} that can be tuned to control galaxy formation and evolution. We show the names of the parameters in column 1, and \citet{Henriques_2015} and \citet{Henriques_2020} values tuned to $z<3$ observations in columns 2 and 3. Our cosmological models (fiducial, $h$ high, $\sigma_8$ low and $\sigma_8$ high) tuned to UVLF observations at $z = 10$ and $9$ are shown in columns 4, 5, 6 and 7 respectively. We highlight the parameters that differ from \citet{Henriques_2015,Henriques_2020} values in bold and fix the values of the unchanged parameters to the latter.
For the unconstrained case (see text for more details), we use the fiducial model parameters for all our $N$-body simulations.}
\begin{tabular}{lllllll}
\hline 
Parameter & \citet{Henriques_2015} & \citet{Henriques_2020} & fiducial & $h$ high & $\sigma_8$ low & $\sigma_8$ high \\
\hline
\vspace{0.2em}
$\alpha_{\rm SF}$ & 0.025 & 0.060 & \textbf{0.20} & \textbf{0.20} & \textbf{0.50} & \textbf{0.20} \\
\vspace{0.2em}
$\alpha_{\rm SF,burst}$ & 0.60 & 0.50 & \textbf{0.80} & \textbf{0.80} & \textbf{0.90} & \textbf{0.80} \\
\vspace{0.2em}
$\beta_{\rm SF,burst}$ & 1.90 & 0.38 & 0.38 & 0.38 & 0.38 & 0.38 \\
\vspace{0.2em}
$k_{\rm AGN}~[\rm 10^{-3} M_{\odot}~yr^{-1}]$ & 5.3 & 2.5 & \textbf{2.0} & \textbf{2.0} & \textbf{2.0} & \textbf{2.0} \\
$f_{\rm BH}$ & 0.041 & 0.066 & 0.066 & 0.066 & 0.066 & 0.066 \\
\vspace{0.2em}
$V_{\rm BH}~[\rm km~s^{-1}]$ & 750 & 700 & 700 & 700 & 700 & 700 \\
\vspace{0.2em}
$M_{\rm r.p.}~[\rm 10^{14} M_{\odot}]$ & 1.2 & 5.1 & 5.1 & 5.1 & 5.1 & 5.1 \\
\vspace{0.2em}
$\alpha_{\rm dyn.fric.}$ & 2.5 & 1.8 & 1.8 & 1.8 & 1.8 & 1.8\\
\vspace{0.2em}
$\epsilon_{\rm reheat}$ & 2.6 & 5.6 & \textbf{8.0} & \textbf{8.0} & \textbf{8.0} & \textbf{8.0} \\
\vspace{0.2em}
$V_{\rm reheat}~[\rm km~s^{-1}]$ & 480 & 110 & \textbf{250} & \textbf{250} & \textbf{250} & \textbf{250} \\
\vspace{0.2em}
$\beta_{\rm reheat}$ & 0.72 & 2.90 & 2.90 & 2.90 & 2.90 & 2.90 \\
\vspace{0.2em}
$\eta_{\rm eject}$ & 0.62 & 5.50 & 5.50 & 5.50 & 5.50 & 5.50 \\
\vspace{0.2em}
$V_{\rm eject}~[\rm km~s^{-1}]$ & 100 & 220 & 220 & 220 & 220 & 220 \\
\vspace{0.2em}
$\beta_{\rm eject}$ & 0.8 & 2.0 & 2.0 & 2.0 & 2.0 & 2.0 \\
\vspace{0.2em}
$\gamma_{\rm reinc}~[\rm 10^{10}~yr^{-1}]$ & 3.0 & 1.2 & 1.2 & 1.2 & 1.2 & 1.2 \\
$E_{\rm SN}~[\rm 10^{51}~erg]$ & 1.0 & 1.0 & \textbf{0.80} & 1.0 & \textbf{0.15} & \textbf{2.00} \\
$Z_{\rm yield}$ & 0.046 & 0.030 & 0.030 & 0.030 & 0.030 & 0.030 \\
\hline
\end{tabular}
\label{table:LGal_all}
\end{table*}

\begin{figure}
        \centering
        \includegraphics[width=\columnwidth,keepaspectratio]{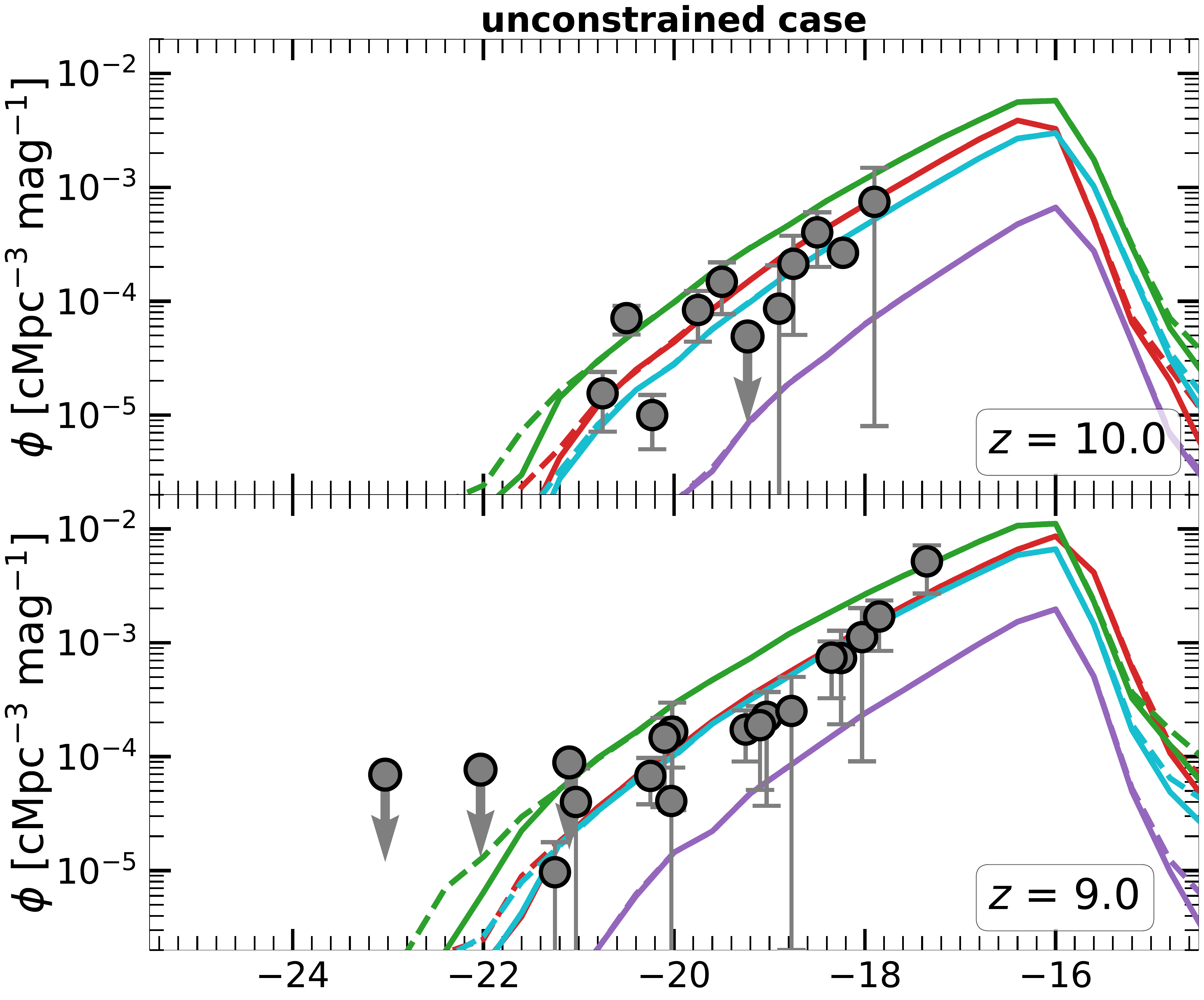}
        \centering
        \includegraphics[width=\columnwidth,keepaspectratio]{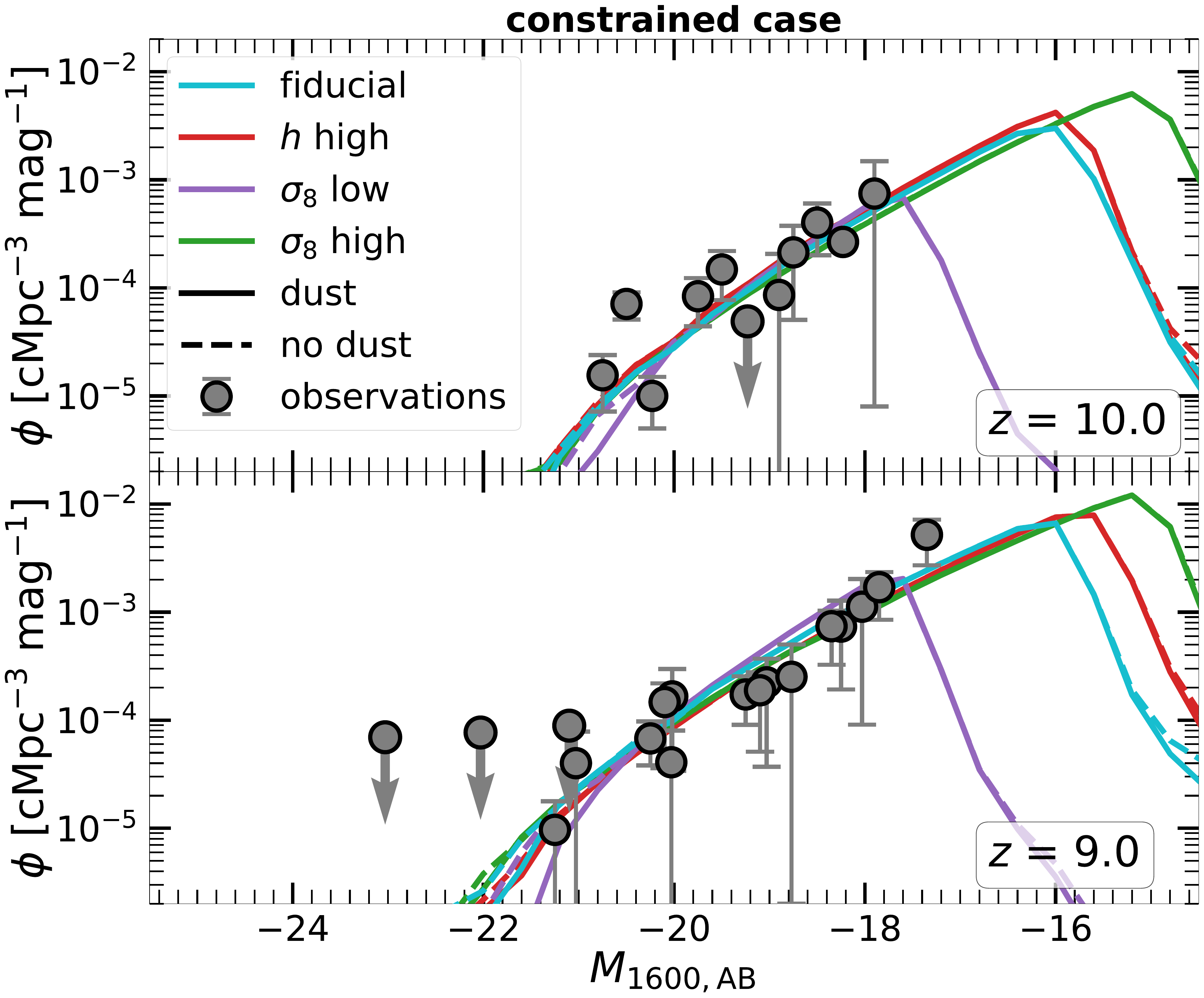}
    \caption{A and B-series averaged UVLFs for the unconstrained (top) and constrained (bottom) cases at  $z = 10$ and 9 for the fiducial (cyan), $h$ high (red), $\sigma_8$ low (purple) and $\sigma_8$ high (green). The averaging of the F\&P pairs allows to compensate for stochasticity in underfilled bins of the UVLFs, by suppressing the impact of cosmic variance. We also show the dust attenuated (solid) and unattenuated  (dashed) UVLFs, and JWST and HST observations \citep[][grey circles]{Finkelstein_2015,Bouwens_2015,Bouwens_2021,Harikane_2022,Bouwens_2023a,Bouwens_2023b,Harikane_2023,Leung_2023,McLeod_2024,Adams_2024}.
    }
    \label{fig_uvlf_combined}
\end{figure}

\subsection{UV luminosity functions}\label{sec:compareobs}

Because \citet{Henriques_2015,Henriques_2020} tuned the {\sc L-Galaxies} parameters to the low-redshift Universe ($z <3$), it is necessary to adapt them to match the redshift regime that we are interested in \citep[see][on discrepancies from observations at higher redshifts]{Vani_2024}. In particular, for the LOFAR EoR KSP team, focussing on  $10 > z > 8.5$ is crucial, as this is the regime in which the LOFAR telescope is most sensitive. 
Thus, we change the values of the parameters listed in Section~\ref{sec:lgalaxies} to match the observations of 
the UVLFs from HST legacy fields and JWST programs \citep{Finkelstein_2015, Bouwens_2015, Bouwens_2021, Harikane_2022, Bouwens_2023a, Bouwens_2023b, Harikane_2023, Leung_2023, McLeod_2024, Adams_2024} at $z = 10$ and 9 by employing the least square minimisation\footnote{In the least square minimisation the square of the residuals between UVLF observations and models is summed for both redshifts over a range of different values for the parameters. The values that lead to the minimum sum of the squares of the residuals are accepted.}. We iteratively vary parameters one at a time, with the range going from 1/100th of the lower value to 100 times the higher value between the fitted values obtained by \citet{Henriques_2015} and \citet{Henriques_2020} for each parameter. While typically the average energy released by each supernova is expected to be $\approx 10^{51}$ ergs \citep[see][]{Jones_1998}, we find that a mild reduction to $0.8 \times 10^{51}$ ergs provides a better fit, while still being consistent with observations \citep{Rubin_2016, Kozyreva_2025}. Further, the resulting values for $\alpha_{\rm SF}$ and $\alpha_{\rm SF, burst}$ suggest higher star formation efficiency, in agreement with past observations \citep{Stefanon_2021}, abundance matching estimates \citep{Tacchella_2018} and models \citep{Sun_2016, Sipple_2024}. 
The mild reduction of $k_{\rm AGN}$ with respect to the values in \citet{Henriques_2020} leads to lower gas accretion by AGNs, allowing for more star formation and thus more galaxies at the brightest magnitudes ($M_{\rm 1600, AB} < -20$), which is necessary to match recent observations. Similarly, the small differences in the supernova feedback-based heating efficiency parameters $\epsilon_{\rm reheat}$ and $V_{\rm reheat}$ are required to improve the agreement for $M_{\rm 1600, AB} > -19$. Indeed, as supernovae can effectively disrupt smaller galaxies, reducing the impact of their heating efficiency allows for more cold gas to form, and thus more star formation, further increasing the UVLF at the low brightness end and providing a closer match with observations.

We have also varied other parameters affecting star formation ($\beta_{\rm SF, burst}$), the heating of cold gas by stars and supernova feedback (such as $\beta_{\rm reheat}$, $\eta_{\rm eject}$, $V_{\rm eject}$, $\beta_{\rm eject}$, $\gamma_{\rm reinc}$), black hole properties ($f_{\rm BH}$, $V_{\rm BH}$), and galaxy mergers ($M_{\rm r.p.}$, $\alpha_{\rm dyn.fric.}$), but their impact on the UVLFs is minimal \citepalias[see][for a detailed analysis of the impact of different parameters]{Ma_2023}. For such parameters, we fix their value to that obtained by \citet{Henriques_2020}. In future work, we will broaden the range of astrophysical parameters and analyse their effect on other observables. Additionally, we also plan to employ more rigorous MCMC-based techniques for fitting parameters \citep[as shown in ][ submitted]{Ma_2025}.

So far, we only considered the fiducial model. 
While the values of the parameters obtained may be sensitive to the method used to derive them (in this case, the least squares minimization), the qualitative difference in their values for the different cosmological models introduced in Section~\ref{sec:dmsims} can still be explored. Thus, we consider two cases. In the first one, the same {\sc L-Galaxies} parameters of the fiducial model above are adopted for all $N$-body simulations to investigate the impact of different cosmologies on the 21-cm signal independently from observations of galaxies at high redshifts. We refer to this model as ``unconstrained''. This case enables us to gain a clearer understanding of the impact of changing cosmologies on the 21-cm signal, without astrophysical processes potentially cancelling out their effects on observables. However, not changing the astrophysical parameters
for the different cosmological models may produce UVLFs that are inconsistent with observations. Thus, we also consider another scenario, called the ``constrained'' case. Here, the {\sc L-Galaxies} parameters (specifically $\alpha_{\rm SF}$, $\alpha_{\rm SF, burst}$, and $E_{\rm SN}$) are changed for each cosmological model to match the observed UVLFs. As the average energy released per supernova is the most sensitive parameter in scaling the UVLF while maintaining its overall shape, we first focus on tuning it. For the $h$ high model, we find that fixing it at $10^{51}$ ergs and using the fiducial model's values for $\alpha_{\rm SF}$ and $\alpha_{\rm SF, burst}$ is enough to get the best fit via the least squares minimisation described above. The increase in $E_{\rm SN}$ as compared to the fiducial model is required to reduce the amount of cold gas in the galaxy available for star formation, balancing the {increased} clustering of matter due to the higher Hubble constant. Similarly, due to even stronger matter clustering due to a larger $\sigma_8$ in the $\sigma_8$ high model, we note that a value of $2 \times 10^{51}$ ergs for $E_{\rm SN}$ provides the best fit. Because of this, we expect that just reducing $E_{\rm SN}$ for the $\sigma_8$ low model to improve star formation efficiency is enough despite the lower matter clustering. However, we note that in addition to reducing the average supernova energy to $0.15 \times 10^{51}$ ergs, it is also {necessary} to increase the parameters regulating the star formation efficiency, $\alpha_{\rm SF}$, and the burstiness of star formation efficiency during mergers, $\alpha_{\rm SF, burst}$. Further raising the values {beyond 0.50 for $\alpha_{\rm SF}$ and 0.90 for $\alpha_{\rm SF, burst}$} does not induce any appreciable difference in the sum of the squares of the residuals. 

These differences in the astrophysical parameters for various cosmologies are potentially more interesting, as they allow for a joint constraint of astrophysical and cosmological parameters based on multi-frequency observations. The list of all the {\sc L-Galaxies} parameters adopted for the four models is provided in Table~\ref{table:LGal_all}.

We show the UVLFs for the unconstrained (top) and constrained (bottom) case in Figure~\ref{fig_uvlf_combined} for all cosmological models, along with observations at $z = 10$ and 9. In the unconstrained case, we note that the fiducial and $h$ high models have similar UVLFs, while the $\sigma_8$ low model underpredicts the UVLF as compared to observations. On the other hand, the UVLF in the $\sigma_8$ high model is mildly overpredicted. These results are in agreement with the impact that the parameters are expected to have on matter clustering, as discussed above. In the constrained case, the boost in star formation due to higher $\alpha_{\rm SF}$ and $\alpha_{\rm SF, burst}$ values and a lower $E_{\rm SN}$ value, allows the $\sigma_8$ low model to match the bright end of the UVLF. However, due to lower matter clustering, it does not produce as many faint galaxies as the other cases, and sees a steep drop at $M_{\rm 1600, AB} > -18$. On the other hand, the high $E_{\rm SN}$ in the $\sigma_8$ high model strongly suppresses star formation and thus matches other cosmological models, as well as JWST and HST observations. Additionally, it produces more faint galaxies even at  $M_{\rm 1600, AB} > -16$. Between the $h$ high and fiducial models, the difference is minimal at $z = 10$ due to the mildly higher $E_{\rm SN}$ in the $h$ high model. However, at $z = 9$ the impact of higher matter clustering in the $h$ high model shows up in the form of more galaxies at the low luminosity end.

Lastly, we note that in the fiducial model the brightest galaxy at $z=14$ has $M_{\rm 1600, AB} = -20.36$, while at $z =13$ it has $M_{\rm 1600, AB} = -20.60$, which is in agreement with the recent spectroscopically confirmed galaxies at $z = 14.32$ and $13.90$ \citep{Carniani_2024}. For completeness,  in Appendix~\ref{appendix_uvlf} we also show the UVLFs in the range $12 > z > 5$ for the constrained case. We find that our models reasonably agree with UVLF observations across this redshift range.

\subsection{Reionization history}\label{sec:reiohist}

\begin{figure*}
\centering
\includegraphics[width=2\columnwidth,keepaspectratio]{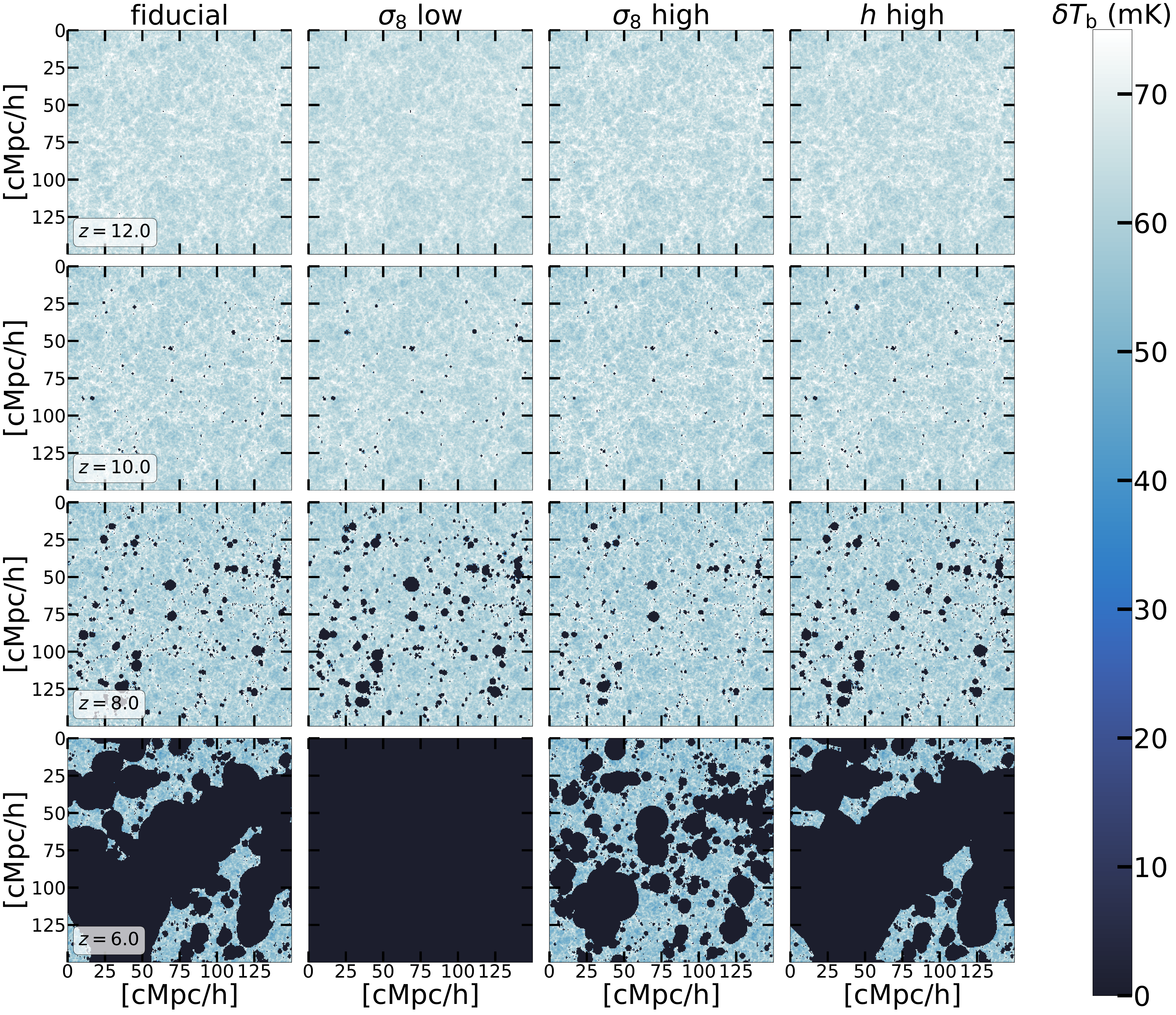}
\caption{Maps of $\delta T_{\rm b}$ of the A-series middle slices of single cell thickness (i.e., $\approx 600~h^{-1}~\rm ckpc$) for the four cosmological models in the constrained case at $z = 12, 10, 8,~\rm and~6$ (from top to bottom). Here the dark areas represent the ionized regions with $\delta T_{\rm b} = 0$. 
Note that $\delta T_{\rm b}$ cannot have negative values due to the assumption of $T_{\rm S} \gg T_{\rm CMB}$.
} 
\label{fig_midslice_fconst}
\end{figure*}

\begin{figure}
\centering
\includegraphics[width=\columnwidth,keepaspectratio]{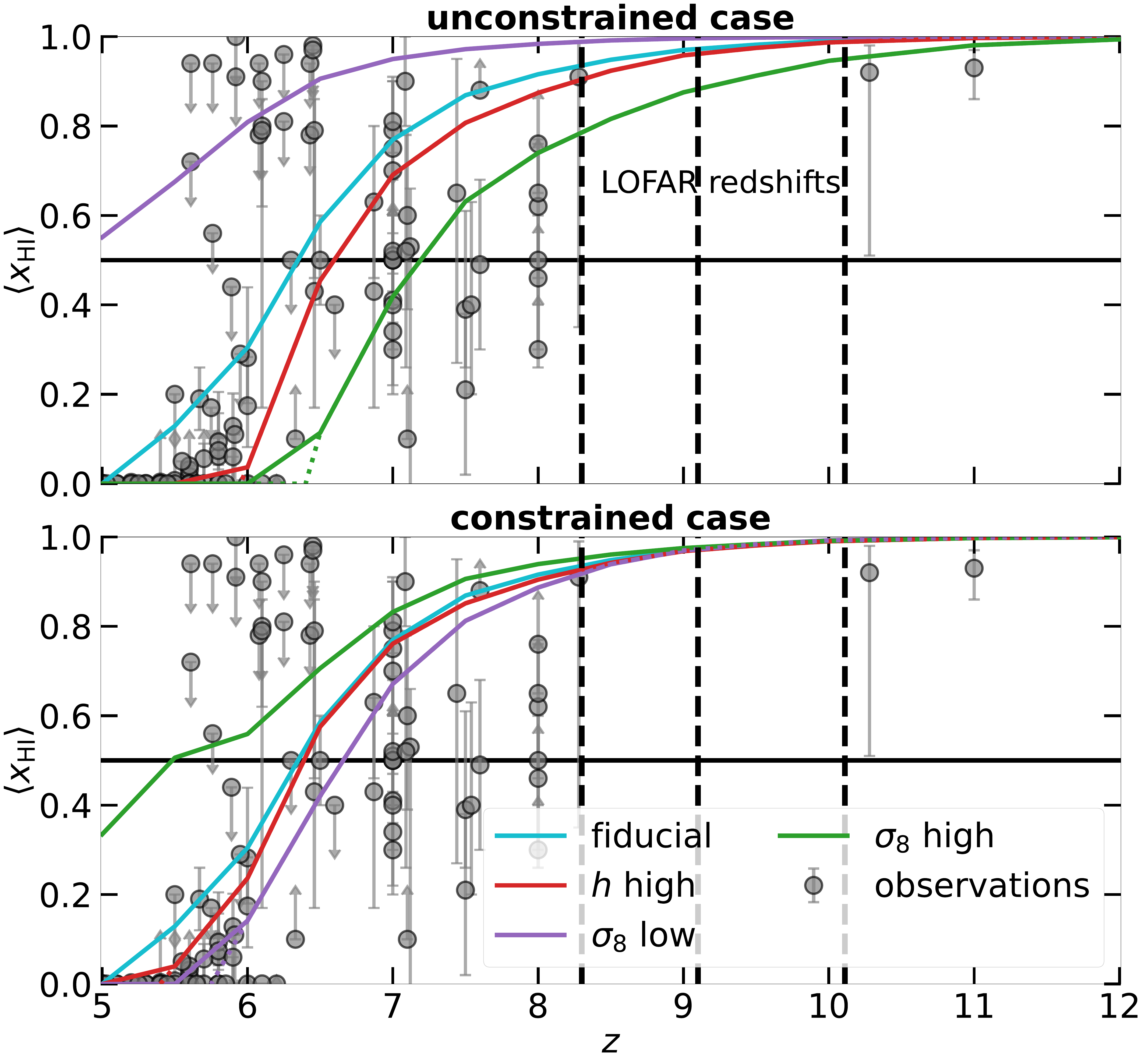}
\caption{Redshift evolution of the average of the A and B-series volume-averaged neutral hydrogen fraction $\langle x_{\rm HI} \rangle$  for the fiducial (cyan), $h$ high (red), $\sigma_8$ low (purple) and $\sigma_8$ high (green) models for the unconstrained (top) and constrained (bottom) cases. Dotted lines refer to a fit to the curves, which is used for a better estimate of the redshift of reionization when the redshift resolution is too coarse.
The vertical grey dashed lines indicate the redshifts observationally relevant for LOFAR ($z = 10.11$, 9.16 and 8.3), and the black solid line is drawn at $\langle x_{\rm HI} \rangle = 0.5$ to guide the eye. Grey circles are a collection of observational constraints \citep[][collected in the {\sc CoReCon} module, \citealt{Garaldi_2023}]{Fan_2006b, Totani_2006, Ota_2008, Ouchi_2010, Bolton_2011, Dijkstra_2011, McGreer_2011, Mortlock_2011, Ono_2012, Chornock_2013, Jensen_2013, Robertson_2013, Schroeder_2013, Pentericci_2014, Schenker_2014, McGreer_2015, Sobacchi_2015, Choudhury_2015, Mesinger_2015, Greig_2017, Davies_2018, Mason_2018,  Hoag_2019, Greig_2019, Jones_2024}.
} 
\label{fig_xhi_both}
\end{figure}

Before discussing the reionization history in more detail, we note that the escape fraction of UV photons had been set to $f_{\rm esc} = 12.5 \%$ in order for the fiducial model to reionize completely (i.e. the average neutral fraction $\langle x_{\rm HI} \rangle$ becomes zero) by $z = 5$. For the other models, though, full reionization is reached at different redshifts. Further, we note that this is a simplified assumption, as $f_{\rm esc}$ is expected to vary with galaxy and halo properties \citep[as shown by][]{Kostyuk_2023, Ma_2020fire} and maybe with redshift (while e.g. \citealt{Ma_2020fire} and \citealt{Kostyuk_2025} find redshift variation, \citealt{Gorce_2018} favours $f_{\rm esc}$ to be constant with redshift), and thus can be strongly affected by modelling choices and approximations. Nevertheless, our choice aligns with recent estimates from observations \citep[][which report median values of 10-12\%]{Mascia_2023, Bosman_2024, Jaskot_2024}. In the future, we plan to consider $f_{\rm esc}$ as a free parameter as well, and, as a reference, in Appendix~\ref{appendix_highfesc} we show the reionization history of the fiducial model with $f_{\rm esc} = 25 \%$. This is of particular interest, as observations like \citet{Jaskot_2024} have reported higher escape fractions ($f_{\rm esc} > 20 \%$) for $\geq 33\%$ of observed galaxies.

For a qualitative view of the reionization history, in Figure~\ref{fig_midslice_fconst} we show the A-series (i.e. with ``fixed'' initial conditions in the F\&P pair of simulations) middle slices of the $\delta T_{\rm b}$ cubes for the four cosmological models in the constrained case at $z = 12, 10, 8~\rm and~6$. For completeness, in Appendix~\ref{appendix_pairconstrained_midslice} we also show the corresponding B-series slices, and in Appendix~\ref{appendix_bothunconstrained_midslice} both the A and B-series middle slices for the unconstrained case.
We note that there are no major differences until $z = 8$, when more and larger ionized regions appear in the $\sigma_8$ low model, while fewer and smaller ones are present in the $\sigma_8$ high model. This leads to the $\sigma_8$ low model being almost completely ionized by $z = 6$ (note that while in Figure~\ref{fig_midslice_fconst}, the shown slice at $z = 6$ happens to be fully ionized, there are still some neutral regions in the full simulation box. The same is shown below by computing the volume-averaged neutral hydrogen fraction), while the $\sigma_8$ high model still has large neutral regions. This is because in these cases the choice of the parameters controlling the astrophysical processes at high redshift is balanced by the impact on the matter clustering due to the choice of $\sigma_8$, but at lower redshift, once enough halos have formed, the astrophysical processes dominate. In particular, a higher (lower) average energy per supernova in the $\sigma_8$ high (low) case significantly impedes (bolsters) star formation as too much cold gas is blown away (collapsed). However, the differences between the astrophysical parameters of the $h$ high and the fiducial model are smaller, so the differences in the size of the ionized regions from $z = 10$ are mainly driven by matter clustering. Thus, overall, the choice of both astrophysical and cosmological parameters leads to significantly different speeds at which reionization happens across all four models. Indeed, despite the relatively similar maps observed at $z=8$, by $z=6$ the differences are much more evident.

In Figure~\ref{fig_xhi_both}, we present the redshift evolution of the average of the A and B-series volume-averaged neutral hydrogen fraction, $\langle x_{\rm HI} \rangle$, in the unconstrained (top) and constrained (bottom) cases, together with observational constraints. As in some models the redshift of reionization is not properly captured due to the coarse redshift resolution\footnote{While we have a large number of snapshots as discussed in Section~\ref{sec:dmsims} for the sake of simplicity, to run {\sc Polar} we adopt a uniform step size of $\Delta z = 0.5$.}, we additionally provide a fit to these curves (shown as a dotted line) to estimate more accurately the end of reionization. We note that in the unconstrained case, all models reionize at different redshifts.  Reionization is the fastest (at $z=6.4$) in the $\sigma_8$ high model, as structure formation happens earlier. The $\sigma_8$ low model is still only 45\% reionized by $z=5$ because of the lack of sources, and shows the poorest agreement with observational constraints. In the constrained case, the differences are reduced, but still significant. In particular, we note that the $\sigma_8$ low model reionizes the fastest at $z = 5.75$, while the $\sigma_8$ high model is only $\approx 65\%$ ionized by $z = 5$. This reversal of reionization histories with respect to the unconstrained case is due to the impact of different astrophysical parameters. Specifically, the significantly higher $E_{\rm SN}$ in the $\sigma_8$ high model blows away gas and thus suppresses star formation, which consequently reduces the production of ionizing photons. On the other hand, the higher $\alpha_{\rm SF}$ and $\alpha_{\rm SF, burst}$ in the $\sigma_8$ low model along with the lower $E_{\rm SN}$ significantly boosts star formation.

As the four models reionize at different redshifts and the end of the EoR can be observationally constrained  (e.g., detections of Gunn-Peterson troughs in the Lyman-$\alpha$ forest as shown in \citealt{Becker_2015}, \citealt{Qin_2021} and \citealt{Bosman_2022}), this in turn limits the possible choice of astrophysical and cosmological parameters. We note that to model more realistically the final phases of reionization, the unresolved Lyman limit systems which govern absorption during the final stages of the reionization process should be accounted for  \citep[see][]{Georgiev_2024, Giri_2024}. These, though, are not included here. 

Finally, the four models produce a  Thomson scattering optical depth in the range (0.041-0.067) and (0.048- 0.060) for the unconstrained and constrained case respectively, assuming an Helium (He) II fraction equal to the HII fraction, and an instantaneous  HeII reionization at $z\,=\,3$. These numbers are in agreement with \textit{Planck} observations \citep{Planck_2020, deBelsunce_2021, Giare_2024}.

\section{Results}\label{sec:results}
In this section, we present the results from our simulations with respect to various galactic and IGM properties for the unconstrained and constrained case.  To reduce the effect of cosmic variance, we always consider the A and B-series averages when comparing to observations.

\subsection{Unconstrained case}\label{sec:uncon_results}

In the unconstrained case, we discuss the galactic and IGM observables for models with the UVLFs reported in the top panel of Figure~\ref{fig_uvlf_combined}. 

\subsubsection{Galactic properties}\label{sec:gal_uncon}

\begin{figure}
    \centering
    
    \begin{subfigure}{0.49\textwidth}
        \centering
        \includegraphics[width=\textwidth,keepaspectratio]{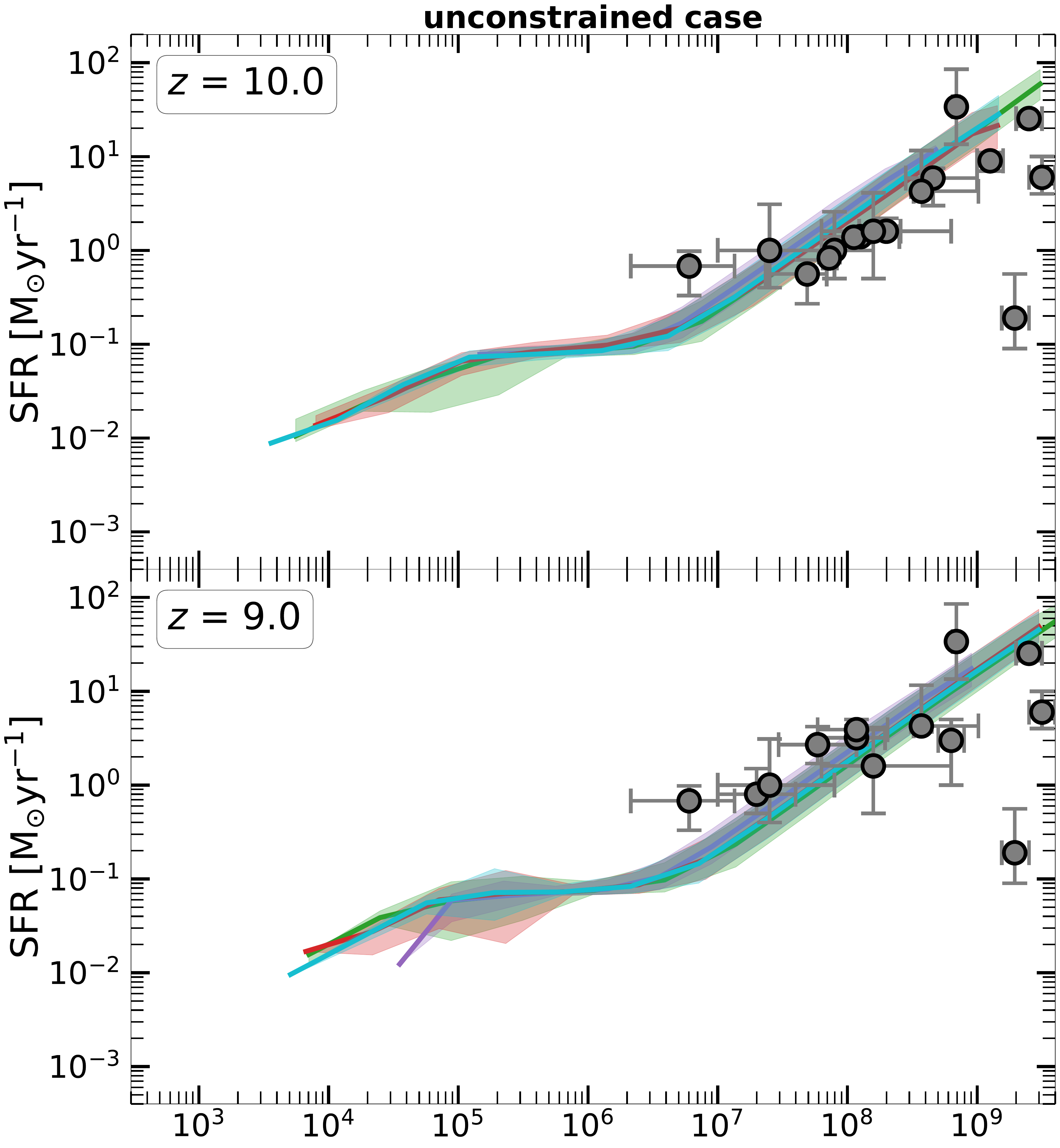}
    \end{subfigure}
    \hfill
    \begin{subfigure}{0.49\textwidth}
        \centering
        \includegraphics[width=\textwidth,keepaspectratio]{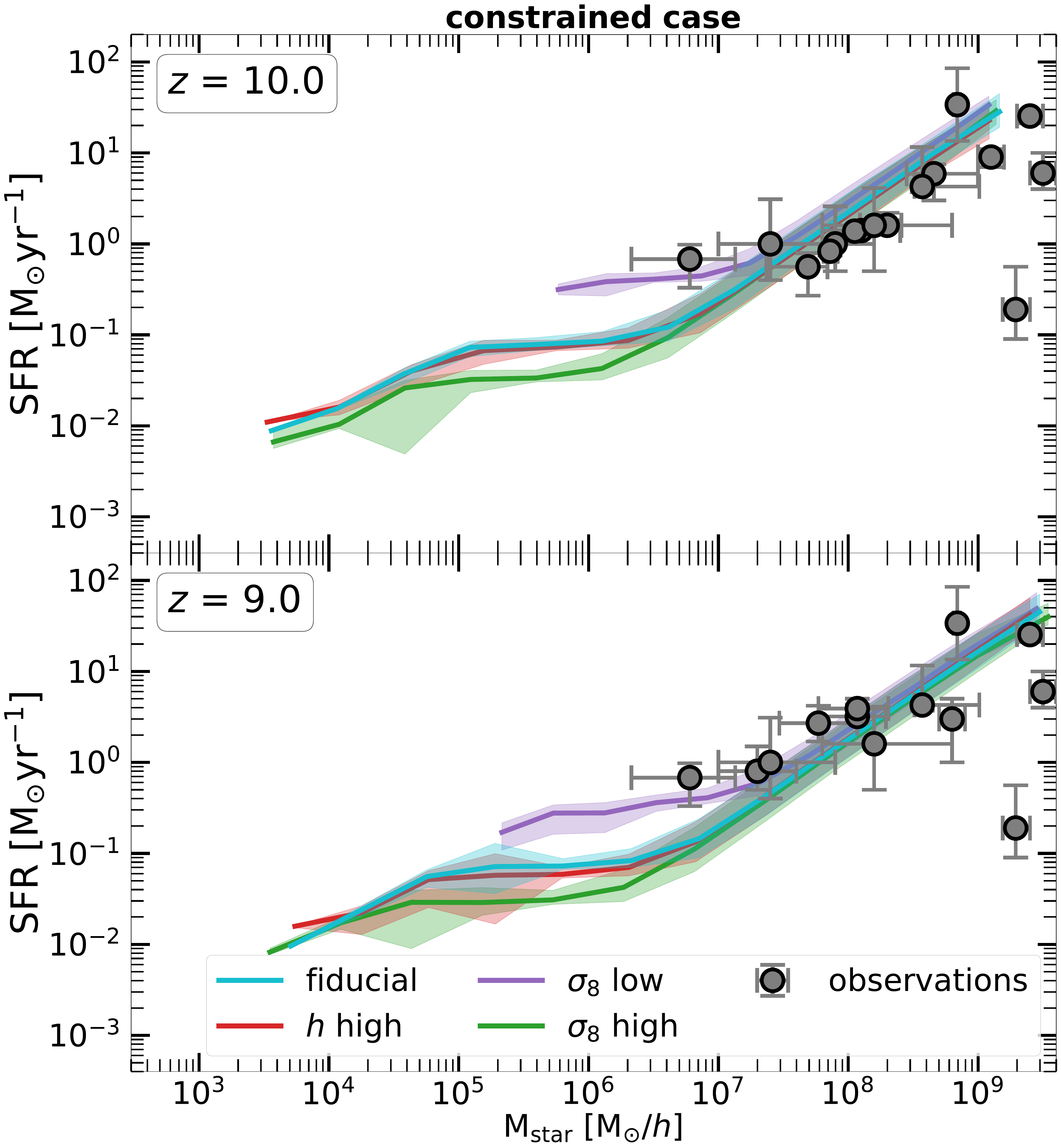}
    \end{subfigure}
    
    \caption{A and B-series averaged star formation rate (SFR) as a function of stellar mass for the unconstrained (top) and constrained (bottom) cases, for the fiducial (cyan), $h$ high (red), $\sigma_8$ low (purple) and $\sigma_8$ high (green) models at  $z = 10$ and 9. Solid lines refer to the median star formation rate in each mass bin, with 16th and 84th percentiles shown as shaded regions. Grey circles are observations from various JWST programs \citep{Treu_2023,Fujimoto_2023,Looser_2023,Bouwens_2023b,Papovich_2023,Arrabal_2023,Arrabal_2023b,Long_2023,Leethochawalit_2023,Atek_2023,Robertson_2023,Heintz_2023,Heintz_2023b,Jin_2023,Helton_2024,Jung_2024}.
    }
    \label{fig_sfr_combined}
\end{figure}

In Figure~\ref{fig_sfr_combined}, we present the mass-binned star formation rate (SFR) of galaxies in the four models at $z = 10$ and 9, with the top panel showing the unconstrained case. The solid lines are the median SFR, with the shaded regions referring to the 16th to 84th percentiles. The star formation main sequence (SFMS) in each case is additionally compared with results from various recent JWST programs represented by grey circles with a binning of $\pm 0.25$ around the redshift. As all parameters affecting the star formation rate were kept the same across all four models, we note that the overall trends are similar, and agree with observations. The impact of matter clustering is seen at the low-mass end, where a larger clustering allows for the formation of a statistically significant sample of smaller mass galaxies in the $\sigma_8$ high and $h$ high models, while the smallest mass galaxies formed in the $\sigma_8$ low case are an order of magnitude more massive. Further, at the high mass end, we note that the $\sigma_8$ high model produces the most massive galaxies, which also have the highest SFR.

\begin{figure}
    \centering
    
    \begin{subfigure}{0.49\textwidth}
        \centering
        \includegraphics[width=\columnwidth,keepaspectratio]{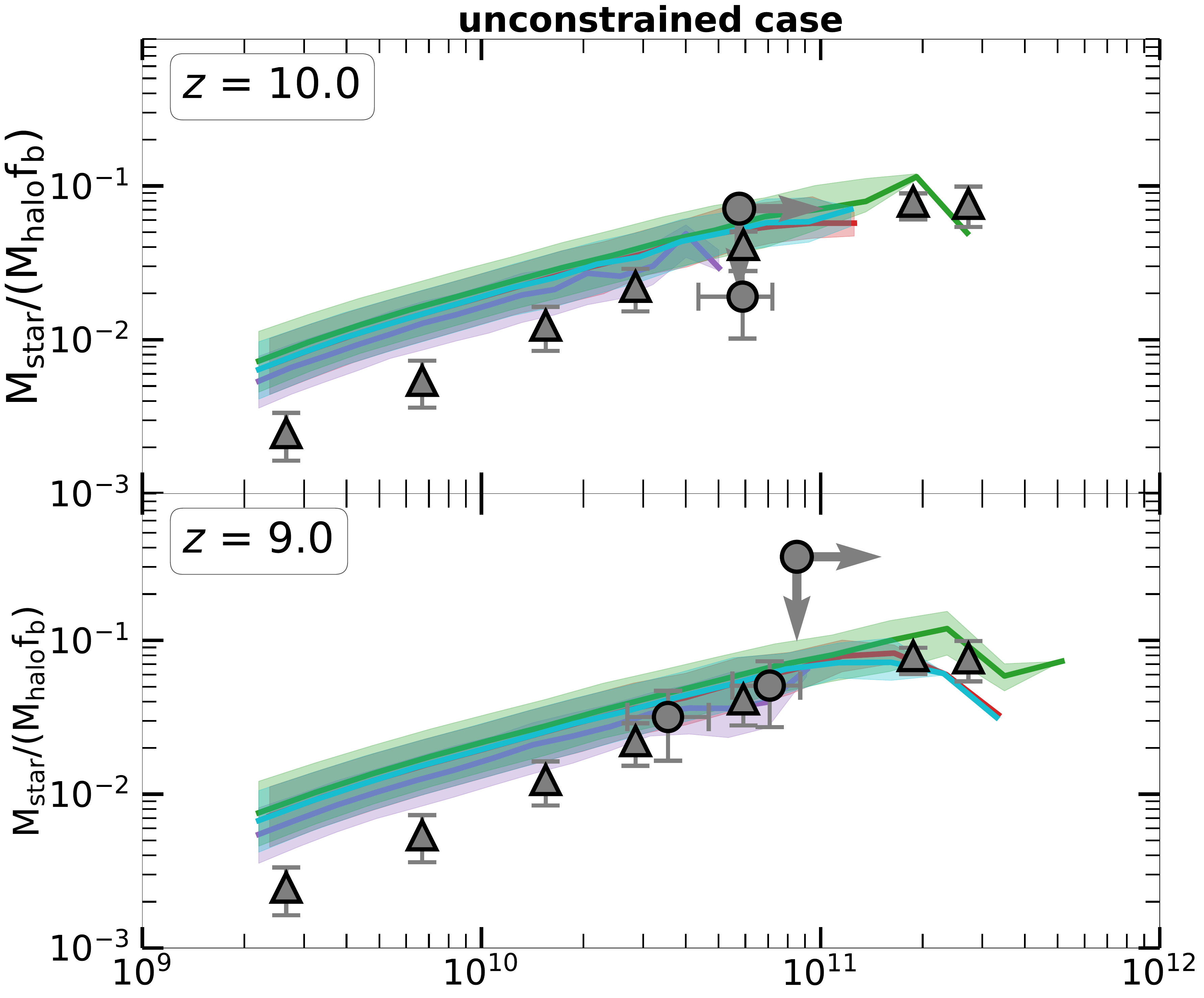}
    \end{subfigure}
    \hfill
    \begin{subfigure}{0.49\textwidth}
        \centering
        \includegraphics[width=\columnwidth,keepaspectratio]{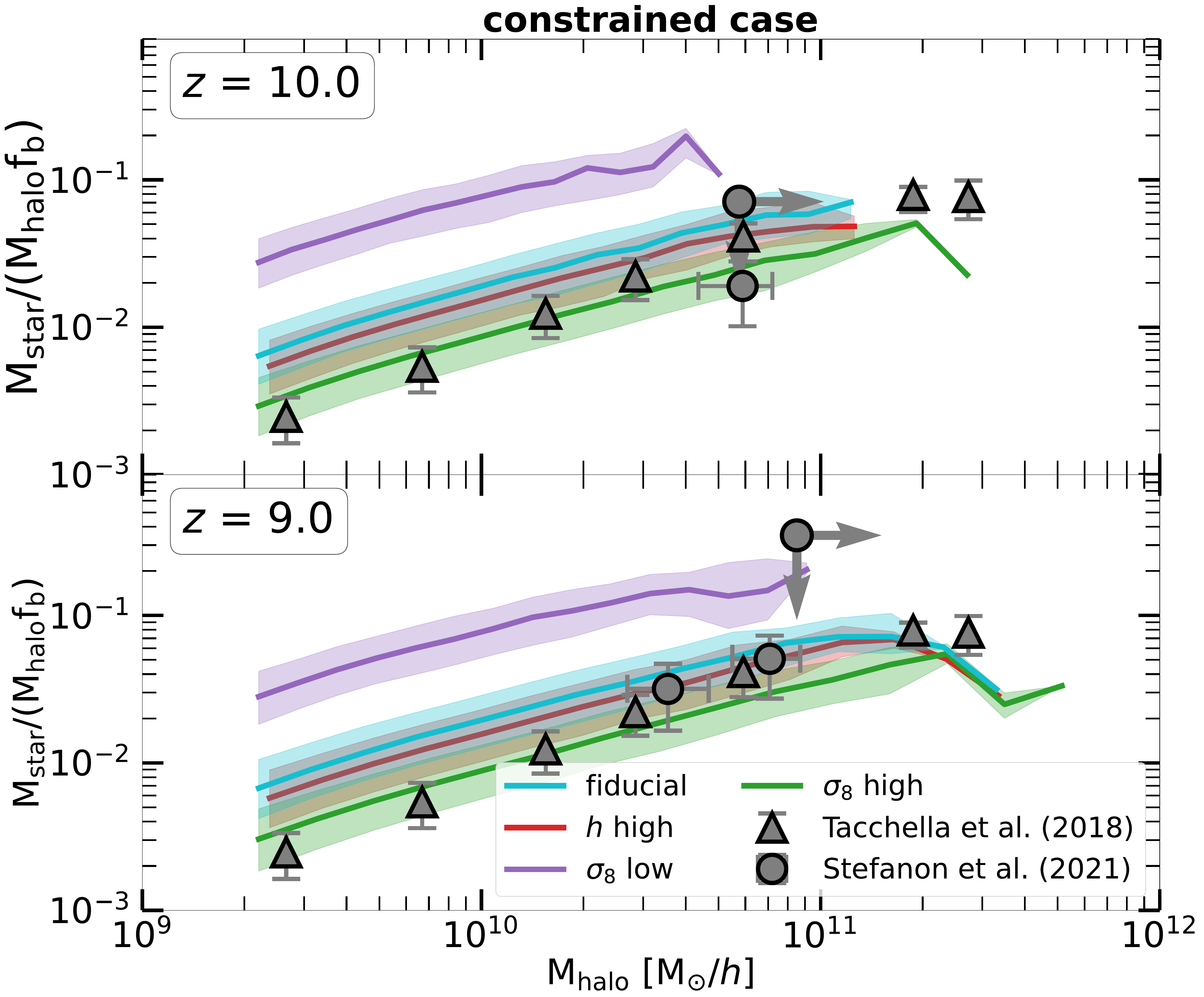}
    \end{subfigure}
    \caption{A and B-series averaged stellar fraction (which serves as a proxy for the star formation efficiency) as a function of halo mass for the unconstrained (top) and constrained (bottom) cases, for the fiducial (cyan), $h$ high (red), $\sigma_8$ low (purple) and $\sigma_8$ high (green) models at $z = 10$ and 9. Solid lines refer to the median stellar fraction in each mass bin, with 16$^{\rm th}$ and 84$^{\rm th}$ percentiles shown as shaded regions. Grey circles are Spitzer observations from \citet{Stefanon_2021}, and grey triangles are abundance matching estimates from \citet{Tacchella_2018}.
    }
    \label{fig_sfe_combined}
\end{figure}
To confirm the agreement in star formation rates, in Figure~\ref{fig_sfe_combined} we additionally look at the stellar fraction, which serves as a proxy for the global star formation efficiency (SFE) at the same redshifts with the top panel showing the unconstrained case. We define the stellar fraction as the ratio of the stellar and halo mass scaled by $f_{\rm b} = \Omega_{\rm b}/\Omega_{\rm m}$, i.e. ${\rm M}_{\rm star}/{\rm M}_{\rm halo} f_{\rm b}$, as a function of the halo mass ${\rm M}_{\rm halo}$. We also compare this to Spitzer observations at $z = 7$ \citep{Stefanon_2021} and to abundance matching estimates from \citet{Tacchella_2018}. As mentioned above, the parameters controlling the star formation efficiency are the same across all models, and thus the corresponding stellar fraction is the same as well. The only difference is in the largest galaxies formed, which is directly proportional to $\sigma_8$ as it controls the clustering of mass. We note that the stellar fraction is marginally higher than the abundance matching estimates for the lower mass halos, but the difference is $\lesssim 0.3$ dex. Further, the weak mass dependence of their trend is in agreement with zoom-in hydrodynamical simulations like FIREbox$^{\rm HR}$ \citep{Feldmann_2024} and {\sc ColdSIM} \citep{Maio_2023}.

\subsubsection{21-cm signal from the IGM}\label{sec:igm_uncon}

\begin{figure}
    \centering
    
    \begin{subfigure}{0.49\textwidth}
        \centering
        \includegraphics[width=\columnwidth,keepaspectratio]{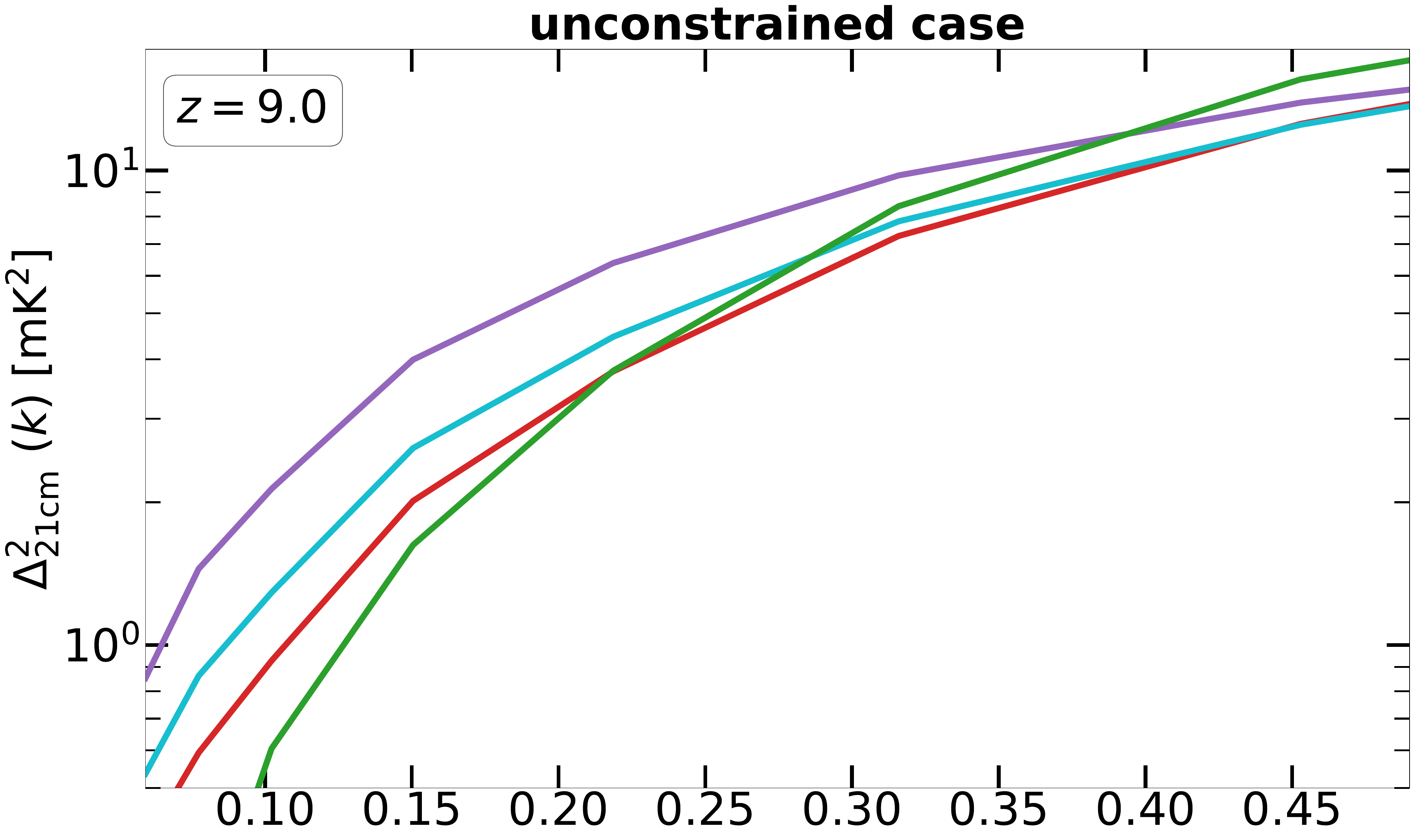}
    \end{subfigure}
    \hfill
    \begin{subfigure}{0.49\textwidth}
        \centering
        \includegraphics[width=\columnwidth,keepaspectratio]{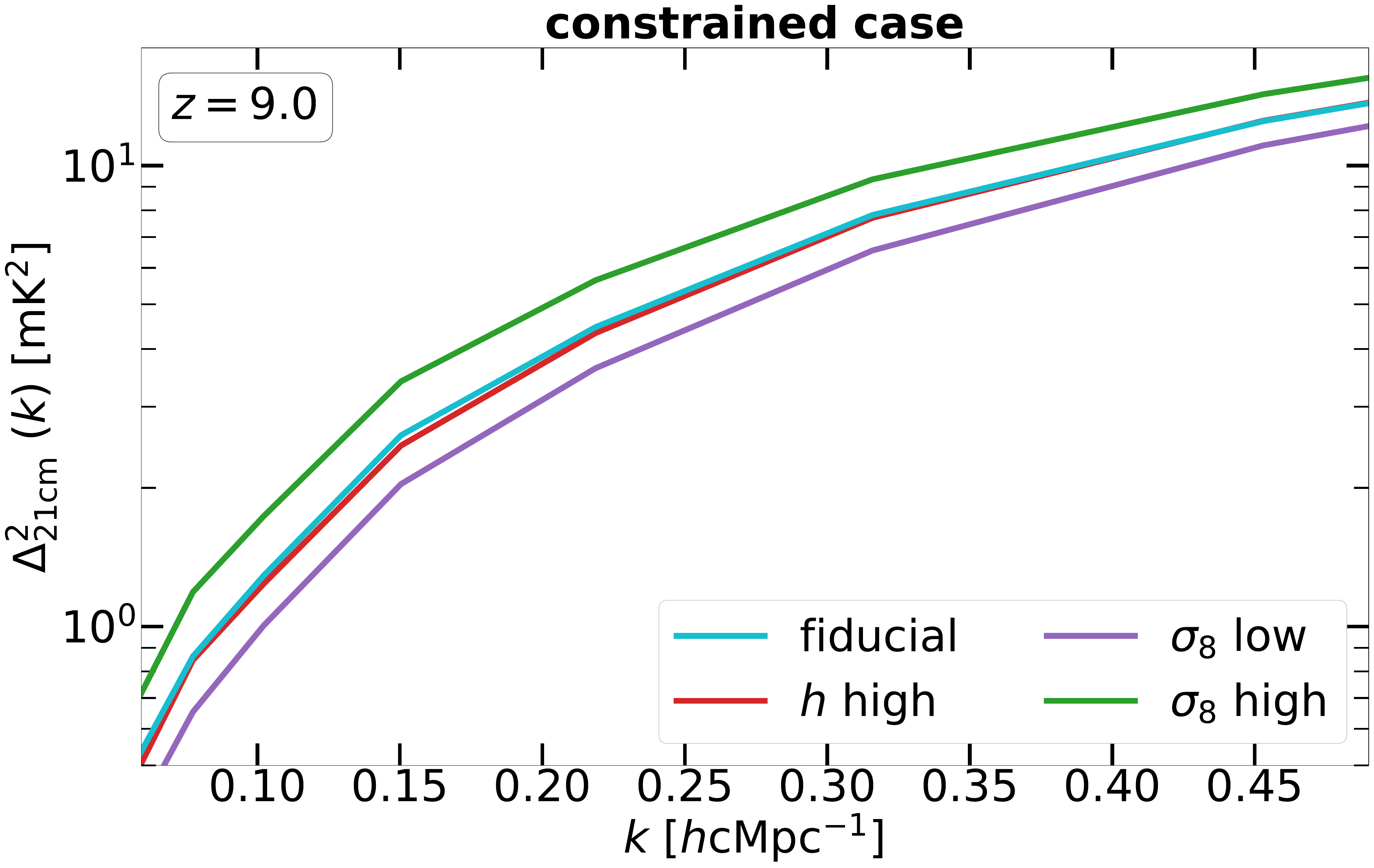}
    \end{subfigure}
    \caption{A and B-series averaged normalized 21-cm signal power spectrum ($\Delta^{2}_{\rm 21cm}$) for the unconstrained (top) and constrained (bottom) cases, for the fiducial (cyan), $h$ high (red), $\sigma_8$ low (purple) and $\sigma_8$ high (green) models at $z = 9$.}
    \label{fig_ps_combined}
\end{figure}

We note that despite the disagreement in UVLFs for the four cosmological models shown in the top panel of Figure~\ref{fig_uvlf_combined}, the models still reproduce fairly well observable quantities such as the SFR and the stellar fraction, as shown in Section~\ref{sec:gal_uncon}. Further, they are broadly in agreement with available data. 
We now look at IGM observables, and more specifically at the 21-cm signal power spectrum (calculated according to Equation~\ref{eq:psnorm}), shown in the top panel of Figure~\ref{fig_ps_combined} for the unconstrained case at $z = 9$. In particular, we focus on the $0.05 < k < 0.5 h \rm cMpc^{-1}$ regime, which is where the LOFAR telescope is the most sensitive. At redshift $z = 9$, we note from the top panel of Figure~\ref{fig_xhi_both} that the matter clustering primarily governs the reionization history. Thus, the $\sigma_8$ low model having the highest $\langle x_{\rm HI} \rangle$ suggests that the power amplitude up to $k = 0.40~h \rm cMpc^{-1}$ for it is higher than the other three models, as it is primarily driven by the matter power spectrum. Additionally, the power at these scales is further boosted, because while $\langle x_{\rm HI} \rangle \approx 1$, there are still voxels where it is $\ll1$ as they are included in ionized regions. These ionized regions in the $\sigma_8$ low model are small and few enough to lead to higher contrast on average in the differential brightness temperature across all scales greater than the size of the ionized bubbles. In other words, as matter clustering translates into the neutral hydrogen fraction term of Equation~\ref{eq:dtb} as well as the overdensity term, in a scenario where the ionized regions are not large and numerous enough to wash out contrast, it leads to a marginal boost in power.

The lower $\langle x_{\rm HI} \rangle$ in the other three models suggests suppression of the power spectrum for the 21-cm signal as compared to the matter power spectrum due to more numerous and larger ionized regions, especially in the $\sigma_8$ high and $h$ high models. This is due to the washing out of the contrast driven purely by the large-scale structure. However, the more numerous ionized regions present at smaller scales induce on average a greater contrast, and thus the $h$ high model reaches values similar to those of the fiducial model, and the $\sigma_8$ high model has the highest power overall at $k > 0.40~h \rm cMpc^{-1}$.

\begin{figure}
    \centering
    \begin{subfigure}{0.49\textwidth}
        \centering
        \includegraphics[width=\columnwidth,keepaspectratio]{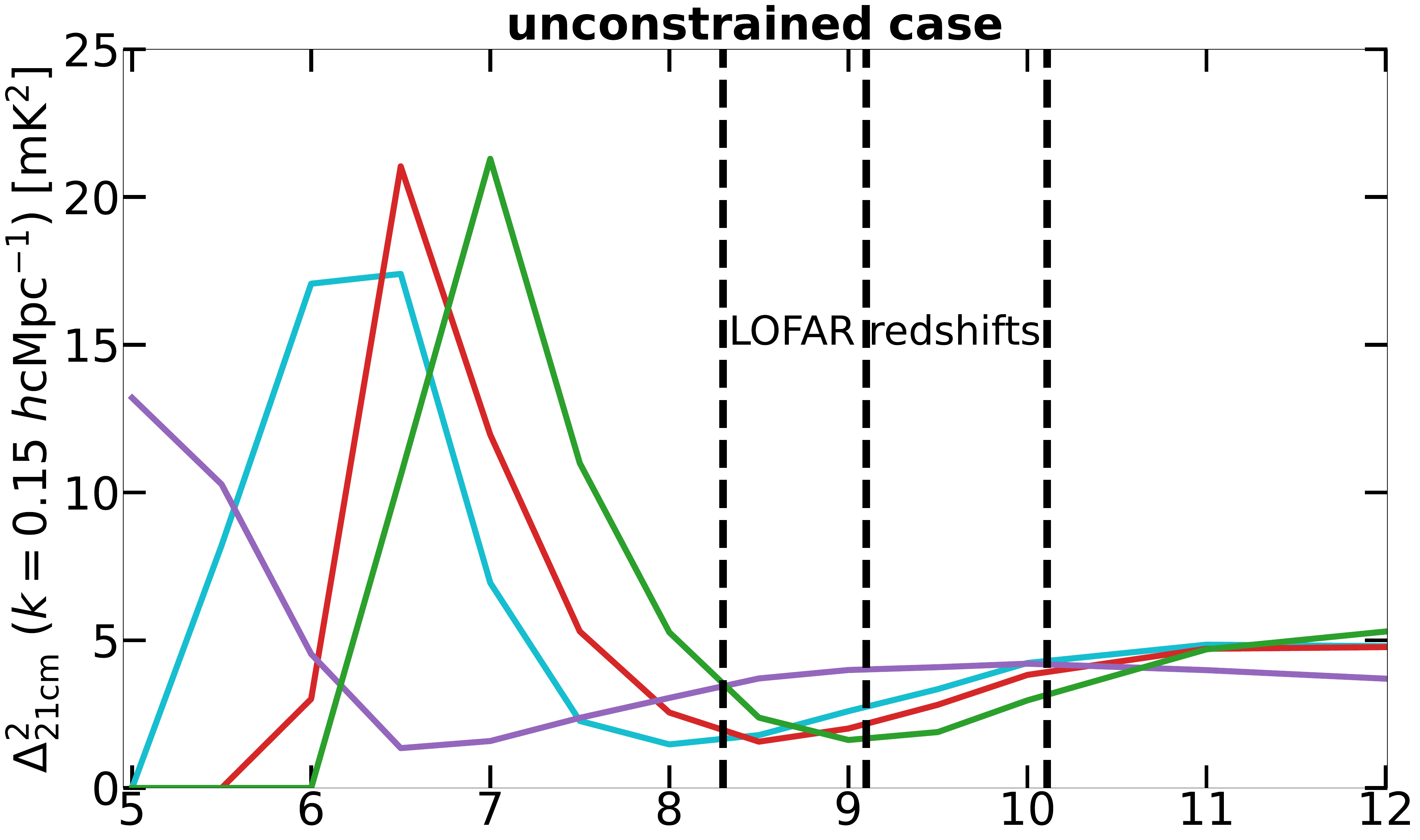}
    \end{subfigure}
    \hfill
    \begin{subfigure}{0.49\textwidth}
        \centering
        \includegraphics[width=\columnwidth,keepaspectratio]{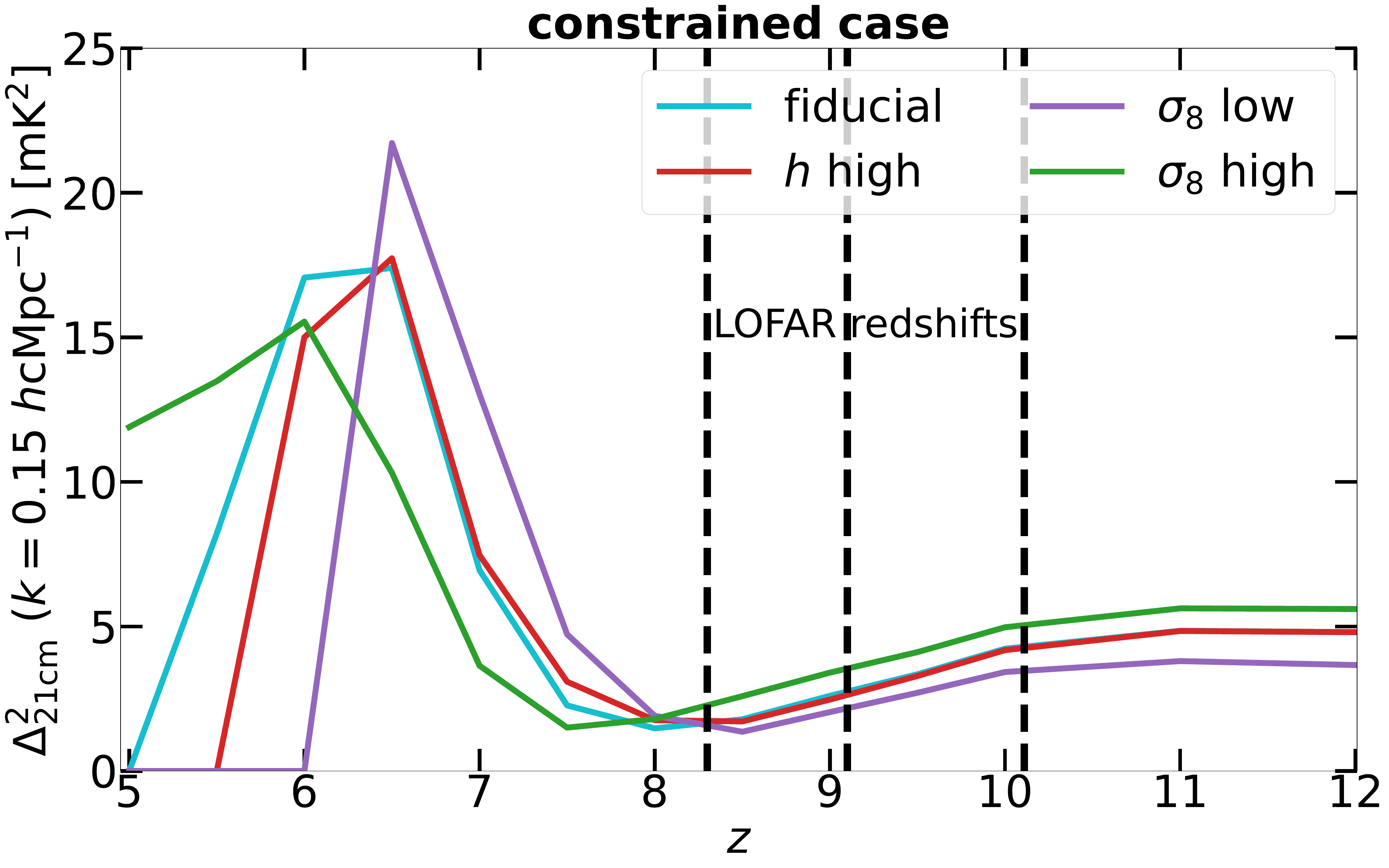}
    \end{subfigure}
    \caption{Redshift evolution of the A and B-series averaged normalized 21-cm signal power spectrum ($\Delta^{2}_{\rm 21cm}$) at $k = 0.15~h\rm cMpc^{-1}$  for the unconstrained (top) and constrained (bottom) cases, for the fiducial (cyan), $h$ high (red), $\sigma_8$ low (purple) and $\sigma_8$ high (green) models. The vertical dashed lines indicate the redshifts relevant for LOFAR, i.e. $z = 10.11$, 9.16, and 8.3.}
    \label{fig_ps0.1_combined}
\end{figure}

For a better understanding of the 21-cm signal power spectrum evolution, in Figure~\ref{fig_ps0.1_combined} we additionally present its redshift dependence at large scales, i.e. $k = 0.15~h\rm cMpc^{-1}$, with the top panel showing the unconstrained case. This allows us to compare how the 21-cm signal evolves in the four cosmological models across redshifts relevant to LOFAR observations. As suggested by the reionization histories, we note that the $\sigma_8$ low model has a larger contrast due to slower reionization, and thus has a higher amplitude of $\Delta^{2}_{\rm 21cm}~(k = 0.15~h\rm cMpc^{-1})$ at the beginning of reionization. However, the $\sigma_8$ high and $h$ high models reach a higher amplitude  by $z \approx 8$ due to the formation of larger ionized regions.

\subsection{Constrained case}\label{sec:con_results}
As in this case we have also changed parameters that control star formation efficiency and supernova feedback, it is crucial to understand their impact on observables other than the UVLFs shown in the bottom panel of Figure~\ref{fig_uvlf_combined}.

\subsubsection{Galactic properties}\label{sec:gal_con}

As in Section~\ref{sec:gal_uncon}, in the bottom panel of Figure~\ref{fig_sfr_combined} we look at the A and B-series averaged SFMS and compare them with observations from JWST at  $ z = 10$ and 9 for the constrained case. We note that despite differences in the energy of supernovae ($\alpha_{\rm SF}$ and $\alpha_{\rm SF, burst}$ also differ for the $\sigma_8$ low model), the high mass ends of the SFMS of the four models still agree with each other and with observations. 
This is because the most massive galaxies are also the most luminous, and matching their UVLFs in Section~\ref{sec:compareobs} consequently matches their star formation properties. However, at the low-mass end, the SFMS of the four models shows some deviation at both redshifts. This is because at lower masses, the impact of astrophysical parameters starts to dominate, and while the $\sigma_8$ high model forms galaxies with masses smaller than the $\sigma_8$ low model, the median SFR of these low mass galaxies is an order of magnitude smaller than the lowest mass galaxies in the $\sigma_8$ low model, because the impact of the energy released per supernova is sufficient to clear out gas and suppress star formation in the low-mass galaxies.

To further explore the impact of the energy released by supernovae, in the bottom panel of Figure~\ref{fig_sfe_combined}, we also look at the global SFE for the constrained case by again considering the stellar fraction. Here, we notice that, despite similar SFRs at the high-mass end, the efficiency of star formation in each case differs. The stellar fraction for the $\sigma_8$ low model is significantly boosted compared to observations, while the fiducial and $h$ high models only mildly overpredict the observed stellar fraction. The $\sigma_8$ high case, on the other hand, is the best match. We note that the stellar fraction is inversely proportional to $E_{\rm SN}$ for all halo masses. The weak mass dependence trend of the stellar fraction is the same compared to the unconstrained case, and in agreement with the trends of the FIREbox$^{\rm HR}$ simulations \citep{Feldmann_2024} and {\sc ColdSIM} \citep{Maio_2023} for the fiducial, $\sigma_8$ high and $h$ high models, while the $\sigma_8$ low model is an order of magnitude higher but with the same slope.

\subsubsection{21-cm signal from the IGM}\label{sec:igm_con}

While the UVLFs at $M_{\rm 1600, AB} < -18$ in the constrained case are similar for all four cosmological models, the SFR and stellar fraction show differences, and in some cases do not match observations. In such a scenario, it becomes interesting to analyze the impact on the 21-cm signal power spectrum, which is presented in the bottom panel of Figure~\ref{fig_ps_combined} at  $z = 9$ for the constrained case. Unlike the unconstrained case where the differences in matter clustering contributed to differences not only in the overdensity term but also in the neutral hydrogen fraction term in Equation~\ref{eq:dtb}, here the changes in astrophysical parameters also contribute to the neutral hydrogen fraction and reduce the impact of matter clustering, leading to a drastic reduction in the differences between the value of $\langle x_{\rm HI} \rangle$ across the four models. Thus, the differences in the 21-cm signal power spectra of the four models are governed primarily by the overdensity term. This leads to a higher power across all wave modes for the $\sigma_8$ high model due to greater matter clustering.  
While one would expect the $h$ high model to have slightly higher power than the fiducial one due to the dependence on overdensity, the neutral-fraction term in Equation~\ref{eq:dtb} reduces the contrast due to a marginally faster reionization process, so that the power in the two cases becomes similar. On the other hand, due to lower matter clustering, the $\sigma_8$ low model has a consistently lower power amplitude as compared to the fiducial model.

In the bottom panel of Figure~\ref{fig_ps0.1_combined}, we present the redshift evolution of the 21-cm signal power spectrum at $k = 0.15~h\rm cMpc^{-1}$ for the constrained case and notice a significant reduction in the differences of $\Delta^2_{\rm 21cm} (k = 0.15~h\rm cMpc^{-1})$ values between the various cosmological models at $12 > z > 5$ compared to the extent of differences in the unconstrained case (top panel of Figure~\ref{fig_ps0.1_combined}). Indeed, at $10>z>8.5$, where LOFAR observations are focused, variations in the $\sigma_8$ parameter lead to differences in the power spectrum of only a few mK$^2$ (in contrast, even with improvements in signal extraction, with current LOFAR data, \citealt{Acharya_2024gprletter} reported the lowest 2$\sigma$ measurement error of $\approx 25^2$ mK$^2$ on the recovered power spectrum). While the exact magnitude of the difference is subject to the position of the peak of $\Delta^2_{\rm 21cm} (k = 0.15~h\rm cMpc^{-1})$, which in turn is governed by modelling assumptions like the choice of $f_{\rm esc}$ (see Appendix~\ref{appendix_highfesc} for a comparison between  $f_{\rm esc} = 12.5\%$ and 25\%), the key takeaway is that \textit{significantly different choices of cosmological and astrophysical parameters can still lead to similar 21-cm signal observables, even when constrained by the UVLFs from JWST and HST observations}. Thus, to explore the parameter space of models that agree with the upper limits of 21-cm signal power spectrum observations at the redshifts of interest to LOFAR, one should still consider various cosmological and astrophysical parameters as free, unlike what has been done so far for parameter inference not just for observations from LOFAR, but also from HERA and MWA. However, jointly constraining these models with a more diverse set of galactic observables from JWST and upcoming surveys with Euclid and SPHEREx can allow us to narrow down the choice of viable ones. Additionally, tomographic imaging of the distribution of neutral Hydrogen with the upcoming SKA-Low could also be used for additional constraints on IGM properties from images such as those shown in Figure~\ref{fig_midslice_fconst} \citep[see for e.g.][]{Bianco_2021}.

\section{Discussion}\label{sec:discuss}
By analyzing various galactic and IGM observables, we note that the interplay of astrophysical and cosmological parameters can lead to similarities as well as differences. For example, in the constrained case, while agreement with observed UVLFs leads to similar SFR for the high-mass galaxies across all models, there are significant differences in the SFR and stellar fraction for low-mass galaxies. On the other hand, while in the unconstrained case, the SFR and stellar fraction show a very similar dependence on stellar mass in all models because the parameters that regulate such quantities are the same, the UVLFs exhibit significant differences.

For the IGM, we note that the 21-cm signal power spectrum at the redshifts of interest for LOFAR is very similar across all four models in the constrained case (bottom panel, Figure~\ref{fig_ps_combined}). This is due to the choice of astrophysical parameters, which reduces the impact of the cosmological ones, and consequently the differences between the reionization histories in comparison to the unconstrained case. In the unconstrained case, we note significant differences in reionization histories simply because the choice of cosmological parameters changes the overall matter clustering. 

These similarities in terms of 21-cm observables have a profound impact on inference modeling, as models with very different cosmological and astrophysical parameters may still produce 21-cm signal power spectra in agreement with current upper limits \citep[for example from][]{Mertens2020, Trott_2020, HERA_2023, Acharya_2024gprletter}, or possibly also with an eventual measurement. This is because, while the end of reionization may differ among models, at $z > 7$ the extent and speed of reionization are still largely similar. This means that fixing a priori any of these parameters could exclude viable models. While introducing cosmological and astrophysical parameters as free parameters during inference modeling has been attempted in several earlier studies \citep[see e.g.][]{Santos_2006, Liu_2016, Kern_2017, Hassan_2020, Schneider_2023}, these have used only approximate reionization simulation codes due to the significant computational costs involved in building such a high-dimensional parameter space. Indeed, in a more physical approach, it is necessary to run a large number of dark-matter only simulations with different cosmologies, each post-processed with a large set of astrophysical parameters. The problem worsens if one would like to boost the resolution of the simulations to take into account the role of mini halos \citep[see][]{Haiman_2001, Iliev_2005}, dwarf galaxies \citep{Wu_2024}, or Lyman Limit Systems \citep{Georgiev_2024, Giri_2024} which likely play a significant role in the EoR. While faster semi-numerical codes and emulators can be employed, they are not ideal tools to model galaxy-scale physics and radiative transfer effects.

Because of this, boosting the resolution of low-resolution simulations using analytic techniques based on smaller boxes with higher resolution has been proposed \citep[see for example][]{Nasirudin_2020,Barsode_2024}. Newer methods based on Machine Learning, such as Generative Adversarial Networks \citep[GANs, as done in][]{Zhang_2024}, bypass the requirement of a higher resolution simulation. However, it needs to be explored whether such methods are robust enough to factor in small differences in individual parameters. In subsequent work with {\sc Polar}, we intend to include resolution-boosting techniques to resolve halos and galaxies down by at least two orders of magnitude in mass.
The development and implementation of such techniques will additionally allow us to create more diverse training sets of power spectra for building Machine Learning kernels for Gaussian Process Regression based signal extraction, as shown by \citet{Mertens_2024} and \citet{Acharya_2024gpr, Acharya_2024gprletter}.

\section{Summary}\label{sec:summary}

In this work, we have investigated the impact of different cosmological models on the 21-cm signal and some galactic properties using {\sc Polar}, which combines $N$-body dark matter simulations run with {\sc Gadget-4} with the semi-analytic model of galaxy formation {\sc L-Galaxies}, and the 1-D radiative transfer code {\sc Grizzly}. We have applied the framework to four different cosmological models: a ``fiducial'' model, which adopts values of the cosmological parameters from \citet{Planck_2020}; a ``$h$ high'' model, with $h = 0.733$ based on results from studies of Cepheid variables in the host galaxies of 42 Type Ia supernovae  \citep{Riess_2022}; a ``$\sigma_8$ low'' case, with $\sigma_8 = 0.702$ from an anisotropic galaxy clustering measurement analysis done by \citet{Troster_2020}; and a ``$\sigma_8$ high'' case, with $\sigma_8 = 0.880$ according to recent eROSITA results \citep{Ghirardini_2024}. We additionally used the Fixed \& Paired approach \citep{Angulo_2016} to suppress cosmic variance by boosting the effective volumes of the simulations. For all the quantities analyzed, we took an average of the fixed initial conditions case (which we refer to as the A-series) and its corresponding pair (referred to as the B-series).

We then choose astrophysical parameters in {\sc L-Galaxies} such that the fiducial model matches UV luminosity functions from HST legacy fields and JWST programs at $z = 10$ and 9. In particular, we focus on parameters affecting the star formation efficiency ($\alpha_{\rm SF}$), star formation efficiency during galaxy mergers ($\alpha_{\rm SF, burst}$), AGN accretion rate ($k_{\rm AGN}$), reheating of cold gas by star formation ($\epsilon_{\rm reheat}$, $V_{\rm reheat}$), and the energy released by each supernova ($E_{\rm SN}$).

To investigate the effect of adopting different values for $h$ and $\sigma_8$ on other galactic and IGM observables, we use the same astrophysical parameters also for the other three cosmological models. We refer to this as the ``unconstrained'' case, as by not changing astrophysical parameters for the different cosmological models the resulting UVLFs will not necessarily be consistent with the observed ones. We also built a ``constrained'' case, where we instead choose the astrophysical parameters such that the UVLFs obtained in each cosmological model are consistent with those observed at  $z = 10$ and 9. For this, we increased the star formation efficiency and reduced the energy released per supernova in the $\sigma_8$ low model to boost star formation and, in turn, its UVLF. On the other hand, we increased the energy released per supernova for the $h$ high and $\sigma_8$ high models to suppress star formation and mitigate the impact of increased matter clustering as compared to the fiducial model. For the radiative transfer calculations, we only took into account the stellar sources while modeling the spectral energy distributions and chose a global escape fraction of 12.5\% across all redshifts for all four models.

Our results can be summarized as follows:
\begin{itemize}
    \item \textbf{Reionization history:} In the unconstrained case, the values of the cosmological parameters strongly influence the matter clustering, which in turn leads to significant differences in how reionization progresses in each model. While the $\sigma_8$ high model reionizes by $z \approx 6.4$, the $\sigma_8$ low model is only 45\% reionized even by $z = 5$. In the constrained case, instead, the impact of astrophysical parameters is significant, and the $\sigma_8$ low model is the first to reionize,  at $z = 5.75$, while the $\sigma_8$ high model is only 70\% reionized by $z = 5$.
    \item \textbf{Star Formation Rate (SFR):} In the unconstrained case, no significant differences are observed as the parameters controlling star formation have been kept the same. However, in the constrained case the SFR in low-mass galaxies is impacted by the required energy released per supernova, and thus the SFR is boosted for the $\sigma_8$ low model and suppressed for the $\sigma_8$ high model. In both cases, all models broadly agree with observations at $z = 10$ and 9.
    \item \textbf{Global Star Formation Efficiency (SFE):} As the parameters controlling star formation are the same in the unconstrained case, the global SFE estimated via the stellar fraction at all masses is similar in all models. The stellar fraction is also in agreement with observations and abundance matching estimates. However, in the constrained case, due to a higher value of the parameter regulating the star formation efficiency, the stellar fraction in the $\sigma_8$ low model is also higher. The other three models show minor differences, governed by the differences in the energy released per supernova.
    \item \textbf{21-cm signal power spectrum:} In the unconstrained case, the differences between the four models are not only due to the difference in overdensity but also to those in the neutral fraction and its redshift evolution. However, in the constrained case, the neutral fraction at $z > 8$ is similar in all models, as the impact of matter clustering on the neutral fraction is cancelled out by the impact of setting different values for the astrophysical parameters controlling star formation. Thus, the 21-cm signal power spectrum at higher redshifts is mostly dictated by the overdensity term, and because of this it shows smaller differences among the four models.
\end{itemize}

Overall, we conclude that jointly varying cosmological and astrophysical parameters can lead to differences in some observables (e.g. low-mass SFR and stellar fraction), while others stay similar (e.g. UVLFs and 21-cm signal power spectrum with the latter remaining well within the expected observational errors despite minor differences). In particular, we note that despite significantly different galactic processes and reionization histories, the 21-cm power spectra are very similar in power across $k$-bins and in agreement with current observational upper limits \citep{Acharya_2024gprletter, Mertens2020, Trott_2020, HERA_2023}. Due to this, when doing inference modeling, it is essential to consider all cosmological and astrophysical parameters as free parameters, with other observational constraints serving as priors. While a limited exploration of astrophysical parameters can be done with excursion set algorithms, a semi-analytic model provides a significantly more rigorous and physical approach to modeling galactic properties. For varying cosmological parameters though, it is necessary to run a large number of $N$-body simulations to populate the prior parameter space. However, running so many high-resolution simulations would be prohibitively expensive in terms of computational resources and time. To address this issue,  in future work we intend to incorporate techniques for boosting the resolution of less costly, low-resolution simulations using either analytic or machine learning techniques. Such an implementation would also additionally allow us to build broader training sets of power spectra for the Machine Learning kernels used with Gaussian Process Regression by the LOFAR EoR Key Science Project team \citep[as done by][]{Acharya_2024gpr}.

Lastly, to achieve a stronger constraining power on a broadened parameter space, it is essential to jointly employ a variety of galactic observables, e.g. from JWST, Euclid, and upcoming SPHEREx, as well as tomographic imaging of neutral Hydrogen in the IGM with the future SKA-Low.

\section*{Acknowledgements}
AA thanks the EoR research group at MPA, Volker Springel and R{\"u}diger Pakmor for helpful discussions. AA acknowledges Nordita's PhD Fellows Visitor program for funding a part of this work, and thanks Nordita for their hospitality. Nordita is supported in part by NordForsk. RG acknowledges support from SERB, DST Ramanujan Fellowship no. RJF/2022/000141. GM is supported by the Swedish Research Council project grant 2020-04691\_VR. SZ acknowledges the Alexander von Humboldt Foundation for the Humboldt Research Award and the Max Planck Institute for Astrophysics for its hospitality. The post-doctoral contract of IH was funded by Sorbonne Université in the framework of the Initiative Physique des Infinis (IDEX SUPER).  LVEK acknowledges the financial support from the European Research Council (ERC) under the European Union's Horizon 2020 research and innovation programme (Grant agreement No. 884760, `CoDEX').

\noindent This work used the color map ``arctic'' for slice images from the publicly available Python package {\sc CMasher} \citep{cmasher_2020} and the Python package {\sc CoReCon} \citep{Garaldi_2023} for collating neutral fraction observations. Additionally, this work made extensive use of commonly used packages like {\sc NumPy} \citep{Harris_2020}, {\sc Matplotlib} \citep{Hunter_2007} and {\sc SciPy} \citep{Virtanen_2020}.

\section*{Data Availability}
The simulation data, and post-analysis
scripts used in this work can be shared upon reasonable request to the corresponding author.


\bibliographystyle{mnras}
\bibliography{altcosmo} 

\begin{thebibliography}{}
\makeatletter
\relax
\def\mn@urlcharsother{\let\do\@makeother \do\$\do\&\do\#\do\^\do\_\do\%\do\~}
\def\mn@doi{\begingroup\mn@urlcharsother \@ifnextchar [ {\mn@doi@}
  {\mn@doi@[]}}
\def\mn@doi@[#1]#2{\def\@tempa{#1}\ifx\@tempa\@empty \href
  {http://dx.doi.org/#2} {doi:#2}\else \href {http://dx.doi.org/#2} {#1}\fi
  \endgroup}
\def\mn@eprint#1#2{\mn@eprint@#1:#2::\@nil}
\def\mn@eprint@arXiv#1{\href {http://arxiv.org/abs/#1} {{\tt arXiv:#1}}}
\def\mn@eprint@dblp#1{\href {http://dblp.uni-trier.de/rec/bibtex/#1.xml}
  {dblp:#1}}
\def\mn@eprint@#1:#2:#3:#4\@nil{\def\@tempa {#1}\def\@tempb {#2}\def\@tempc
  {#3}\ifx \@tempc \@empty \let \@tempc \@tempb \let \@tempb \@tempa \fi \ifx
  \@tempb \@empty \def\@tempb {arXiv}\fi \@ifundefined
  {mn@eprint@\@tempb}{\@tempb:\@tempc}{\expandafter \expandafter \csname
  mn@eprint@\@tempb\endcsname \expandafter{\@tempc}}}

\bibitem[\protect\citeauthoryear{{Abdurashidova} et~al.,}{{Abdurashidova}
  et~al.}{2022}]{Abdurashidova_2022}
{Abdurashidova} Z.,  et~al., 2022, \mn@doi [\apj] {10.3847/1538-4357/ac2ffc},
  \href {https://ui.adsabs.harvard.edu/abs/2022ApJ...924...51A} {924, 51}

\bibitem[\protect\citeauthoryear{{Acharya} et~al.,}{{Acharya}
  et~al.}{2024a}]{Acharya_2024gpr}
{Acharya} A.,  et~al., 2024a, \mn@doi [\mnras] {10.1093/mnras/stad3701}, \href
  {https://ui.adsabs.harvard.edu/abs/2024MNRAS.527.7835A} {527, 7835}

\bibitem[\protect\citeauthoryear{{Acharya}, {Garaldi}, {Ciardi}  \&
  {Ma}}{{Acharya} et~al.}{2024b}]{Acharya_2024}
{Acharya} A.,  {Garaldi} E.,  {Ciardi} B.,   {Ma} Q.-b.,  2024b, \mn@doi
  [\mnras] {10.1093/mnras/stae782}, \href
  {https://ui.adsabs.harvard.edu/abs/2024MNRAS.529.3793A} {529, 3793}

\bibitem[\protect\citeauthoryear{{Acharya}, {Mertens}, {Ciardi}, {Ghara},
  {Koopmans}  \& {Zaroubi}}{{Acharya} et~al.}{2024c}]{Acharya_2024gprletter}
{Acharya} A.,  {Mertens} F.,  {Ciardi} B.,  {Ghara} R.,  {Koopmans} L. V.~E.,
  {Zaroubi} S.,  2024c, \mn@doi [\mnras] {10.1093/mnrasl/slae078}, \href
  {https://ui.adsabs.harvard.edu/abs/2024MNRAS.534L..30A} {534, L30}

\bibitem[\protect\citeauthoryear{{Adams} et~al.,}{{Adams}
  et~al.}{2024}]{Adams_2024}
{Adams} N.~J.,  et~al., 2024, \mn@doi [\apj] {10.3847/1538-4357/ad2a7b}, \href
  {https://ui.adsabs.harvard.edu/abs/2024ApJ...965..169A} {965, 169}

\bibitem[\protect\citeauthoryear{{Amon} et~al.,}{{Amon}
  et~al.}{2023}]{Amon_2023}
{Amon} A.,  et~al., 2023, \mn@doi [\mnras] {10.1093/mnras/stac2938}, \href
  {https://ui.adsabs.harvard.edu/abs/2023MNRAS.518..477A} {518, 477}

\bibitem[\protect\citeauthoryear{{Anderson}, {Pontzen}, {Font-Ribera},
  {Villaescusa-Navarro}, {Rogers}  \& {Genel}}{{Anderson}
  et~al.}{2019}]{Anderson_2019}
{Anderson} L.,  {Pontzen} A.,  {Font-Ribera} A.,  {Villaescusa-Navarro} F.,
  {Rogers} K.~K.,   {Genel} S.,  2019, \mn@doi [\apj]
  {10.3847/1538-4357/aaf576}, \href
  {https://ui.adsabs.harvard.edu/abs/2019ApJ...871..144A} {871, 144}

\bibitem[\protect\citeauthoryear{{Angulo} \& {Pontzen}}{{Angulo} \&
  {Pontzen}}{2016}]{Angulo_2016}
{Angulo} R.~E.,  {Pontzen} A.,  2016, \mn@doi [\mnras] {10.1093/mnrasl/slw098},
  \href {https://ui.adsabs.harvard.edu/abs/2016MNRAS.462L...1A} {462, L1}

\bibitem[\protect\citeauthoryear{{Arrabal Haro} et~al.,}{{Arrabal Haro}
  et~al.}{2023a}]{Arrabal_2023b}
{Arrabal Haro} P.,  et~al., 2023a, \mn@doi [\nat] {10.1038/s41586-023-06521-7},
  \href {https://ui.adsabs.harvard.edu/abs/2023Natur.622..707A} {622, 707}

\bibitem[\protect\citeauthoryear{{Arrabal Haro} et~al.,}{{Arrabal Haro}
  et~al.}{2023b}]{Arrabal_2023}
{Arrabal Haro} P.,  et~al., 2023b, \mn@doi [\apjl] {10.3847/2041-8213/acdd54},
  \href {https://ui.adsabs.harvard.edu/abs/2023ApJ...951L..22A} {951, L22}

\bibitem[\protect\citeauthoryear{{Atek} et~al.,}{{Atek}
  et~al.}{2023}]{Atek_2023}
{Atek} H.,  et~al., 2023, \mn@doi [\mnras] {10.1093/mnras/stad1998}, \href
  {https://ui.adsabs.harvard.edu/abs/2023MNRAS.524.5486A} {524, 5486}

\bibitem[\protect\citeauthoryear{{Balu}, {Greig}, {Qiu}, {Power}, {Qin},
  {Mutch}  \& {Wyithe}}{{Balu} et~al.}{2023}]{Balu_2023}
{Balu} S.,  {Greig} B.,  {Qiu} Y.,  {Power} C.,  {Qin} Y.,  {Mutch} S.,
  {Wyithe} J. S.~B.,  2023, \mn@doi [\mnras] {10.1093/mnras/stad281}, \href
  {https://ui.adsabs.harvard.edu/abs/2023MNRAS.520.3368B} {520, 3368}

\bibitem[\protect\citeauthoryear{{Barkana} \& {Loeb}}{{Barkana} \&
  {Loeb}}{2001}]{Barkana_2001}
{Barkana} R.,  {Loeb} A.,  2001, \mn@doi [\physrep]
  {10.1016/S0370-1573(01)00019-9}, \href
  {https://ui.adsabs.harvard.edu/abs/2001PhR...349..125B} {349, 125}

\bibitem[\protect\citeauthoryear{{Barrera} et~al.,}{{Barrera}
  et~al.}{2023}]{Barrera_2023}
{Barrera} M.,  et~al., 2023, \mn@doi [\mnras] {10.1093/mnras/stad2688}, \href
  {https://ui.adsabs.harvard.edu/abs/2023MNRAS.525.6312B} {525, 6312}

\bibitem[\protect\citeauthoryear{{Barsode} \& {Choudhury}}{{Barsode} \&
  {Choudhury}}{2024}]{Barsode_2024}
{Barsode} A.,  {Choudhury} T.~R.,  2024, arXiv e-prints, \href
  {https://ui.adsabs.harvard.edu/abs/2024arXiv240710585B} {p. arXiv:2407.10585}

\bibitem[\protect\citeauthoryear{{Basu}, {Garaldi}  \& {Ciardi}}{{Basu}
  et~al.}{2024}]{Basu_2024}
{Basu} A.,  {Garaldi} E.,   {Ciardi} B.,  2024, \mn@doi [\mnras]
  {10.1093/mnras/stae1488}, \href
  {https://ui.adsabs.harvard.edu/abs/2024MNRAS.532..841B} {532, 841}

\bibitem[\protect\citeauthoryear{{Becker}, {Bolton}, {Madau}, {Pettini},
  {Ryan-Weber}  \& {Venemans}}{{Becker} et~al.}{2015}]{Becker_2015}
{Becker} G.~D.,  {Bolton} J.~S.,  {Madau} P.,  {Pettini} M.,  {Ryan-Weber}
  E.~V.,   {Venemans} B.~P.,  2015, \mn@doi [\mnras] {10.1093/mnras/stu2646},
  \href {https://ui.adsabs.harvard.edu/abs/2015MNRAS.447.3402B} {447, 3402}

\bibitem[\protect\citeauthoryear{{Berti}, {Spinelli}  \& {Viel}}{{Berti}
  et~al.}{2024}]{Berti_2024}
{Berti} M.,  {Spinelli} M.,   {Viel} M.,  2024, \mn@doi [\mnras]
  {10.1093/mnras/stae755}, \href
  {https://ui.adsabs.harvard.edu/abs/2024MNRAS.529.4803B} {529, 4803}

\bibitem[\protect\citeauthoryear{{Bhagwat}, {Costa}, {Ciardi}, {Pakmor}  \&
  {Garaldi}}{{Bhagwat} et~al.}{2024}]{Bhagwat_2024}
{Bhagwat} A.,  {Costa} T.,  {Ciardi} B.,  {Pakmor} R.,   {Garaldi} E.,  2024,
  \mn@doi [\mnras] {10.1093/mnras/stae1125}, \href
  {https://ui.adsabs.harvard.edu/abs/2024MNRAS.531.3406B} {531, 3406}

\bibitem[\protect\citeauthoryear{{Bianco}, {Giri}, {Iliev}  \&
  {Mellema}}{{Bianco} et~al.}{2021}]{Bianco_2021}
{Bianco} M.,  {Giri} S.~K.,  {Iliev} I.~T.,   {Mellema} G.,  2021, \mn@doi
  [\mnras] {10.1093/mnras/stab1518}, \href
  {https://ui.adsabs.harvard.edu/abs/2021MNRAS.505.3982B} {505, 3982}

\bibitem[\protect\citeauthoryear{{Bird}, {Ni}, {Di Matteo}, {Croft}, {Feng}  \&
  {Chen}}{{Bird} et~al.}{2022}]{Bird_2022}
{Bird} S.,  {Ni} Y.,  {Di Matteo} T.,  {Croft} R.,  {Feng} Y.,   {Chen} N.,
  2022, \mn@doi [\mnras] {10.1093/mnras/stac648}, \href
  {https://ui.adsabs.harvard.edu/abs/2022MNRAS.512.3703B} {512, 3703}

\bibitem[\protect\citeauthoryear{{Bolton}, {Haehnelt}, {Warren}, {Hewett},
  {Mortlock}, {Venemans}, {McMahon}  \& {Simpson}}{{Bolton}
  et~al.}{2011}]{Bolton_2011}
{Bolton} J.~S.,  {Haehnelt} M.~G.,  {Warren} S.~J.,  {Hewett} P.~C.,
  {Mortlock} D.~J.,  {Venemans} B.~P.,  {McMahon} R.~G.,   {Simpson} C.,  2011,
  \mn@doi [\mnras] {10.1111/j.1745-3933.2011.01100.x}, \href
  {https://ui.adsabs.harvard.edu/abs/2011MNRAS.416L..70B} {416, L70}

\bibitem[\protect\citeauthoryear{{Bosman} \& {Davies}}{{Bosman} \&
  {Davies}}{2024}]{Bosman_2024}
{Bosman} S.~E.~I.,  {Davies} F.~B.,  2024, \mn@doi [\aap]
  {10.1051/0004-6361/202451463}, \href
  {https://ui.adsabs.harvard.edu/abs/2024A&A...690A.391B} {690, A391}

\bibitem[\protect\citeauthoryear{{Bosman} et~al.,}{{Bosman}
  et~al.}{2022}]{Bosman_2022}
{Bosman} S. E.~I.,  et~al., 2022, \mn@doi [\mnras] {10.1093/mnras/stac1046},
  \href {https://ui.adsabs.harvard.edu/abs/2022MNRAS.514...55B} {514, 55}

\bibitem[\protect\citeauthoryear{{Bouwens} et~al.,}{{Bouwens}
  et~al.}{2015}]{Bouwens_2015}
{Bouwens} R.~J.,  et~al., 2015, \mn@doi [\apj] {10.1088/0004-637X/803/1/34},
  \href {https://ui.adsabs.harvard.edu/abs/2015ApJ...803...34B} {803, 34}

\bibitem[\protect\citeauthoryear{{Bouwens} et~al.,}{{Bouwens}
  et~al.}{2021}]{Bouwens_2021}
{Bouwens} R.~J.,  et~al., 2021, \mn@doi [\aj] {10.3847/1538-3881/abf83e}, \href
  {https://ui.adsabs.harvard.edu/abs/2021AJ....162...47B} {162, 47}

\bibitem[\protect\citeauthoryear{{Bouwens}, {Illingworth}, {Oesch}, {Stefanon},
  {Naidu}, {van Leeuwen}  \& {Magee}}{{Bouwens} et~al.}{2023a}]{Bouwens_2023a}
{Bouwens} R.,  {Illingworth} G.,  {Oesch} P.,  {Stefanon} M.,  {Naidu} R.,
  {van Leeuwen} I.,   {Magee} D.,  2023a, \mn@doi [\mnras]
  {10.1093/mnras/stad1014}, \href
  {https://ui.adsabs.harvard.edu/abs/2023MNRAS.523.1009B} {523, 1009}

\bibitem[\protect\citeauthoryear{{Bouwens} et~al.,}{{Bouwens}
  et~al.}{2023b}]{Bouwens_2023b}
{Bouwens} R.~J.,  et~al., 2023b, \mn@doi [\mnras] {10.1093/mnras/stad1145},
  \href {https://ui.adsabs.harvard.edu/abs/2023MNRAS.523.1036B} {523, 1036}

\bibitem[\protect\citeauthoryear{{Bromm} \& {Larson}}{{Bromm} \&
  {Larson}}{2004}]{Bromm_2004}
{Bromm} V.,  {Larson} R.~B.,  2004, \mn@doi [\araa]
  {10.1146/annurev.astro.42.053102.134034}, \href
  {https://ui.adsabs.harvard.edu/abs/2004ARA&A..42...79B} {42, 79}

\bibitem[\protect\citeauthoryear{{Carniani} et~al.,}{{Carniani}
  et~al.}{2024}]{Carniani_2024}
{Carniani} S.,  et~al., 2024, \mn@doi [arXiv e-prints]
  {10.48550/arXiv.2405.18485}, \href
  {https://ui.adsabs.harvard.edu/abs/2024arXiv240518485C} {p. arXiv:2405.18485}

\bibitem[\protect\citeauthoryear{{Casavecchia}, {Maio}, {P{\'e}roux}  \&
  {Ciardi}}{{Casavecchia} et~al.}{2024}]{Casavecchia_2024}
{Casavecchia} B.,  {Maio} U.,  {P{\'e}roux} C.,   {Ciardi} B.,  2024, \mn@doi
  [\aap] {10.1051/0004-6361/202450332}, \href
  {https://ui.adsabs.harvard.edu/abs/2024A&A...689A.106C} {689, A106}

\bibitem[\protect\citeauthoryear{{Chartier}, {Wandelt}, {Akrami}  \&
  {Villaescusa-Navarro}}{{Chartier} et~al.}{2021}]{Chartier_2021}
{Chartier} N.,  {Wandelt} B.,  {Akrami} Y.,   {Villaescusa-Navarro} F.,  2021,
  \mn@doi [\mnras] {10.1093/mnras/stab430}, \href
  {https://ui.adsabs.harvard.edu/abs/2021MNRAS.503.1897C} {503, 1897}

\bibitem[\protect\citeauthoryear{{Chornock}, {Berger}, {Fox}, {Lunnan},
  {Drout}, {Fong}, {Laskar}  \& {Roth}}{{Chornock}
  et~al.}{2013}]{Chornock_2013}
{Chornock} R.,  {Berger} E.,  {Fox} D.~B.,  {Lunnan} R.,  {Drout} M.~R.,
  {Fong} W.-f.,  {Laskar} T.,   {Roth} K.~C.,  2013, \mn@doi [\apj]
  {10.1088/0004-637X/774/1/26}, \href
  {https://ui.adsabs.harvard.edu/abs/2013ApJ...774...26C} {774, 26}

\bibitem[\protect\citeauthoryear{{Choudhury}, {Puchwein}, {Haehnelt}  \&
  {Bolton}}{{Choudhury} et~al.}{2015}]{Choudhury_2015}
{Choudhury} T.~R.,  {Puchwein} E.,  {Haehnelt} M.~G.,   {Bolton} J.~S.,  2015,
  \mn@doi [\mnras] {10.1093/mnras/stv1250}, \href
  {https://ui.adsabs.harvard.edu/abs/2015MNRAS.452..261C} {452, 261}

\bibitem[\protect\citeauthoryear{{Choudhury} et~al.,}{{Choudhury}
  et~al.}{2024}]{Choudhury_2024}
{Choudhury} M.,  et~al., 2024, \mn@doi [arXiv e-prints]
  {10.48550/arXiv.2407.03523}, \href
  {https://ui.adsabs.harvard.edu/abs/2024arXiv240703523C} {p. arXiv:2407.03523}

\bibitem[\protect\citeauthoryear{{Ciardi} \& {Madau}}{{Ciardi} \&
  {Madau}}{2003}]{Ciardi_2003}
{Ciardi} B.,  {Madau} P.,  2003, \mn@doi [\apj] {10.1086/377634}, \href
  {https://ui.adsabs.harvard.edu/abs/2003ApJ...596....1C} {596, 1}

\bibitem[\protect\citeauthoryear{{Dark Energy Survey and Kilo-Degree Survey
  Collaboration} et~al.,}{{Dark Energy Survey and Kilo-Degree Survey
  Collaboration} et~al.}{2023}]{DES_2023}
{Dark Energy Survey and Kilo-Degree Survey Collaboration} et~al., 2023, \mn@doi
  [The Open Journal of Astrophysics] {10.21105/astro.2305.17173}, \href
  {https://ui.adsabs.harvard.edu/abs/2023OJAp....6E..36D} {6, 36}

\bibitem[\protect\citeauthoryear{{Davies} et~al.,}{{Davies}
  et~al.}{2018}]{Davies_2018}
{Davies} F.~B.,  et~al., 2018, \mn@doi [\apj] {10.3847/1538-4357/aad6dc}, \href
  {https://ui.adsabs.harvard.edu/abs/2018ApJ...864..142D} {864, 142}

\bibitem[\protect\citeauthoryear{{Dijkstra}, {Mesinger}  \&
  {Wyithe}}{{Dijkstra} et~al.}{2011}]{Dijkstra_2011}
{Dijkstra} M.,  {Mesinger} A.,   {Wyithe} J. S.~B.,  2011, \mn@doi [\mnras]
  {10.1111/j.1365-2966.2011.18530.x}, \href
  {https://ui.adsabs.harvard.edu/abs/2011MNRAS.414.2139D} {414, 2139}

\bibitem[\protect\citeauthoryear{{Dixon}, {Iliev}, {Mellema}, {Ahn}  \&
  {Shapiro}}{{Dixon} et~al.}{2016}]{Dixon_2016}
{Dixon} K.~L.,  {Iliev} I.~T.,  {Mellema} G.,  {Ahn} K.,   {Shapiro} P.~R.,
  2016, \mn@doi [\mnras] {10.1093/mnras/stv2887}, \href
  {https://ui.adsabs.harvard.edu/abs/2016MNRAS.456.3011D} {456, 3011}

\bibitem[\protect\citeauthoryear{{Eide}, {Graziani}, {Ciardi}, {Feng},
  {Kakiichi}  \& {Di Matteo}}{{Eide} et~al.}{2018}]{Eide_2018}
{Eide} M.~B.,  {Graziani} L.,  {Ciardi} B.,  {Feng} Y.,  {Kakiichi} K.,   {Di
  Matteo} T.,  2018, \mn@doi [\mnras] {10.1093/mnras/sty272}, \href
  {https://ui.adsabs.harvard.edu/abs/2018MNRAS.476.1174E} {476, 1174}

\bibitem[\protect\citeauthoryear{{Eide}, {Ciardi}, {Graziani}, {Busch}, {Feng}
  \& {Di Matteo}}{{Eide} et~al.}{2020}]{Eide_2020}
{Eide} M.~B.,  {Ciardi} B.,  {Graziani} L.,  {Busch} P.,  {Feng} Y.,   {Di
  Matteo} T.,  2020, \mn@doi [\mnras] {10.1093/mnras/staa2774}, \href
  {https://ui.adsabs.harvard.edu/abs/2020MNRAS.498.6083E} {498, 6083}

\bibitem[\protect\citeauthoryear{{Esmerian} \& {Gnedin}}{{Esmerian} \&
  {Gnedin}}{2021}]{Esmerian_2021}
{Esmerian} C.~J.,  {Gnedin} N.~Y.,  2021, \mn@doi [\apj]
  {10.3847/1538-4357/abe869}, \href
  {https://ui.adsabs.harvard.edu/abs/2021ApJ...910..117E} {910, 117}

\bibitem[\protect\citeauthoryear{{Esmerian} \& {Gnedin}}{{Esmerian} \&
  {Gnedin}}{2022}]{Esmerian_2022}
{Esmerian} C.~J.,  {Gnedin} N.~Y.,  2022, \mn@doi [\apj]
  {10.3847/1538-4357/ac9612}, \href
  {https://ui.adsabs.harvard.edu/abs/2022ApJ...940...74E} {940, 74}

\bibitem[\protect\citeauthoryear{{Esmerian} \& {Gnedin}}{{Esmerian} \&
  {Gnedin}}{2024}]{Esmerian_2024}
{Esmerian} C.~J.,  {Gnedin} N.~Y.,  2024, \mn@doi [\apj]
  {10.3847/1538-4357/ad410f}, \href
  {https://ui.adsabs.harvard.edu/abs/2024ApJ...968..113E} {968, 113}

\bibitem[\protect\citeauthoryear{{Fan} et~al.,}{{Fan} et~al.}{2006}]{Fan_2006b}
{Fan} X.,  et~al., 2006, \mn@doi [\aj] {10.1086/504836}, \href
  {https://ui.adsabs.harvard.edu/abs/2006AJ....132..117F} {132, 117}

\bibitem[\protect\citeauthoryear{{Feldmann} et~al.,}{{Feldmann}
  et~al.}{2024}]{Feldmann_2024}
{Feldmann} R.,  et~al., 2024, \mn@doi [arXiv e-prints]
  {10.48550/arXiv.2407.02674}, \href
  {https://ui.adsabs.harvard.edu/abs/2024arXiv240702674F} {p. arXiv:2407.02674}

\bibitem[\protect\citeauthoryear{{Field}}{{Field}}{1959}]{Field_1959}
{Field} G.~B.,  1959, \mn@doi [\apj] {10.1086/146654}, \href
  {https://ui.adsabs.harvard.edu/abs/1959ApJ...129..551F} {129, 551}

\bibitem[\protect\citeauthoryear{{Finkelstein} et~al.,}{{Finkelstein}
  et~al.}{2015}]{Finkelstein_2015}
{Finkelstein} S.~L.,  et~al., 2015, \mn@doi [\apj]
  {10.1088/0004-637X/810/1/71}, \href
  {https://ui.adsabs.harvard.edu/abs/2015ApJ...810...71F} {810, 71}

\bibitem[\protect\citeauthoryear{{Finlator}, {Keating}, {Oppenheimer},
  {Dav{\'e}}  \& {Zackrisson}}{{Finlator} et~al.}{2018}]{Finlator_2018}
{Finlator} K.,  {Keating} L.,  {Oppenheimer} B.~D.,  {Dav{\'e}} R.,
  {Zackrisson} E.,  2018, \mn@doi [\mnras] {10.1093/mnras/sty1949}, \href
  {https://ui.adsabs.harvard.edu/abs/2018MNRAS.480.2628F} {480, 2628}

\bibitem[\protect\citeauthoryear{{Fujimoto} et~al.,}{{Fujimoto}
  et~al.}{2023}]{Fujimoto_2023}
{Fujimoto} S.,  et~al., 2023, \mn@doi [\apjl] {10.3847/2041-8213/acd2d9}, \href
  {https://ui.adsabs.harvard.edu/abs/2023ApJ...949L..25F} {949, L25}

\bibitem[\protect\citeauthoryear{{Furlanetto}, {Oh}  \&
  {Pierpaoli}}{{Furlanetto} et~al.}{2006}]{Furlanetto_2006eq}
{Furlanetto} S.~R.,  {Oh} S.~P.,   {Pierpaoli} E.,  2006, \mn@doi [\prd]
  {10.1103/PhysRevD.74.103502}, \href
  {https://ui.adsabs.harvard.edu/abs/2006PhRvD..74j3502F} {74, 103502}

\bibitem[\protect\citeauthoryear{{Garaldi}}{{Garaldi}}{2023}]{Garaldi_2023}
{Garaldi} E.,  2023, \mn@doi [The Journal of Open Source Software]
  {10.21105/joss.05407}, \href
  {https://ui.adsabs.harvard.edu/abs/2023JOSS....8.5407G} {8, 5407}

\bibitem[\protect\citeauthoryear{{Garaldi}, {Kannan}, {Smith}, {Springel},
  {Pakmor}, {Vogelsberger}  \& {Hernquist}}{{Garaldi}
  et~al.}{2022}]{Garaldi_2022}
{Garaldi} E.,  {Kannan} R.,  {Smith} A.,  {Springel} V.,  {Pakmor} R.,
  {Vogelsberger} M.,   {Hernquist} L.,  2022, \mn@doi [\mnras]
  {10.1093/mnras/stac257}, \href
  {https://ui.adsabs.harvard.edu/abs/2022MNRAS.512.4909G} {512, 4909}

\bibitem[\protect\citeauthoryear{{Garaldi} et~al.,}{{Garaldi}
  et~al.}{2024}]{Garaldi_2024}
{Garaldi} E.,  et~al., 2024, \mn@doi [\mnras] {10.1093/mnras/stae839}, \href
  {https://ui.adsabs.harvard.edu/abs/2024MNRAS.530.3765G} {530, 3765}

\bibitem[\protect\citeauthoryear{{Georgiev}, {Mellema}  \& {Giri}}{{Georgiev}
  et~al.}{2024}]{Georgiev_2024}
{Georgiev} I.,  {Mellema} G.,   {Giri} S.~K.,  2024, \mn@doi [arXiv e-prints]
  {10.48550/arXiv.2405.04273}, \href
  {https://ui.adsabs.harvard.edu/abs/2024arXiv240504273G} {p. arXiv:2405.04273}

\bibitem[\protect\citeauthoryear{{Ghara}, {Choudhury}  \& {Datta}}{{Ghara}
  et~al.}{2015}]{Ghara_2015}
{Ghara} R.,  {Choudhury} T.~R.,   {Datta} K.~K.,  2015, \mn@doi [\mnras]
  {10.1093/mnras/stu2512}, \href
  {https://ui.adsabs.harvard.edu/abs/2015MNRAS.447.1806G} {447, 1806}

\bibitem[\protect\citeauthoryear{{Ghara}, {Mellema}, {Giri}, {Choudhury},
  {Datta}  \& {Majumdar}}{{Ghara} et~al.}{2018}]{Ghara_2018}
{Ghara} R.,  {Mellema} G.,  {Giri} S.~K.,  {Choudhury} T.~R.,  {Datta} K.~K.,
  {Majumdar} S.,  2018, \mn@doi [\mnras] {10.1093/mnras/sty314}, \href
  {https://ui.adsabs.harvard.edu/abs/2018MNRAS.476.1741G} {476, 1741}

\bibitem[\protect\citeauthoryear{{Ghara} et~al.,}{{Ghara}
  et~al.}{2020}]{Ghara2020}
{Ghara} R.,  et~al., 2020, \mn@doi [\mnras] {10.1093/mnras/staa487}, \href
  {https://ui.adsabs.harvard.edu/abs/2020MNRAS.493.4728G} {493, 4728}

\bibitem[\protect\citeauthoryear{{Ghara} et~al.,}{{Ghara}
  et~al.}{2024}]{Ghara_2024}
{Ghara} R.,  et~al., 2024, \mn@doi [\aap] {10.1051/0004-6361/202449444}, \href
  {https://ui.adsabs.harvard.edu/abs/2024A&A...687A.252G} {687, A252}

\bibitem[\protect\citeauthoryear{{Ghirardini} et~al.,}{{Ghirardini}
  et~al.}{2024}]{Ghirardini_2024}
{Ghirardini} V.,  et~al., 2024, \mn@doi [\aap] {10.1051/0004-6361/202348852},
  \href {https://ui.adsabs.harvard.edu/abs/2024A&A...689A.298G} {689, A298}

\bibitem[\protect\citeauthoryear{{Giar{\`e}}, {Di Valentino}  \&
  {Melchiorri}}{{Giar{\`e}} et~al.}{2024}]{Giare_2024}
{Giar{\`e}} W.,  {Di Valentino} E.,   {Melchiorri} A.,  2024, \mn@doi [\prd]
  {10.1103/PhysRevD.109.103519}, \href
  {https://ui.adsabs.harvard.edu/abs/2024PhRvD.109j3519G} {109, 103519}

\bibitem[\protect\citeauthoryear{{Giri} \& {Mellema}}{{Giri} \&
  {Mellema}}{2021}]{Giri_2021}
{Giri} S.~K.,  {Mellema} G.,  2021, \mn@doi [\mnras] {10.1093/mnras/stab1320},
  \href {https://ui.adsabs.harvard.edu/abs/2021MNRAS.505.1863G} {505, 1863}

\bibitem[\protect\citeauthoryear{{Giri}, {Mellema}, {Aldheimer}, {Dixon}  \&
  {Iliev}}{{Giri} et~al.}{2019a}]{Giri_2019b}
{Giri} S.~K.,  {Mellema} G.,  {Aldheimer} T.,  {Dixon} K.~L.,   {Iliev} I.~T.,
  2019a, \mn@doi [\mnras] {10.1093/mnras/stz2224}, \href
  {https://ui.adsabs.harvard.edu/abs/2019MNRAS.489.1590G} {489, 1590}

\bibitem[\protect\citeauthoryear{{Giri}, {D'Aloisio}, {Mellema}, {Komatsu},
  {Ghara}  \& {Majumdar}}{{Giri} et~al.}{2019b}]{Giri_2019}
{Giri} S.~K.,  {D'Aloisio} A.,  {Mellema} G.,  {Komatsu} E.,  {Ghara} R.,
  {Majumdar} S.,  2019b, \mn@doi [\jcap] {10.1088/1475-7516/2019/02/058}, \href
  {https://ui.adsabs.harvard.edu/abs/2019JCAP...02..058G} {2019, 058}

\bibitem[\protect\citeauthoryear{{Giri}, {Schneider}, {Maion}  \&
  {Angulo}}{{Giri} et~al.}{2023}]{Giri_2023}
{Giri} S.~K.,  {Schneider} A.,  {Maion} F.,   {Angulo} R.~E.,  2023, \mn@doi
  [\aap] {10.1051/0004-6361/202244986}, \href
  {https://ui.adsabs.harvard.edu/abs/2023A&A...669A...6G} {669, A6}

\bibitem[\protect\citeauthoryear{{Giri}, {Bianco}, {Schaeffer}, {Iliev},
  {Mellema}  \& {Schneider}}{{Giri} et~al.}{2024}]{Giri_2024}
{Giri} S.~K.,  {Bianco} M.,  {Schaeffer} T.,  {Iliev} I.~T.,  {Mellema} G.,
  {Schneider} A.,  2024, \mn@doi [\mnras] {10.1093/mnras/stae1999}, \href
  {https://ui.adsabs.harvard.edu/abs/2024MNRAS.533.2364G} {533, 2364}

\bibitem[\protect\citeauthoryear{{Gnedin}}{{Gnedin}}{2014}]{Gnedin_2014}
{Gnedin} N.~Y.,  2014, \mn@doi [\apj] {10.1088/0004-637X/793/1/29}, \href
  {https://ui.adsabs.harvard.edu/abs/2014ApJ...793...29G} {793, 29}

\bibitem[\protect\citeauthoryear{{Gnedin} \& {Kaurov}}{{Gnedin} \&
  {Kaurov}}{2014}]{Gnedin_2014b}
{Gnedin} N.~Y.,  {Kaurov} A.~A.,  2014, \mn@doi [\apj]
  {10.1088/0004-637X/793/1/30}, \href
  {https://ui.adsabs.harvard.edu/abs/2014ApJ...793...30G} {793, 30}

\bibitem[\protect\citeauthoryear{{Gorce}, {Douspis}, {Aghanim}  \&
  {Langer}}{{Gorce} et~al.}{2018}]{Gorce_2018}
{Gorce} A.,  {Douspis} M.,  {Aghanim} N.,   {Langer} M.,  2018, \mn@doi [\aap]
  {10.1051/0004-6361/201629661}, \href
  {https://ui.adsabs.harvard.edu/abs/2018A&A...616A.113G} {616, A113}

\bibitem[\protect\citeauthoryear{{Greig}, {Mesinger}, {Haiman}  \&
  {Simcoe}}{{Greig} et~al.}{2017}]{Greig_2017}
{Greig} B.,  {Mesinger} A.,  {Haiman} Z.,   {Simcoe} R.~A.,  2017, \mn@doi
  [\mnras] {10.1093/mnras/stw3351}, \href
  {https://ui.adsabs.harvard.edu/abs/2017MNRAS.466.4239G} {466, 4239}

\bibitem[\protect\citeauthoryear{{Greig}, {Mesinger}  \& {Ba{\~n}ados}}{{Greig}
  et~al.}{2019}]{Greig_2019}
{Greig} B.,  {Mesinger} A.,   {Ba{\~n}ados} E.,  2019, \mn@doi [\mnras]
  {10.1093/mnras/stz230}, \href
  {https://ui.adsabs.harvard.edu/abs/2019MNRAS.484.5094G} {484, 5094}

\bibitem[\protect\citeauthoryear{{Greig}, {Trott}, {Barry}, {Mutch}, {Pindor},
  {Webster}  \& {Wyithe}}{{Greig} et~al.}{2021a}]{Greig_2021}
{Greig} B.,  {Trott} C.~M.,  {Barry} N.,  {Mutch} S.~J.,  {Pindor} B.,
  {Webster} R.~L.,   {Wyithe} J. S.~B.,  2021a, \mn@doi [\mnras]
  {10.1093/mnras/staa3494}, \href
  {https://ui.adsabs.harvard.edu/abs/2021MNRAS.500.5322G} {500, 5322}

\bibitem[\protect\citeauthoryear{{Greig} et~al.,}{{Greig}
  et~al.}{2021b}]{Greig_2021lofar}
{Greig} B.,  et~al., 2021b, \mn@doi [\mnras] {10.1093/mnras/staa3593}, \href
  {https://ui.adsabs.harvard.edu/abs/2021MNRAS.501....1G} {501, 1}

\bibitem[\protect\citeauthoryear{{Greig}, {Prelogovi{\'c}}, {Qin}, {Ting}  \&
  {Mesinger}}{{Greig} et~al.}{2024}]{Greig_2024}
{Greig} B.,  {Prelogovi{\'c}} D.,  {Qin} Y.,  {Ting} Y.-S.,   {Mesinger} A.,
  2024, \mn@doi [\mnras] {10.1093/mnras/stae1984}, \href
  {https://ui.adsabs.harvard.edu/abs/2024MNRAS.533.2530G} {533, 2530}

\bibitem[\protect\citeauthoryear{{HERA Collaboration} et~al.,}{{HERA
  Collaboration} et~al.}{2023}]{HERA_2023}
{HERA Collaboration} et~al., 2023, \mn@doi [\apj] {10.3847/1538-4357/acaf50},
  \href {https://ui.adsabs.harvard.edu/abs/2023ApJ...945..124H} {945, 124}

\bibitem[\protect\citeauthoryear{{Haiman}, {Abel}  \& {Madau}}{{Haiman}
  et~al.}{2001}]{Haiman_2001}
{Haiman} Z.,  {Abel} T.,   {Madau} P.,  2001, \mn@doi [\apj] {10.1086/320232},
  \href {https://ui.adsabs.harvard.edu/abs/2001ApJ...551..599H} {551, 599}

\bibitem[\protect\citeauthoryear{{Han}, {Cole}, {Frenk}, {Benitez-Llambay}  \&
  {Helly}}{{Han} et~al.}{2018}]{Han_2018}
{Han} J.,  {Cole} S.,  {Frenk} C.~S.,  {Benitez-Llambay} A.,   {Helly} J.,
  2018, \mn@doi [\mnras] {10.1093/mnras/stx2792}, \href
  {https://ui.adsabs.harvard.edu/abs/2018MNRAS.474..604H} {474, 604}

\bibitem[\protect\citeauthoryear{{Harikane} et~al.,}{{Harikane}
  et~al.}{2022}]{Harikane_2022}
{Harikane} Y.,  et~al., 2022, \mn@doi [\apjs] {10.3847/1538-4365/ac3dfc}, \href
  {https://ui.adsabs.harvard.edu/abs/2022ApJS..259...20H} {259, 20}

\bibitem[\protect\citeauthoryear{{Harikane} et~al.,}{{Harikane}
  et~al.}{2023}]{Harikane_2023}
{Harikane} Y.,  et~al., 2023, \mn@doi [\apjs] {10.3847/1538-4365/acaaa9}, \href
  {https://ui.adsabs.harvard.edu/abs/2023ApJS..265....5H} {265, 5}

\bibitem[\protect\citeauthoryear{Harris et~al.,}{Harris
  et~al.}{2020}]{Harris_2020}
Harris C.~R.,  et~al., 2020, \mn@doi [Nature] {10.1038/s41586-020-2649-2}, 585,
  357

\bibitem[\protect\citeauthoryear{{Hassan}, {Andrianomena}  \&
  {Doughty}}{{Hassan} et~al.}{2020}]{Hassan_2020}
{Hassan} S.,  {Andrianomena} S.,   {Doughty} C.,  2020, \mn@doi [\mnras]
  {10.1093/mnras/staa1151}, \href
  {https://ui.adsabs.harvard.edu/abs/2020MNRAS.494.5761H} {494, 5761}

\bibitem[\protect\citeauthoryear{{Heger} \& {Woosley}}{{Heger} \&
  {Woosley}}{2002}]{Heger_2002}
{Heger} A.,  {Woosley} S.~E.,  2002, \mn@doi [\apj] {10.1086/338487}, \href
  {https://ui.adsabs.harvard.edu/abs/2002ApJ...567..532H} {567, 532}

\bibitem[\protect\citeauthoryear{{Heintz} et~al.,}{{Heintz}
  et~al.}{2023a}]{Heintz_2023}
{Heintz} K.~E.,  et~al., 2023a, \mn@doi [arXiv e-prints]
  {10.48550/arXiv.2306.00647}, \href
  {https://ui.adsabs.harvard.edu/abs/2023arXiv230600647H} {p. arXiv:2306.00647}

\bibitem[\protect\citeauthoryear{{Heintz} et~al.,}{{Heintz}
  et~al.}{2023b}]{Heintz_2023b}
{Heintz} K.~E.,  et~al., 2023b, \mn@doi [Nature Astronomy]
  {10.1038/s41550-023-02078-7}, \href
  {https://ui.adsabs.harvard.edu/abs/2023NatAs...7.1517H} {7, 1517}

\bibitem[\protect\citeauthoryear{{Helton} et~al.,}{{Helton}
  et~al.}{2024}]{Helton_2024}
{Helton} J.~M.,  et~al., 2024, \mn@doi [\apj] {10.3847/1538-4357/ad0da7}, \href
  {https://ui.adsabs.harvard.edu/abs/2024ApJ...962..124H} {962, 124}

\bibitem[\protect\citeauthoryear{{Henriques}, {White}, {Thomas}, {Angulo},
  {Guo}, {Lemson}, {Springel}  \& {Overzier}}{{Henriques}
  et~al.}{2015}]{Henriques_2015}
{Henriques} B. M.~B.,  {White} S. D.~M.,  {Thomas} P.~A.,  {Angulo} R.,  {Guo}
  Q.,  {Lemson} G.,  {Springel} V.,   {Overzier} R.,  2015, \mn@doi [\mnras]
  {10.1093/mnras/stv705}, \href
  {https://ui.adsabs.harvard.edu/abs/2015MNRAS.451.2663H} {451, 2663}

\bibitem[\protect\citeauthoryear{{Henriques}, {Yates}, {Fu}, {Guo},
  {Kauffmann}, {Srisawat}, {Thomas}  \& {White}}{{Henriques}
  et~al.}{2020}]{Henriques_2020}
{Henriques} B. M.~B.,  {Yates} R.~M.,  {Fu} J.,  {Guo} Q.,  {Kauffmann} G.,
  {Srisawat} C.,  {Thomas} P.~A.,   {White} S. D.~M.,  2020, \mn@doi [\mnras]
  {10.1093/mnras/stz3233}, \href
  {https://ui.adsabs.harvard.edu/abs/2020MNRAS.491.5795H} {491, 5795}

\bibitem[\protect\citeauthoryear{{Hern{\'a}ndez-Aguayo}
  et~al.,}{{Hern{\'a}ndez-Aguayo} et~al.}{2023}]{Hernandez_2023}
{Hern{\'a}ndez-Aguayo} C.,  et~al., 2023, \mn@doi [\mnras]
  {10.1093/mnras/stad1657}, \href
  {https://ui.adsabs.harvard.edu/abs/2023MNRAS.524.2556H} {524, 2556}

\bibitem[\protect\citeauthoryear{{Hirling}, {Bianco}, {Giri}, {Iliev},
  {Mellema}  \& {Kneib}}{{Hirling} et~al.}{2024}]{Hirling_2024}
{Hirling} P.,  {Bianco} M.,  {Giri} S.~K.,  {Iliev} I.~T.,  {Mellema} G.,
  {Kneib} J.~P.,  2024, \mn@doi [Astronomy and Computing]
  {10.1016/j.ascom.2024.100861}, \href
  {https://ui.adsabs.harvard.edu/abs/2024A&C....4800861H} {48, 100861}

\bibitem[\protect\citeauthoryear{{Hoag} et~al.,}{{Hoag}
  et~al.}{2019}]{Hoag_2019}
{Hoag} A.,  et~al., 2019, \mn@doi [\apj] {10.3847/1538-4357/ab1de7}, \href
  {https://ui.adsabs.harvard.edu/abs/2019ApJ...878...12H} {878, 12}

\bibitem[\protect\citeauthoryear{{Hogan} \& {Rees}}{{Hogan} \&
  {Rees}}{1979}]{Hogan_1979}
{Hogan} C.~J.,  {Rees} M.~J.,  1979, \mn@doi [\mnras]
  {10.1093/mnras/188.4.791}, \href
  {https://ui.adsabs.harvard.edu/abs/1979MNRAS.188..791H} {188, 791}

\bibitem[\protect\citeauthoryear{{Hothi}, {Allys}, {Semelin}  \&
  {Boulanger}}{{Hothi} et~al.}{2024}]{Hothi_2024}
{Hothi} I.,  {Allys} E.,  {Semelin} B.,   {Boulanger} F.,  2024, \mn@doi [\aap]
  {10.1051/0004-6361/202348444}, \href
  {https://ui.adsabs.harvard.edu/abs/2024A&A...686A.212H} {686, A212}

\bibitem[\protect\citeauthoryear{Hunter}{Hunter}{2007}]{Hunter_2007}
Hunter J.~D.,  2007, \mn@doi [Computing in Science \& Engineering]
  {10.1109/MCSE.2007.55}, 9, 90

\bibitem[\protect\citeauthoryear{{Hutter}, {Dayal}, {Yepes}, {Gottl{\"o}ber},
  {Legrand}  \& {Ucci}}{{Hutter} et~al.}{2021}]{Hutter_2021}
{Hutter} A.,  {Dayal} P.,  {Yepes} G.,  {Gottl{\"o}ber} S.,  {Legrand} L.,
  {Ucci} G.,  2021, \mn@doi [\mnras] {10.1093/mnras/stab602}, \href
  {https://ui.adsabs.harvard.edu/abs/2021MNRAS.503.3698H} {503, 3698}

\bibitem[\protect\citeauthoryear{{Hutter}, {Cueto}, {Dayal}, {Gottl{\"o}ber},
  {Trebitsch}  \& {Yepes}}{{Hutter} et~al.}{2024}]{Hutter_2024}
{Hutter} A.,  {Cueto} E.~R.,  {Dayal} P.,  {Gottl{\"o}ber} S.,  {Trebitsch} M.,
    {Yepes} G.,  2024, \mn@doi [arXiv e-prints] {10.48550/arXiv.2410.00730},
  \href {https://ui.adsabs.harvard.edu/abs/2024arXiv241000730H} {p.
  arXiv:2410.00730}

\bibitem[\protect\citeauthoryear{{Iliev}, {Shapiro}  \& {Raga}}{{Iliev}
  et~al.}{2005}]{Iliev_2005}
{Iliev} I.~T.,  {Shapiro} P.~R.,   {Raga} A.~C.,  2005, \mn@doi [\mnras]
  {10.1111/j.1365-2966.2005.09155.x}, \href
  {https://ui.adsabs.harvard.edu/abs/2005MNRAS.361..405I} {361, 405}

\bibitem[\protect\citeauthoryear{{Iliev}, {Mellema}, {Ahn}, {Shapiro}, {Mao}
  \& {Pen}}{{Iliev} et~al.}{2014}]{Iliev_2014}
{Iliev} I.~T.,  {Mellema} G.,  {Ahn} K.,  {Shapiro} P.~R.,  {Mao} Y.,   {Pen}
  U.-L.,  2014, \mn@doi [\mnras] {10.1093/mnras/stt2497}, \href
  {https://ui.adsabs.harvard.edu/abs/2014MNRAS.439..725I} {439, 725}

\bibitem[\protect\citeauthoryear{{Jaskot} et~al.,}{{Jaskot}
  et~al.}{2024}]{Jaskot_2024}
{Jaskot} A.~E.,  et~al., 2024, \mn@doi [\apj] {10.3847/1538-4357/ad5557}, \href
  {https://ui.adsabs.harvard.edu/abs/2024ApJ...973..111J} {973, 111}

\bibitem[\protect\citeauthoryear{{Jensen}, {Laursen}, {Mellema}, {Iliev},
  {Sommer-Larsen}  \& {Shapiro}}{{Jensen} et~al.}{2013}]{Jensen_2013}
{Jensen} H.,  {Laursen} P.,  {Mellema} G.,  {Iliev} I.~T.,  {Sommer-Larsen} J.,
    {Shapiro} P.~R.,  2013, \mn@doi [\mnras] {10.1093/mnras/sts116}, \href
  {https://ui.adsabs.harvard.edu/abs/2013MNRAS.428.1366J} {428, 1366}

\bibitem[\protect\citeauthoryear{{Jin} et~al.,}{{Jin} et~al.}{2023}]{Jin_2023}
{Jin} S.,  et~al., 2023, \mn@doi [\aap] {10.1051/0004-6361/202245724}, \href
  {https://ui.adsabs.harvard.edu/abs/2023A&A...670L..11J} {670, L11}

\bibitem[\protect\citeauthoryear{{Jones} et~al.,}{{Jones}
  et~al.}{1998}]{Jones_1998}
{Jones} T.~W.,  et~al., 1998, \mn@doi [\pasp] {10.1086/316122}, \href
  {https://ui.adsabs.harvard.edu/abs/1998PASP..110..125J} {110, 125}

\bibitem[\protect\citeauthoryear{{Jones} et~al.,}{{Jones}
  et~al.}{2024}]{Jones_2024}
{Jones} G.~C.,  et~al., 2024, \mn@doi [arXiv e-prints]
  {10.48550/arXiv.2409.06405}, \href
  {https://ui.adsabs.harvard.edu/abs/2024arXiv240906405J} {p. arXiv:2409.06405}

\bibitem[\protect\citeauthoryear{{Jung} et~al.,}{{Jung}
  et~al.}{2024}]{Jung_2024}
{Jung} I.,  et~al., 2024, \mn@doi [\apj] {10.3847/1538-4357/ad3913}, \href
  {https://ui.adsabs.harvard.edu/abs/2024ApJ...967...73J} {967, 73}

\bibitem[\protect\citeauthoryear{{Kannan}, {Garaldi}, {Smith}, {Pakmor},
  {Springel}, {Vogelsberger}  \& {Hernquist}}{{Kannan}
  et~al.}{2022}]{Kannan_2022}
{Kannan} R.,  {Garaldi} E.,  {Smith} A.,  {Pakmor} R.,  {Springel} V.,
  {Vogelsberger} M.,   {Hernquist} L.,  2022, \mn@doi [\mnras]
  {10.1093/mnras/stab3710}, \href
  {https://ui.adsabs.harvard.edu/abs/2022MNRAS.511.4005K} {511, 4005}

\bibitem[\protect\citeauthoryear{{Kaur}, {Gillet}  \& {Mesinger}}{{Kaur}
  et~al.}{2020}]{Kaur_2020}
{Kaur} H.~D.,  {Gillet} N.,   {Mesinger} A.,  2020, \mn@doi [\mnras]
  {10.1093/mnras/staa1323}, \href
  {https://ui.adsabs.harvard.edu/abs/2020MNRAS.495.2354K} {495, 2354}

\bibitem[\protect\citeauthoryear{{Kern}, {Liu}, {Parsons}, {Mesinger}  \&
  {Greig}}{{Kern} et~al.}{2017}]{Kern_2017}
{Kern} N.~S.,  {Liu} A.,  {Parsons} A.~R.,  {Mesinger} A.,   {Greig} B.,  2017,
  \mn@doi [\apj] {10.3847/1538-4357/aa8bb4}, \href
  {https://ui.adsabs.harvard.edu/abs/2017ApJ...848...23K} {848, 23}

\bibitem[\protect\citeauthoryear{{Klypin}, {Prada}  \& {Byun}}{{Klypin}
  et~al.}{2020}]{Klypin_2020}
{Klypin} A.,  {Prada} F.,   {Byun} J.,  2020, \mn@doi [\mnras]
  {10.1093/mnras/staa734}, \href
  {https://ui.adsabs.harvard.edu/abs/2020MNRAS.496.3862K} {496, 3862}

\bibitem[\protect\citeauthoryear{{Koopmans} et~al.,}{{Koopmans}
  et~al.}{2015}]{Koopmans_2015}
{Koopmans} L.,  et~al., 2015, in Advancing Astrophysics with the Square
  Kilometre Array (AASKA14). p.~1 (\mn@eprint {arXiv} {1505.07568}),
  \mn@doi{10.22323/1.215.0001}

\bibitem[\protect\citeauthoryear{{Kostyuk}, {Nelson}, {Ciardi}, {Glatzle}  \&
  {Pillepich}}{{Kostyuk} et~al.}{2023}]{Kostyuk_2023}
{Kostyuk} I.,  {Nelson} D.,  {Ciardi} B.,  {Glatzle} M.,   {Pillepich} A.,
  2023, \mn@doi [\mnras] {10.1093/mnras/stad677}, \href
  {https://ui.adsabs.harvard.edu/abs/2023MNRAS.521.3077K} {521, 3077}

\bibitem[\protect\citeauthoryear{{Kostyuk}, {Ciardi}  \& {Ferrara}}{{Kostyuk}
  et~al.}{2025}]{Kostyuk_2025}
{Kostyuk} I.,  {Ciardi} B.,   {Ferrara} A.,  2025, \mn@doi [\aap]
  {10.1051/0004-6361/202449997}, \href
  {https://ui.adsabs.harvard.edu/abs/2025A&A...695A..32K} {695, A32}

\bibitem[\protect\citeauthoryear{{Kozyreva}, {Caputo}, {Baklanov}, {Mironov}
  \& {Janka}}{{Kozyreva} et~al.}{2025}]{Kozyreva_2025}
{Kozyreva} A.,  {Caputo} A.,  {Baklanov} P.,  {Mironov} A.,   {Janka} H.-T.,
  2025, \mn@doi [\aap] {10.1051/0004-6361/202452758}, \href
  {https://ui.adsabs.harvard.edu/abs/2025A&A...694A.319K} {694, A319}

\bibitem[\protect\citeauthoryear{{La Plante}, {Mirocha}, {Gorce}, {Lidz}  \&
  {Parsons}}{{La Plante} et~al.}{2023}]{LaPlante_2023}
{La Plante} P.,  {Mirocha} J.,  {Gorce} A.,  {Lidz} A.,   {Parsons} A.,  2023,
  \mn@doi [\apj] {10.3847/1538-4357/acaeb0}, \href
  {https://ui.adsabs.harvard.edu/abs/2023ApJ...944...59L} {944, 59}

\bibitem[\protect\citeauthoryear{{Leethochawalit} et~al.,}{{Leethochawalit}
  et~al.}{2023}]{Leethochawalit_2023}
{Leethochawalit} N.,  et~al., 2023, \mn@doi [\apjl] {10.3847/2041-8213/ac959b},
  \href {https://ui.adsabs.harvard.edu/abs/2023ApJ...942L..26L} {942, L26}

\bibitem[\protect\citeauthoryear{{Leung} et~al.,}{{Leung}
  et~al.}{2023}]{Leung_2023}
{Leung} G. C.~K.,  et~al., 2023, \mn@doi [\apjl] {10.3847/2041-8213/acf365},
  \href {https://ui.adsabs.harvard.edu/abs/2023ApJ...954L..46L} {954, L46}

\bibitem[\protect\citeauthoryear{{Lewis} et~al.,}{{Lewis}
  et~al.}{2022}]{Lewis_2022}
{Lewis} J. S.~W.,  et~al., 2022, \mn@doi [\mnras] {10.1093/mnras/stac2383},
  \href {https://ui.adsabs.harvard.edu/abs/2022MNRAS.516.3389L} {516, 3389}

\bibitem[\protect\citeauthoryear{{Lidz}, {Zahn}, {Furlanetto}, {McQuinn},
  {Hernquist}  \& {Zaldarriaga}}{{Lidz} et~al.}{2009}]{Lidz_2009}
{Lidz} A.,  {Zahn} O.,  {Furlanetto} S.~R.,  {McQuinn} M.,  {Hernquist} L.,
  {Zaldarriaga} M.,  2009, \mn@doi [\apj] {10.1088/0004-637X/690/1/252}, \href
  {https://ui.adsabs.harvard.edu/abs/2009ApJ...690..252L} {690, 252}

\bibitem[\protect\citeauthoryear{{Liu} \& {Parsons}}{{Liu} \&
  {Parsons}}{2016}]{Liu_2016}
{Liu} A.,  {Parsons} A.~R.,  2016, \mn@doi [\mnras] {10.1093/mnras/stw071},
  \href {https://ui.adsabs.harvard.edu/abs/2016MNRAS.457.1864L} {457, 1864}

\bibitem[\protect\citeauthoryear{{Long} et~al.,}{{Long}
  et~al.}{2023}]{Long_2023}
{Long} A.~S.,  et~al., 2023, \mn@doi [arXiv e-prints]
  {10.48550/arXiv.2305.04662}, \href
  {https://ui.adsabs.harvard.edu/abs/2023arXiv230504662L} {p. arXiv:2305.04662}

\bibitem[\protect\citeauthoryear{{Longley} et~al.,}{{Longley}
  et~al.}{2023}]{Longley_2023}
{Longley} E.~P.,  et~al., 2023, \mn@doi [\mnras] {10.1093/mnras/stad246}, \href
  {https://ui.adsabs.harvard.edu/abs/2023MNRAS.520.5016L} {520, 5016}

\bibitem[\protect\citeauthoryear{{Looser} et~al.,}{{Looser}
  et~al.}{2023}]{Looser_2023}
{Looser} T.~J.,  et~al., 2023, \mn@doi [arXiv e-prints]
  {10.48550/arXiv.2306.02470}, \href
  {https://ui.adsabs.harvard.edu/abs/2023arXiv230602470L} {p. arXiv:2306.02470}

\bibitem[\protect\citeauthoryear{{Ma} et~al.,}{{Ma} et~al.}{2018}]{Ma_2018fire}
{Ma} X.,  et~al., 2018, \mn@doi [\mnras] {10.1093/mnras/sty1024}, \href
  {https://ui.adsabs.harvard.edu/abs/2018MNRAS.478.1694M} {478, 1694}

\bibitem[\protect\citeauthoryear{{Ma}, {Quataert}, {Wetzel}, {Hopkins},
  {Faucher-Gigu{\`e}re}  \& {Kere{\v{s}}}}{{Ma} et~al.}{2020}]{Ma_2020fire}
{Ma} X.,  {Quataert} E.,  {Wetzel} A.,  {Hopkins} P.~F.,  {Faucher-Gigu{\`e}re}
  C.-A.,   {Kere{\v{s}}} D.,  2020, \mn@doi [\mnras] {10.1093/mnras/staa2404},
  \href {https://ui.adsabs.harvard.edu/abs/2020MNRAS.498.2001M} {498, 2001}

\bibitem[\protect\citeauthoryear{{Ma}, {Ciardi}, {Eide}, {Busch}, {Mao}  \&
  {Zhi}}{{Ma} et~al.}{2021}]{Ma_2021}
{Ma} Q.-B.,  {Ciardi} B.,  {Eide} M.~B.,  {Busch} P.,  {Mao} Y.,   {Zhi} Q.-J.,
   2021, \mn@doi [\apj] {10.3847/1538-4357/abefd5}, \href
  {https://ui.adsabs.harvard.edu/abs/2021ApJ...912..143M} {912, 143}

\bibitem[\protect\citeauthoryear{{Ma}, {Ghara}, {Ciardi}, {Iliev}, {Koopmans},
  {Mellema}, {Mondal}  \& {Zaroubi}}{{Ma} et~al.}{2023}]{Ma_2023}
{Ma} Q.-B.,  {Ghara} R.,  {Ciardi} B.,  {Iliev} I.~T.,  {Koopmans} L. V.~E.,
  {Mellema} G.,  {Mondal} R.,   {Zaroubi} S.,  2023, \mn@doi [\mnras]
  {10.1093/mnras/stad1203}, \href
  {https://ui.adsabs.harvard.edu/abs/2023MNRAS.522.3284M} {522, 3284}

\bibitem[\protect\citeauthoryear{{Ma}, {Chen}, {Li}, {Guo}, {Ciardi}, {Acharya}
   \& {Wang}}{{Ma} et~al.}{2025}]{Ma_2025}
{Ma} Q.-B.,  {Chen} X.-R.,  {Li} M.,  {Guo} Q.,  {Ciardi} B.,  {Acharya} A.,
  {Wang} X.,  2025, arXiv e-prints, \href
  {https://ui.adsabs.harvard.edu/abs/2025arXiv250419422M} {p. arXiv:2504.19422}

\bibitem[\protect\citeauthoryear{{Madau}, {Meiksin}  \& {Rees}}{{Madau}
  et~al.}{1997}]{Madau_1997}
{Madau} P.,  {Meiksin} A.,   {Rees} M.~J.,  1997, \mn@doi [\apj]
  {10.1086/303549}, \href
  {https://ui.adsabs.harvard.edu/abs/1997ApJ...475..429M} {475, 429}

\bibitem[\protect\citeauthoryear{{Maio} \& {Viel}}{{Maio} \&
  {Viel}}{2023}]{Maio_2023}
{Maio} U.,  {Viel} M.,  2023, \mn@doi [\aap] {10.1051/0004-6361/202345851},
  \href {https://ui.adsabs.harvard.edu/abs/2023A&A...672A..71M} {672, A71}

\bibitem[\protect\citeauthoryear{{Maio}, {P{\'e}roux}  \& {Ciardi}}{{Maio}
  et~al.}{2022}]{Maio_2022}
{Maio} U.,  {P{\'e}roux} C.,   {Ciardi} B.,  2022, \mn@doi [\aap]
  {10.1051/0004-6361/202142264}, \href
  {https://ui.adsabs.harvard.edu/abs/2022A&A...657A..47M} {657, A47}

\bibitem[\protect\citeauthoryear{{Maion}, {Angulo}  \& {Zennaro}}{{Maion}
  et~al.}{2022}]{Maion_2022}
{Maion} F.,  {Angulo} R.~E.,   {Zennaro} M.,  2022, \mn@doi [\jcap]
  {10.1088/1475-7516/2022/10/036}, \href
  {https://ui.adsabs.harvard.edu/abs/2022JCAP...10..036M} {2022, 036}

\bibitem[\protect\citeauthoryear{{Mao}, {Tegmark}, {McQuinn}, {Zaldarriaga}  \&
  {Zahn}}{{Mao} et~al.}{2008}]{Mao_2008}
{Mao} Y.,  {Tegmark} M.,  {McQuinn} M.,  {Zaldarriaga} M.,   {Zahn} O.,  2008,
  \mn@doi [\prd] {10.1103/PhysRevD.78.023529}, \href
  {https://ui.adsabs.harvard.edu/abs/2008PhRvD..78b3529M} {78, 023529}

\bibitem[\protect\citeauthoryear{{Mascia} et~al.,}{{Mascia}
  et~al.}{2023}]{Mascia_2023}
{Mascia} S.,  et~al., 2023, \mn@doi [\aap] {10.1051/0004-6361/202345866}, \href
  {https://ui.adsabs.harvard.edu/abs/2023A&A...672A.155M} {672, A155}

\bibitem[\protect\citeauthoryear{{Mason} et~al.,}{{Mason}
  et~al.}{2018}]{Mason_2018}
{Mason} C.~A.,  et~al., 2018, \mn@doi [\apjl] {10.3847/2041-8213/aabbab}, \href
  {https://ui.adsabs.harvard.edu/abs/2018ApJ...857L..11M} {857, L11}

\bibitem[\protect\citeauthoryear{{McGreer}, {Mesinger}  \& {Fan}}{{McGreer}
  et~al.}{2011}]{McGreer_2011}
{McGreer} I.~D.,  {Mesinger} A.,   {Fan} X.,  2011, \mn@doi [\mnras]
  {10.1111/j.1365-2966.2011.18935.x}, \href
  {https://ui.adsabs.harvard.edu/abs/2011MNRAS.415.3237M} {415, 3237}

\bibitem[\protect\citeauthoryear{{McGreer}, {Mesinger}  \&
  {D'Odorico}}{{McGreer} et~al.}{2015}]{McGreer_2015}
{McGreer} I.~D.,  {Mesinger} A.,   {D'Odorico} V.,  2015, \mn@doi [\mnras]
  {10.1093/mnras/stu2449}, \href
  {https://ui.adsabs.harvard.edu/abs/2015MNRAS.447..499M} {447, 499}

\bibitem[\protect\citeauthoryear{{McLeod} et~al.,}{{McLeod}
  et~al.}{2024}]{McLeod_2024}
{McLeod} D.~J.,  et~al., 2024, \mn@doi [\mnras] {10.1093/mnras/stad3471}, \href
  {https://ui.adsabs.harvard.edu/abs/2024MNRAS.527.5004M} {527, 5004}

\bibitem[\protect\citeauthoryear{{McQuinn}, {Zahn}, {Zaldarriaga}, {Hernquist}
  \& {Furlanetto}}{{McQuinn} et~al.}{2006}]{McQuinn_2006}
{McQuinn} M.,  {Zahn} O.,  {Zaldarriaga} M.,  {Hernquist} L.,   {Furlanetto}
  S.~R.,  2006, \mn@doi [\apj] {10.1086/505167}, \href
  {https://ui.adsabs.harvard.edu/abs/2006ApJ...653..815M} {653, 815}

\bibitem[\protect\citeauthoryear{{Mellema}, {Iliev}, {Alvarez}  \&
  {Shapiro}}{{Mellema} et~al.}{2006}]{Mellema_2006}
{Mellema} G.,  {Iliev} I.~T.,  {Alvarez} M.~A.,   {Shapiro} P.~R.,  2006,
  \mn@doi [\na] {10.1016/j.newast.2005.09.004}, \href
  {https://ui.adsabs.harvard.edu/abs/2006NewA...11..374M} {11, 374}

\bibitem[\protect\citeauthoryear{{Mertens} et~al.,}{{Mertens}
  et~al.}{2020}]{Mertens2020}
{Mertens} F.~G.,  et~al., 2020, \mn@doi [\mnras] {10.1093/mnras/staa327}, \href
  {https://ui.adsabs.harvard.edu/abs/2020MNRAS.493.1662M} {493, 1662}

\bibitem[\protect\citeauthoryear{{Mertens}, {Bobin}  \& {Carucci}}{{Mertens}
  et~al.}{2024}]{Mertens_2024}
{Mertens} F.~G.,  {Bobin} J.,   {Carucci} I.~P.,  2024, \mn@doi [\mnras]
  {10.1093/mnras/stad3430}, \href
  {https://ui.adsabs.harvard.edu/abs/2024MNRAS.527.3517M} {527, 3517}

\bibitem[\protect\citeauthoryear{{Mertens} et~al.,}{{Mertens}
  et~al.}{2025}]{Mertens_2025}
{Mertens} F.~G.,  et~al., 2025, \mn@doi [arXiv e-prints]
  {10.48550/arXiv.2503.05576}, \href
  {https://ui.adsabs.harvard.edu/abs/2025arXiv250305576M} {p. arXiv:2503.05576}

\bibitem[\protect\citeauthoryear{{Mesinger} \& {Furlanetto}}{{Mesinger} \&
  {Furlanetto}}{2007}]{Mesinger_2007}
{Mesinger} A.,  {Furlanetto} S.,  2007, \mn@doi [\apj] {10.1086/521806}, \href
  {https://ui.adsabs.harvard.edu/abs/2007ApJ...669..663M} {669, 663}

\bibitem[\protect\citeauthoryear{{Mesinger}, {Furlanetto}  \& {Cen}}{{Mesinger}
  et~al.}{2011}]{Mesinger_2011}
{Mesinger} A.,  {Furlanetto} S.,   {Cen} R.,  2011, \mn@doi [\mnras]
  {10.1111/j.1365-2966.2010.17731.x}, \href
  {https://ui.adsabs.harvard.edu/abs/2011MNRAS.411..955M} {411, 955}

\bibitem[\protect\citeauthoryear{{Mesinger}, {Aykutalp}, {Vanzella},
  {Pentericci}, {Ferrara}  \& {Dijkstra}}{{Mesinger}
  et~al.}{2015}]{Mesinger_2015}
{Mesinger} A.,  {Aykutalp} A.,  {Vanzella} E.,  {Pentericci} L.,  {Ferrara} A.,
    {Dijkstra} M.,  2015, \mn@doi [\mnras] {10.1093/mnras/stu2089}, \href
  {https://ui.adsabs.harvard.edu/abs/2015MNRAS.446..566M} {446, 566}

\bibitem[\protect\citeauthoryear{{Metha} \& {Berger}}{{Metha} \&
  {Berger}}{2024}]{Metha_2024}
{Metha} B.,  {Berger} S.,  2024, \mn@doi [arXiv e-prints]
  {10.48550/arXiv.2407.14068}, \href
  {https://ui.adsabs.harvard.edu/abs/2024arXiv240714068M} {p. arXiv:2407.14068}

\bibitem[\protect\citeauthoryear{{Molaro}, {Dav{\'e}}, {Hassan}, {Santos}  \&
  {Finlator}}{{Molaro} et~al.}{2019}]{Molaro_2019}
{Molaro} M.,  {Dav{\'e}} R.,  {Hassan} S.,  {Santos} M.~G.,   {Finlator} K.,
  2019, \mn@doi [\mnras] {10.1093/mnras/stz2171}, \href
  {https://ui.adsabs.harvard.edu/abs/2019MNRAS.489.5594M} {489, 5594}

\bibitem[\protect\citeauthoryear{{Mondal}, {Bharadwaj}  \& {Majumdar}}{{Mondal}
  et~al.}{2017}]{Mondal_2017}
{Mondal} R.,  {Bharadwaj} S.,   {Majumdar} S.,  2017, \mn@doi [\mnras]
  {10.1093/mnras/stw2599}, \href
  {https://ui.adsabs.harvard.edu/abs/2017MNRAS.464.2992M} {464, 2992}

\bibitem[\protect\citeauthoryear{{Mondal} et~al.,}{{Mondal}
  et~al.}{2020}]{Mondal_2020}
{Mondal} R.,  et~al., 2020, \mn@doi [\mnras] {10.1093/mnras/staa2422}, \href
  {https://ui.adsabs.harvard.edu/abs/2020MNRAS.498.4178M} {498, 4178}

\bibitem[\protect\citeauthoryear{{Mortlock} et~al.,}{{Mortlock}
  et~al.}{2011}]{Mortlock_2011}
{Mortlock} D.~J.,  et~al., 2011, \mn@doi [\nat] {10.1038/nature10159}, \href
  {https://ui.adsabs.harvard.edu/abs/2011Natur.474..616M} {474, 616}

\bibitem[\protect\citeauthoryear{{Murray}}{{Murray}}{2014}]{Murray_2014}
{Murray} S.,  2014, {HMF: Halo Mass Function calculator}, Astrophysics Source
  Code Library, record ascl:1412.006

\bibitem[\protect\citeauthoryear{{Mutch}, {Geil}, {Poole}, {Angel}, {Duffy},
  {Mesinger}  \& {Wyithe}}{{Mutch} et~al.}{2016}]{Mutch_2016}
{Mutch} S.~J.,  {Geil} P.~M.,  {Poole} G.~B.,  {Angel} P.~W.,  {Duffy} A.~R.,
  {Mesinger} A.,   {Wyithe} J. S.~B.,  2016, \mn@doi [\mnras]
  {10.1093/mnras/stw1506}, \href
  {https://ui.adsabs.harvard.edu/abs/2016MNRAS.462..250M} {462, 250}

\bibitem[\protect\citeauthoryear{{Nasirudin}, {Iliev}  \& {Ahn}}{{Nasirudin}
  et~al.}{2020}]{Nasirudin_2020}
{Nasirudin} A.,  {Iliev} I.~T.,   {Ahn} K.,  2020, \mn@doi [\mnras]
  {10.1093/mnras/staa853}, \href
  {https://ui.adsabs.harvard.edu/abs/2020MNRAS.494.3294N} {494, 3294}

\bibitem[\protect\citeauthoryear{{Ocvirk} et~al.,}{{Ocvirk}
  et~al.}{2016}]{Ocvirk_2016}
{Ocvirk} P.,  et~al., 2016, \mn@doi [\mnras] {10.1093/mnras/stw2036}, \href
  {https://ui.adsabs.harvard.edu/abs/2016MNRAS.463.1462O} {463, 1462}

\bibitem[\protect\citeauthoryear{{Ocvirk} et~al.,}{{Ocvirk}
  et~al.}{2020}]{Ocvirk_2020}
{Ocvirk} P.,  et~al., 2020, \mn@doi [\mnras] {10.1093/mnras/staa1266}, \href
  {https://ui.adsabs.harvard.edu/abs/2020MNRAS.496.4087O} {496, 4087}

\bibitem[\protect\citeauthoryear{{Ono} et~al.,}{{Ono} et~al.}{2012}]{Ono_2012}
{Ono} Y.,  et~al., 2012, \mn@doi [\apj] {10.1088/0004-637X/744/2/83}, \href
  {https://ui.adsabs.harvard.edu/abs/2012ApJ...744...83O} {744, 83}

\bibitem[\protect\citeauthoryear{{Ota} et~al.,}{{Ota} et~al.}{2008}]{Ota_2008}
{Ota} K.,  et~al., 2008, \mn@doi [\apj] {10.1086/529006}, \href
  {https://ui.adsabs.harvard.edu/abs/2008ApJ...677...12O} {677, 12}

\bibitem[\protect\citeauthoryear{{Ouchi} et~al.,}{{Ouchi}
  et~al.}{2010}]{Ouchi_2010}
{Ouchi} M.,  et~al., 2010, \mn@doi [\apj] {10.1088/0004-637X/723/1/869}, \href
  {https://ui.adsabs.harvard.edu/abs/2010ApJ...723..869O} {723, 869}

\bibitem[\protect\citeauthoryear{{Papovich} et~al.,}{{Papovich}
  et~al.}{2023}]{Papovich_2023}
{Papovich} C.,  et~al., 2023, \mn@doi [\apjl] {10.3847/2041-8213/acc948}, \href
  {https://ui.adsabs.harvard.edu/abs/2023ApJ...949L..18P} {949, L18}

\bibitem[\protect\citeauthoryear{{Peacock}}{{Peacock}}{1999}]{Peacock_1999}
{Peacock} J.~A.,  1999, {Cosmological Physics}

\bibitem[\protect\citeauthoryear{{Pentericci} et~al.,}{{Pentericci}
  et~al.}{2014}]{Pentericci_2014}
{Pentericci} L.,  et~al., 2014, \mn@doi [\apj] {10.1088/0004-637X/793/2/113},
  \href {https://ui.adsabs.harvard.edu/abs/2014ApJ...793..113P} {793, 113}

\bibitem[\protect\citeauthoryear{{Planck Collaboration} et~al.,}{{Planck
  Collaboration} et~al.}{2020}]{Planck_2020}
{Planck Collaboration} et~al., 2020, \mn@doi [\aap]
  {10.1051/0004-6361/201833910}, \href
  {https://ui.adsabs.harvard.edu/abs/2020A&A...641A...6P} {641, A6}

\bibitem[\protect\citeauthoryear{{Poole}, {Angel}, {Mutch}, {Power}, {Duffy},
  {Geil}, {Mesinger}  \& {Wyithe}}{{Poole} et~al.}{2016}]{Poole_2016}
{Poole} G.~B.,  {Angel} P.~W.,  {Mutch} S.~J.,  {Power} C.,  {Duffy} A.~R.,
  {Geil} P.~M.,  {Mesinger} A.,   {Wyithe} S.~B.,  2016, \mn@doi [\mnras]
  {10.1093/mnras/stw674}, \href
  {https://ui.adsabs.harvard.edu/abs/2016MNRAS.459.3025P} {459, 3025}

\bibitem[\protect\citeauthoryear{{Qin}, {Mesinger}, {Bosman}  \& {Viel}}{{Qin}
  et~al.}{2021}]{Qin_2021}
{Qin} Y.,  {Mesinger} A.,  {Bosman} S. E.~I.,   {Viel} M.,  2021, \mn@doi
  [\mnras] {10.1093/mnras/stab1833}, \href
  {https://ui.adsabs.harvard.edu/abs/2021MNRAS.506.2390Q} {506, 2390}

\bibitem[\protect\citeauthoryear{{Riess} et~al.,}{{Riess}
  et~al.}{2022}]{Riess_2022}
{Riess} A.~G.,  et~al., 2022, \mn@doi [\apjl] {10.3847/2041-8213/ac5c5b}, \href
  {https://ui.adsabs.harvard.edu/abs/2022ApJ...934L...7R} {934, L7}

\bibitem[\protect\citeauthoryear{{Robertson} et~al.,}{{Robertson}
  et~al.}{2013}]{Robertson_2013}
{Robertson} B.~E.,  et~al., 2013, \mn@doi [\apj] {10.1088/0004-637X/768/1/71},
  \href {https://ui.adsabs.harvard.edu/abs/2013ApJ...768...71R} {768, 71}

\bibitem[\protect\citeauthoryear{{Robertson} et~al.,}{{Robertson}
  et~al.}{2023}]{Robertson_2023}
{Robertson} B.~E.,  et~al., 2023, \mn@doi [Nature Astronomy]
  {10.1038/s41550-023-01921-1}, \href
  {https://ui.adsabs.harvard.edu/abs/2023NatAs...7..611R} {7, 611}

\bibitem[\protect\citeauthoryear{{Rosdahl} et~al.,}{{Rosdahl}
  et~al.}{2018}]{Rosdahl_2018}
{Rosdahl} J.,  et~al., 2018, \mn@doi [\mnras] {10.1093/mnras/sty1655}, \href
  {https://ui.adsabs.harvard.edu/abs/2018MNRAS.479..994R} {479, 994}

\bibitem[\protect\citeauthoryear{{Rubin} et~al.,}{{Rubin}
  et~al.}{2016}]{Rubin_2016}
{Rubin} A.,  et~al., 2016, \mn@doi [\apj] {10.3847/0004-637X/820/1/33}, \href
  {https://ui.adsabs.harvard.edu/abs/2016ApJ...820...33R} {820, 33}

\bibitem[\protect\citeauthoryear{{Santos} \& {Cooray}}{{Santos} \&
  {Cooray}}{2006}]{Santos_2006}
{Santos} M.~G.,  {Cooray} A.,  2006, \mn@doi [\prd]
  {10.1103/PhysRevD.74.083517}, \href
  {https://ui.adsabs.harvard.edu/abs/2006PhRvD..74h3517S} {74, 083517}

\bibitem[\protect\citeauthoryear{{Santos}, {Ferramacho}, {Silva}, {Amblard}  \&
  {Cooray}}{{Santos} et~al.}{2010a}]{Santos_2010code}
{Santos} M.,  {Ferramacho} L.,  {Silva} M.,  {Amblard} A.,   {Cooray} A.,
  2010a, {SimFast21: Simulation of the Cosmological 21cm Signal}, Astrophysics
  Source Code Library, record ascl:1010.025

\bibitem[\protect\citeauthoryear{{Santos}, {Ferramacho}, {Silva}, {Amblard}  \&
  {Cooray}}{{Santos} et~al.}{2010b}]{Santos_2010}
{Santos} M.~G.,  {Ferramacho} L.,  {Silva} M.~B.,  {Amblard} A.,   {Cooray} A.,
   2010b, \mn@doi [\mnras] {10.1111/j.1365-2966.2010.16898.x}, \href
  {https://ui.adsabs.harvard.edu/abs/2010MNRAS.406.2421S} {406, 2421}

\bibitem[\protect\citeauthoryear{{Saxena}, {Cole}, {Gazagnes}, {Meerburg},
  {Weniger}  \& {Witte}}{{Saxena} et~al.}{2023}]{Saxena_2023}
{Saxena} A.,  {Cole} A.,  {Gazagnes} S.,  {Meerburg} P.~D.,  {Weniger} C.,
  {Witte} S.~J.,  2023, \mn@doi [\mnras] {10.1093/mnras/stad2659}, \href
  {https://ui.adsabs.harvard.edu/abs/2023MNRAS.525.6097S} {525, 6097}

\bibitem[\protect\citeauthoryear{{Schaeffer}, {Giri}  \&
  {Schneider}}{{Schaeffer} et~al.}{2023}]{Schaeffer_2023}
{Schaeffer} T.,  {Giri} S.~K.,   {Schneider} A.,  2023, \mn@doi [\mnras]
  {10.1093/mnras/stad2937}, \href
  {https://ui.adsabs.harvard.edu/abs/2023MNRAS.526.2942S} {526, 2942}

\bibitem[\protect\citeauthoryear{{Schaerer}}{{Schaerer}}{2002}]{Schaerer_2002}
{Schaerer} D.,  2002, \mn@doi [\aap] {10.1051/0004-6361:20011619}, \href
  {https://ui.adsabs.harvard.edu/abs/2002A&A...382...28S} {382, 28}

\bibitem[\protect\citeauthoryear{{Schenker}, {Ellis}, {Konidaris}  \&
  {Stark}}{{Schenker} et~al.}{2014}]{Schenker_2014}
{Schenker} M.~A.,  {Ellis} R.~S.,  {Konidaris} N.~P.,   {Stark} D.~P.,  2014,
  \mn@doi [\apj] {10.1088/0004-637X/795/1/20}, \href
  {https://ui.adsabs.harvard.edu/abs/2014ApJ...795...20S} {795, 20}

\bibitem[\protect\citeauthoryear{{Schneider}, {Schaeffer}  \&
  {Giri}}{{Schneider} et~al.}{2023}]{Schneider_2023}
{Schneider} A.,  {Schaeffer} T.,   {Giri} S.~K.,  2023, \mn@doi [\prd]
  {10.1103/PhysRevD.108.043030}, \href
  {https://ui.adsabs.harvard.edu/abs/2023PhRvD.108d3030S} {108, 043030}

\bibitem[\protect\citeauthoryear{{Schroeder}, {Mesinger}  \&
  {Haiman}}{{Schroeder} et~al.}{2013}]{Schroeder_2013}
{Schroeder} J.,  {Mesinger} A.,   {Haiman} Z.,  2013, \mn@doi [\mnras]
  {10.1093/mnras/sts253}, \href
  {https://ui.adsabs.harvard.edu/abs/2013MNRAS.428.3058S} {428, 3058}

\bibitem[\protect\citeauthoryear{{Shaver}, {Windhorst}, {Madau}  \& {de
  Bruyn}}{{Shaver} et~al.}{1999}]{Shaver_1999}
{Shaver} P.~A.,  {Windhorst} R.~A.,  {Madau} P.,   {de Bruyn} A.~G.,  1999,
  \aap, \href {https://ui.adsabs.harvard.edu/abs/1999A&A...345..380S} {345,
  380}

\bibitem[\protect\citeauthoryear{{Sipple} \& {Lidz}}{{Sipple} \&
  {Lidz}}{2024}]{Sipple_2024}
{Sipple} J.,  {Lidz} A.,  2024, \mn@doi [\apj] {10.3847/1538-4357/ad06a7},
  \href {https://ui.adsabs.harvard.edu/abs/2024ApJ...961...50S} {961, 50}

\bibitem[\protect\citeauthoryear{{Smith}, {Kannan}, {Garaldi}, {Vogelsberger},
  {Pakmor}, {Springel}  \& {Hernquist}}{{Smith} et~al.}{2022}]{Smith_2022}
{Smith} A.,  {Kannan} R.,  {Garaldi} E.,  {Vogelsberger} M.,  {Pakmor} R.,
  {Springel} V.,   {Hernquist} L.,  2022, \mn@doi [\mnras]
  {10.1093/mnras/stac713}, \href
  {https://ui.adsabs.harvard.edu/abs/2022MNRAS.512.3243S} {512, 3243}

\bibitem[\protect\citeauthoryear{{Sobacchi} \& {Mesinger}}{{Sobacchi} \&
  {Mesinger}}{2015}]{Sobacchi_2015}
{Sobacchi} E.,  {Mesinger} A.,  2015, \mn@doi [\mnras] {10.1093/mnras/stv1751},
  \href {https://ui.adsabs.harvard.edu/abs/2015MNRAS.453.1843S} {453, 1843}

\bibitem[\protect\citeauthoryear{{Springel}, {White}, {Tormen}  \&
  {Kauffmann}}{{Springel} et~al.}{2001}]{Springel_2001}
{Springel} V.,  {White} S. D.~M.,  {Tormen} G.,   {Kauffmann} G.,  2001,
  \mn@doi [\mnras] {10.1046/j.1365-8711.2001.04912.x}, \href
  {https://ui.adsabs.harvard.edu/abs/2001MNRAS.328..726S} {328, 726}

\bibitem[\protect\citeauthoryear{{Springel}, {Pakmor}, {Zier}  \&
  {Reinecke}}{{Springel} et~al.}{2021}]{Springel_2021}
{Springel} V.,  {Pakmor} R.,  {Zier} O.,   {Reinecke} M.,  2021, \mn@doi
  [\mnras] {10.1093/mnras/stab1855}, \href
  {https://ui.adsabs.harvard.edu/abs/2021MNRAS.506.2871S} {506, 2871}

\bibitem[\protect\citeauthoryear{{Stanway} \& {Eldridge}}{{Stanway} \&
  {Eldridge}}{2018}]{Stanway_2018}
{Stanway} E.~R.,  {Eldridge} J.~J.,  2018, \mn@doi [\mnras]
  {10.1093/mnras/sty1353}, \href
  {https://ui.adsabs.harvard.edu/abs/2018MNRAS.479...75S} {479, 75}

\bibitem[\protect\citeauthoryear{{Stefanon}, {Bouwens}, {Labb{\'e}},
  {Illingworth}, {Gonzalez}  \& {Oesch}}{{Stefanon}
  et~al.}{2021}]{Stefanon_2021}
{Stefanon} M.,  {Bouwens} R.~J.,  {Labb{\'e}} I.,  {Illingworth} G.~D.,
  {Gonzalez} V.,   {Oesch} P.~A.,  2021, \mn@doi [\apj]
  {10.3847/1538-4357/ac1bb6}, \href
  {https://ui.adsabs.harvard.edu/abs/2021ApJ...922...29S} {922, 29}

\bibitem[\protect\citeauthoryear{{Sun} \& {Furlanetto}}{{Sun} \&
  {Furlanetto}}{2016}]{Sun_2016}
{Sun} G.,  {Furlanetto} S.~R.,  2016, \mn@doi [\mnras] {10.1093/mnras/stw980},
  \href {https://ui.adsabs.harvard.edu/abs/2016MNRAS.460..417S} {460, 417}

\bibitem[\protect\citeauthoryear{{Tacchella}, {Bose}, {Conroy}, {Eisenstein}
  \& {Johnson}}{{Tacchella} et~al.}{2018}]{Tacchella_2018}
{Tacchella} S.,  {Bose} S.,  {Conroy} C.,  {Eisenstein} D.~J.,   {Johnson}
  B.~D.,  2018, \mn@doi [\apj] {10.3847/1538-4357/aae8e0}, \href
  {https://ui.adsabs.harvard.edu/abs/2018ApJ...868...92T} {868, 92}

\bibitem[\protect\citeauthoryear{{Thomas} et~al.,}{{Thomas}
  et~al.}{2009}]{Thomas_2009}
{Thomas} R.~M.,  et~al., 2009, \mn@doi [\mnras]
  {10.1111/j.1365-2966.2008.14206.x}, \href
  {https://ui.adsabs.harvard.edu/abs/2009MNRAS.393...32T} {393, 32}

\bibitem[\protect\citeauthoryear{{Tinker}, {Kravtsov}, {Klypin}, {Abazajian},
  {Warren}, {Yepes}, {Gottl{\"o}ber}  \& {Holz}}{{Tinker}
  et~al.}{2008}]{Tinker_2008}
{Tinker} J.,  {Kravtsov} A.~V.,  {Klypin} A.,  {Abazajian} K.,  {Warren} M.,
  {Yepes} G.,  {Gottl{\"o}ber} S.,   {Holz} D.~E.,  2008, \mn@doi [\apj]
  {10.1086/591439}, \href
  {https://ui.adsabs.harvard.edu/abs/2008ApJ...688..709T} {688, 709}

\bibitem[\protect\citeauthoryear{{Totani}, {Kawai}, {Kosugi}, {Aoki}, {Yamada},
  {Iye}, {Ohta}  \& {Hattori}}{{Totani} et~al.}{2006}]{Totani_2006}
{Totani} T.,  {Kawai} N.,  {Kosugi} G.,  {Aoki} K.,  {Yamada} T.,  {Iye} M.,
  {Ohta} K.,   {Hattori} T.,  2006, \mn@doi [\pasj] {10.1093/pasj/58.3.485},
  \href {https://ui.adsabs.harvard.edu/abs/2006PASJ...58..485T} {58, 485}

\bibitem[\protect\citeauthoryear{{Tozzi}, {Madau}, {Meiksin}  \&
  {Rees}}{{Tozzi} et~al.}{2000}]{Tozzi_2000}
{Tozzi} P.,  {Madau} P.,  {Meiksin} A.,   {Rees} M.~J.,  2000, \mn@doi [\apj]
  {10.1086/308196}, \href
  {https://ui.adsabs.harvard.edu/abs/2000ApJ...528..597T} {528, 597}

\bibitem[\protect\citeauthoryear{{Trac}, {Chen}, {Holst}, {Alvarez}  \&
  {Cen}}{{Trac} et~al.}{2022}]{Trac_2022}
{Trac} H.,  {Chen} N.,  {Holst} I.,  {Alvarez} M.~A.,   {Cen} R.,  2022,
  \mn@doi [\apj] {10.3847/1538-4357/ac5116}, \href
  {https://ui.adsabs.harvard.edu/abs/2022ApJ...927..186T} {927, 186}

\bibitem[\protect\citeauthoryear{{Treu} et~al.,}{{Treu}
  et~al.}{2023}]{Treu_2023}
{Treu} T.,  et~al., 2023, \mn@doi [\apjl] {10.3847/2041-8213/ac9283}, \href
  {https://ui.adsabs.harvard.edu/abs/2023ApJ...942L..28T} {942, L28}

\bibitem[\protect\citeauthoryear{{Tr{\"o}ster} et~al.,}{{Tr{\"o}ster}
  et~al.}{2020}]{Troster_2020}
{Tr{\"o}ster} T.,  et~al., 2020, \mn@doi [\aap] {10.1051/0004-6361/201936772},
  \href {https://ui.adsabs.harvard.edu/abs/2020A&A...633L..10T} {633, L10}

\bibitem[\protect\citeauthoryear{{Trott} et~al.,}{{Trott}
  et~al.}{2020}]{Trott_2020}
{Trott} C.~M.,  et~al., 2020, \mn@doi [\mnras] {10.1093/mnras/staa414}, \href
  {https://ui.adsabs.harvard.edu/abs/2020MNRAS.493.4711T} {493, 4711}

\bibitem[\protect\citeauthoryear{{Vani}, {Ayromlou}, {Kauffmann}  \&
  {Springel}}{{Vani} et~al.}{2024}]{Vani_2024}
{Vani} A.,  {Ayromlou} M.,  {Kauffmann} G.,   {Springel} V.,  2024, \mn@doi
  [arXiv e-prints] {10.48550/arXiv.2408.00824}, \href
  {https://ui.adsabs.harvard.edu/abs/2024arXiv240800824V} {p. arXiv:2408.00824}

\bibitem[\protect\citeauthoryear{{Villaescusa-Navarro}
  et~al.,}{{Villaescusa-Navarro} et~al.}{2018}]{Villaescusa-Navarro_2018}
{Villaescusa-Navarro} F.,  et~al., 2018, \mn@doi [\apj]
  {10.3847/1538-4357/aae52b}, \href
  {https://ui.adsabs.harvard.edu/abs/2018ApJ...867..137V} {867, 137}

\bibitem[\protect\citeauthoryear{Virtanen et~al.,}{Virtanen
  et~al.}{2020}]{Virtanen_2020}
Virtanen P.,  et~al., 2020, \mn@doi [Nature Methods]
  {10.1038/s41592-019-0686-2}, \href {https://rdcu.be/b08Wh} {17, 261}

\bibitem[\protect\citeauthoryear{{Vrbanec} et~al.,}{{Vrbanec}
  et~al.}{2016}]{Vrbanec_2016}
{Vrbanec} D.,  et~al., 2016, \mn@doi [\mnras] {10.1093/mnras/stv2993}, \href
  {https://ui.adsabs.harvard.edu/abs/2016MNRAS.457..666V} {457, 666}

\bibitem[\protect\citeauthoryear{Watson, Iliev, D’Aloisio, Knebe, Shapiro  \&
  Yepes}{Watson et~al.}{2013}]{Watson_2013}
Watson W.~A.,  Iliev I.~T.,  D’Aloisio A.,  Knebe A.,  Shapiro P.~R.,   Yepes
  G.,  2013, \mn@doi [Monthly Notices of the Royal Astronomical Society]
  {10.1093/mnras/stt791}, 433, 1230–1245

\bibitem[\protect\citeauthoryear{{Wu} \& {Kravtsov}}{{Wu} \&
  {Kravtsov}}{2024}]{Wu_2024}
{Wu} Z.,  {Kravtsov} A.,  2024, \mn@doi [The Open Journal of Astrophysics]
  {10.33232/001c.121193}, \href
  {https://ui.adsabs.harvard.edu/abs/2024OJAp....7E..56W} {7, 56}

\bibitem[\protect\citeauthoryear{Zaroubi}{Zaroubi}{2013}]{Zaroubi_2013}
Zaroubi S.,  2013, The Epoch of Reionization.
Springer Berlin Heidelberg, Berlin, Heidelberg, pp 45--101,
  \mn@doi{10.1007/978-3-642-32362-1_2}, \url
  {https://doi.org/10.1007/978-3-642-32362-1_2}

\bibitem[\protect\citeauthoryear{{Zhang}, {Lachance}, {Ni}, {Li}, {Croft},
  {Matteo}, {Bird}  \& {Feng}}{{Zhang} et~al.}{2024}]{Zhang_2024}
{Zhang} X.,  {Lachance} P.,  {Ni} Y.,  {Li} Y.,  {Croft} R. A.~C.,  {Matteo}
  T.~D.,  {Bird} S.,   {Feng} Y.,  2024, \mn@doi [\mnras]
  {10.1093/mnras/stad3940}, \href
  {https://ui.adsabs.harvard.edu/abs/2024MNRAS.528..281Z} {528, 281}

\bibitem[\protect\citeauthoryear{{de Belsunce}, {Gratton}, {Coulton}  \&
  {Efstathiou}}{{de Belsunce} et~al.}{2021}]{deBelsunce_2021}
{de Belsunce} R.,  {Gratton} S.,  {Coulton} W.,   {Efstathiou} G.,  2021,
  \mn@doi [\mnras] {10.1093/mnras/stab2215}, \href
  {https://ui.adsabs.harvard.edu/abs/2021MNRAS.507.1072D} {507, 1072}

\bibitem[\protect\citeauthoryear{{van der Velden}}{{van der
  Velden}}{2020}]{cmasher_2020}
{van der Velden} E.,  2020, \mn@doi [The Journal of Open Source Software]
  {10.21105/joss.02004}, \href
  {https://ui.adsabs.harvard.edu/abs/2020JOSS....5.2004V} {5, 2004}

\makeatother
\end{thebibliography}


\clearpage
\appendix

\section{UV Luminosity Functions across redshifts}\label{appendix_uvlf}

\noindent\begin{minipage}{\textwidth}
\begin{flushleft}
As we tune the astrophysical parameters for the constrained case only to match the UVLFs at $z = 10$ and 9, it is interesting to explore how the four cosmological models compare to the UVLFs observed at other redshifts. In Figure~\ref{fig_uvlf_all}, we show results at  $z = 12, 10, 9, 8, 7, 5$. We note that while the models mildly overestimate observational data at $z = 5$, this could be because astrophysical parameters may need to be evolved with redshift, while here they are kept constant. Interestingly, the $\sigma_8$ low model tends to reproduce the bright end much better at lower redshifts.
\end{flushleft}
\end{minipage}

\vspace{2em}
\noindent\begin{minipage}{\textwidth}
\centering
\includegraphics[width=\columnwidth,keepaspectratio]{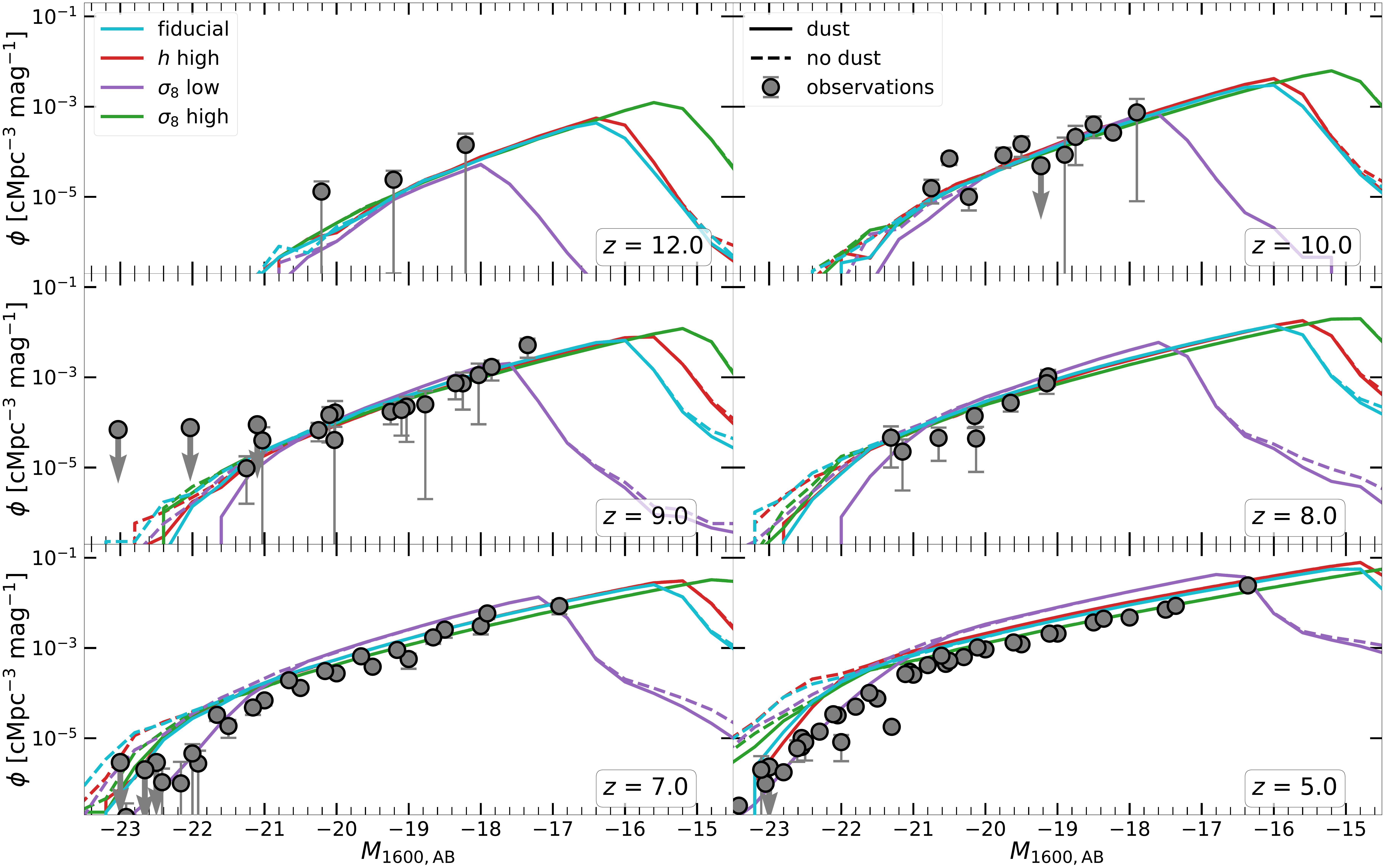}
\captionof{figure}{A and B-series averaged UVLFs for the constrained case at  $z = 12$, 10, 9, 8, 7 and 5 for the fiducial (cyan), $h$ high (red), $\sigma_8$ low (purple) and $\sigma_8$ high (green). We also show the dust attenuated (solid) and unattenuated  (dashed) UVLFs, and JWST and HST observations \citep[][grey circles]{Finkelstein_2015,Bouwens_2015,Bouwens_2021,Harikane_2022,Bouwens_2023a,Bouwens_2023b,Harikane_2023,Leung_2023,McLeod_2024,Adams_2024}.
} 
\label{fig_uvlf_all}
\end{minipage}

\clearpage
\section{Fiducial model with $\text{\lowercase{f}}_{\text{\lowercase{esc}}} = 25\%$}\label{appendix_highfesc}

\noindent\begin{minipage}{\textwidth}
\begin{flushleft}
In Figure~\ref{fig_xhi_highfesc}, we show the redshift evolution of the average of the A and B-series volume-averaged neutral hydrogen fraction $\langle x_{\rm HI} \rangle$ for an escape fraction of $f_{\rm esc} = 12.5\%$ (cyan) and 25\% (brown). We note that as expected, doubling the escape fraction leads to a faster rate of reionization, getting 50\% reionized by $z \approx 7$ for $f_{\rm esc} = 25\%$ as compared to $z \approx 6.4$ for $f_{\rm esc} = 12.5\%$. However, both cases are in agreement with observations, except for redshifts $z \leq 6$, where the lower $f_{\rm esc}$ case has slightly higher values. Nevertheless, it is still within the margin of error of observations.

In Figure~\ref{fig_ps0.1_highfesc} we also show the redshift evolution of the normalized 21-cm signal power spectrum ($\Delta^{2}_{\rm 21cm}$) at $k = 0.15~h\rm cMpc^{-1}$. We note that just like the other parameters as shown in Figure~\ref{fig_ps0.1_combined}, the choice of $f_{\rm esc}$ impacts where $\Delta^{2}_{\rm 21cm} (k = 0.15~h\rm cMpc^{-1})$ peaks as well as its overall trend across redshift.
\end{flushleft}
\end{minipage}

\vspace{2em}
\noindent\begin{minipage}{\textwidth}
\centering
\includegraphics[width=0.5\columnwidth,keepaspectratio]{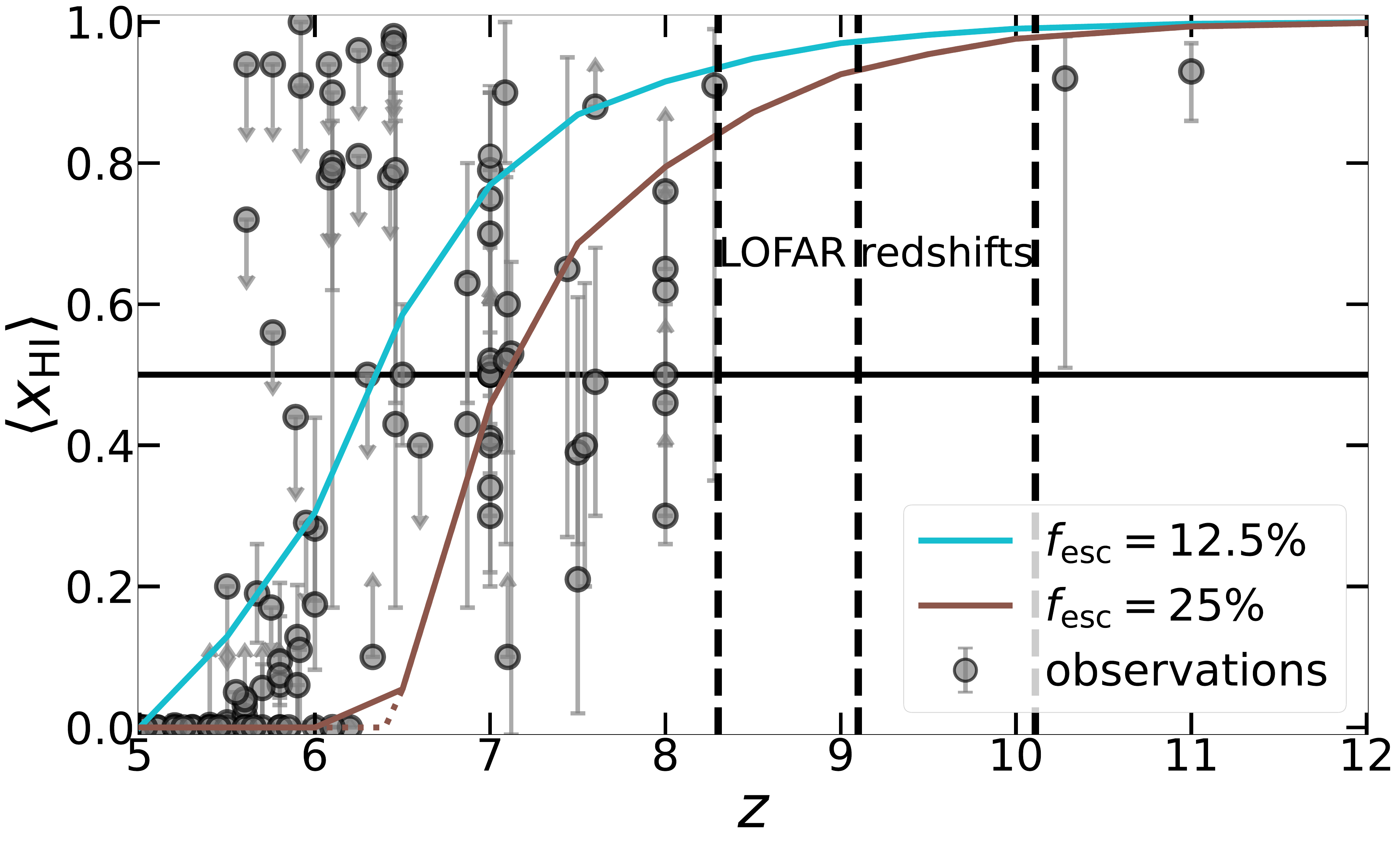}
\captionof{figure}{Redshift evolution of the average of the A and B-series volume-averaged neutral hydrogen fraction $\langle x_{\rm HI} \rangle$  for the fiducial model, for an escape fraction of $f_{\rm esc} = 12.5\%$ (cyan) and 25\% (brown). Dotted lines refer to a fit to the curves, which is used for a better estimate of the redshift of reionization when the redshift resolution is too coarse.
The vertical grey dashed lines indicate the redshifts observationally relevant for LOFAR ($z = 10.11$, 9.16 and 8.3), and the black solid line is drawn at $\langle x_{\rm HI} \rangle = 0.5$ to guide the eye. Grey circles are a collection of observational constraints \citep[][collected in the {\sc CoReCon} module, \citealt{Garaldi_2023}]{Fan_2006b, Totani_2006, Ota_2008, Ouchi_2010, Bolton_2011, Dijkstra_2011, McGreer_2011, Mortlock_2011, Ono_2012, Chornock_2013, Jensen_2013, Robertson_2013, Schroeder_2013, Pentericci_2014, Schenker_2014, McGreer_2015, Sobacchi_2015, Choudhury_2015, Mesinger_2015, Greig_2017, Davies_2018, Mason_2018,  Hoag_2019, Greig_2019, Jones_2024}.
} 
\label{fig_xhi_highfesc}

\includegraphics[width=0.5\columnwidth,keepaspectratio]{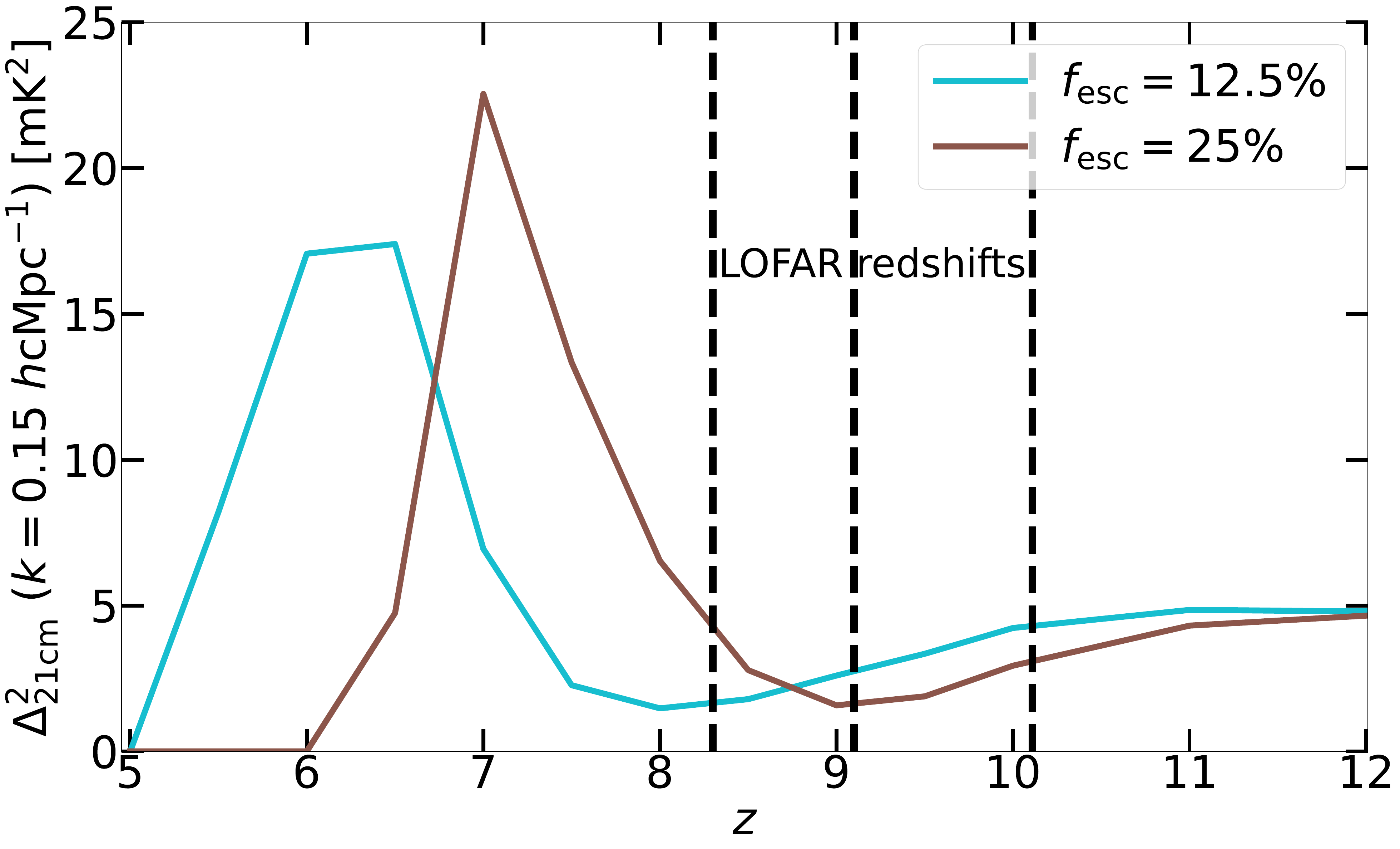}
\captionof{figure}{Redshift evolution of the A and B-series averaged normalized 21-cm signal power spectrum ($\Delta^{2}_{\rm 21cm}$) at $k = 0.15~h\rm cMpc^{-1}$  for the fiducial model, for an escape fraction of $f_{\rm esc} = 12.5\%$ (cyan) and 25\% (brown). The vertical dashed lines indicate the redshifts relevant for LOFAR, i.e. $z = 10.11$, 9.16, and 8.3.
} 
\label{fig_ps0.1_highfesc}
\end{minipage}

\clearpage

\section{Maps of $\delta T_{\rm b}$}\label{appendix_midslice}
\subsection{B-series maps in the constrained case}\label{appendix_pairconstrained_midslice}

\noindent\begin{minipage}{\textwidth}
\begin{flushleft}
In Figure~\ref{fig_midslice_pconst}, we show the maps of $\delta T_{\rm b}$ of the B-series middle slices (the simulations with ``paired'' initial conditions in the F\&P pair of simulations) for the four cosmological models in the constrained case at $z = 12, 10, 8,~\rm and~6$ (from top to bottom). Here the darkest regions represent the ionized regions with $\delta T_{\rm b} = 0$. Note that $\delta T_{\rm b}$ cannot assume negative values, due to the assumption of $T_{\rm S} \gg T_{\rm CMB}$. We note that the ionized regions of the A-series middle slices shown in Figure~\ref{fig_midslice_fconst} correspond to the most neutral regions of Figure~\ref{fig_midslice_pconst}. This is because of the nature of the Fixed \& Paired approach, as regions of matter clustering in the A-series should correspond to voids in the B-series by construction.
\end{flushleft}
\centering
\includegraphics[width=\columnwidth,keepaspectratio]{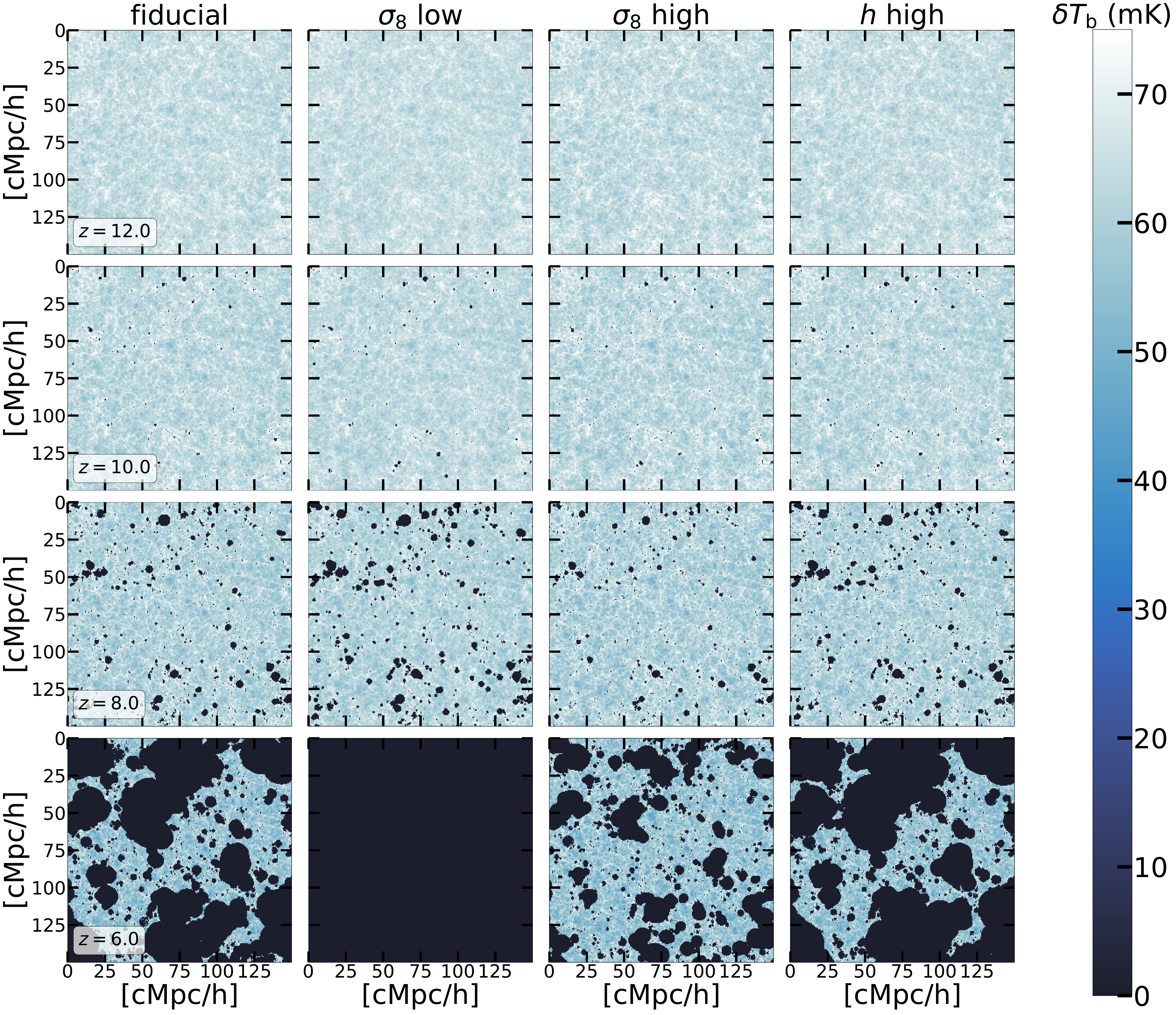}
\captionof{figure}{Maps of $\delta T_{\rm b}$ of the B-series middle slices of single cell thickness (i.e., $\approx 600~h^{-1}~\rm ckpc$) for the four cosmological models in the constrained case at $z = 12, 10, 8,~\rm and~6$ (from top to bottom) corresponding to the A-series slices shown in Figure~\ref{fig_midslice_fconst}. Here the dark areas represent the ionized regions with $\delta T_{\rm b} = 0$. 
Note that $\delta T_{\rm b}$ cannot have negative values due to the assumption of $T_{\rm S} \gg T_{\rm CMB}$.
}
\label{fig_midslice_pconst}
\end{minipage}

\clearpage
\noindent\begin{minipage}{\textwidth}
\begin{flushleft}
In Figure~\ref{fig_midslice_diffconst}, we show the absolute difference between the maps of $\delta T_{\rm b}$ of the A and B-series middle slices for the four cosmological models in the constrained case at $z = 12, 10, 8,~\rm and~6$ (from top to bottom). This makes the difference between the fixed simulation and its pair evident, as well as showing that with the progress of reionization, small-scale stochasticity induced by ionized bubbles is a tracer for deviations from the perfect anti-symmetry that the system started with. That is, until $z = 8$, the middle slices are effectively random noise maps, as they are primarily governed by the underlying matter distribution. However, by $z = 6$, the impact of the growth of ionized bubbles becomes significant and shows up as features in the difference maps.
\end{flushleft}
\centering
\includegraphics[width=\columnwidth,keepaspectratio]{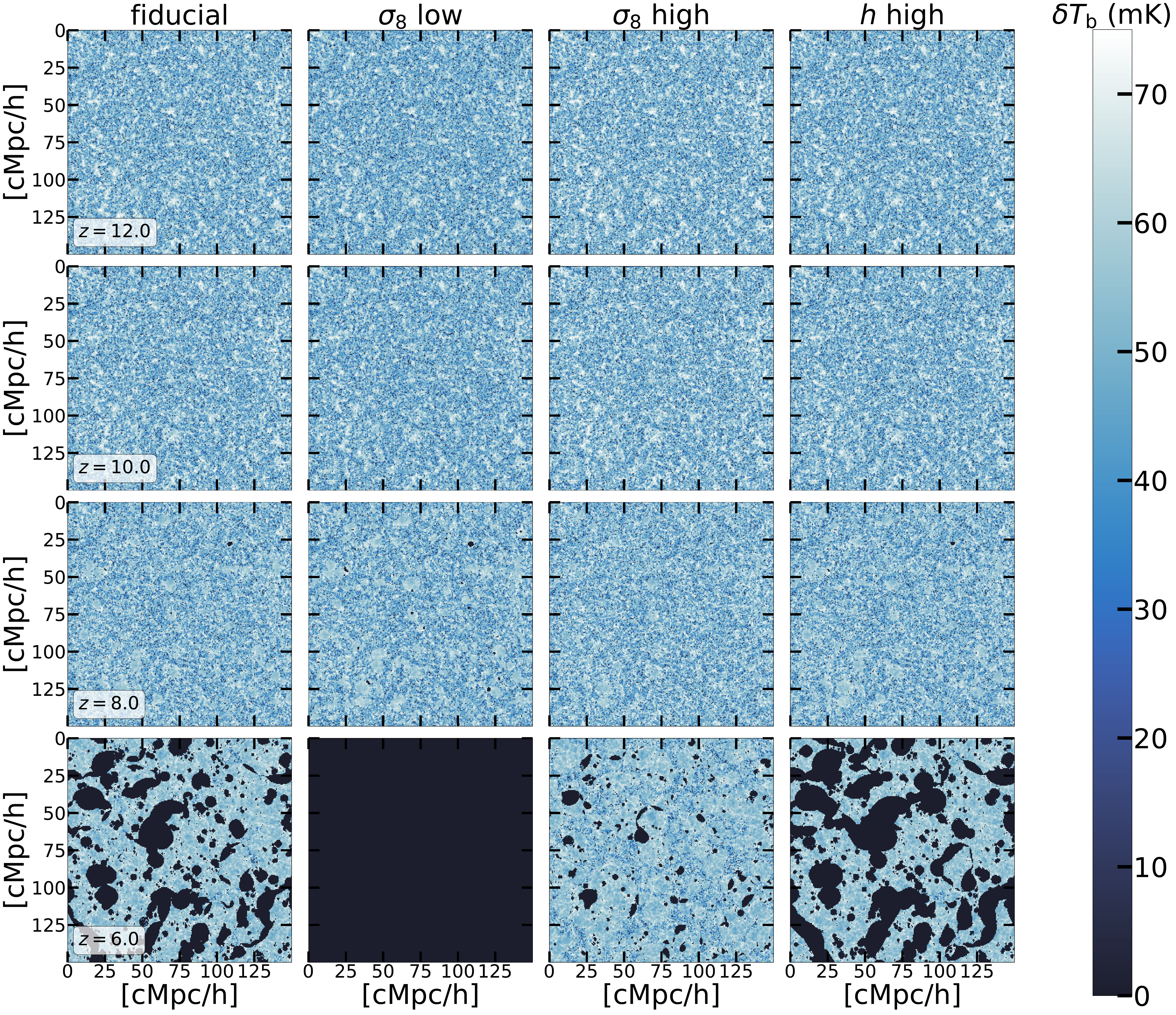}
\captionof{figure}{The absolute difference between the maps of $\delta T_{\rm b}$ of the A and B-series middle slices shown in Figure~\ref{fig_midslice_fconst} and~\ref{fig_midslice_pconst}.
}
\label{fig_midslice_diffconst}
\end{minipage}

\clearpage
\subsection{A and B-series maps in the unconstrained case}\label{appendix_bothunconstrained_midslice}

\noindent\begin{minipage}{\textwidth}
\begin{flushleft}
In Figure~\ref{fig_midslice_funconst} and~\ref{fig_midslice_punconst}, we show the maps of $\delta T_{\rm b}$ of the A and B-series middle slices for the four cosmological models in the unconstrained case at $z = 12, 10, 8,~\rm and~6$ (from top to bottom). Thus, here, we directly see the influence of the choice of cosmological parameters on the 21-cm signal.
\end{flushleft}
\centering
\includegraphics[width=\columnwidth,keepaspectratio]{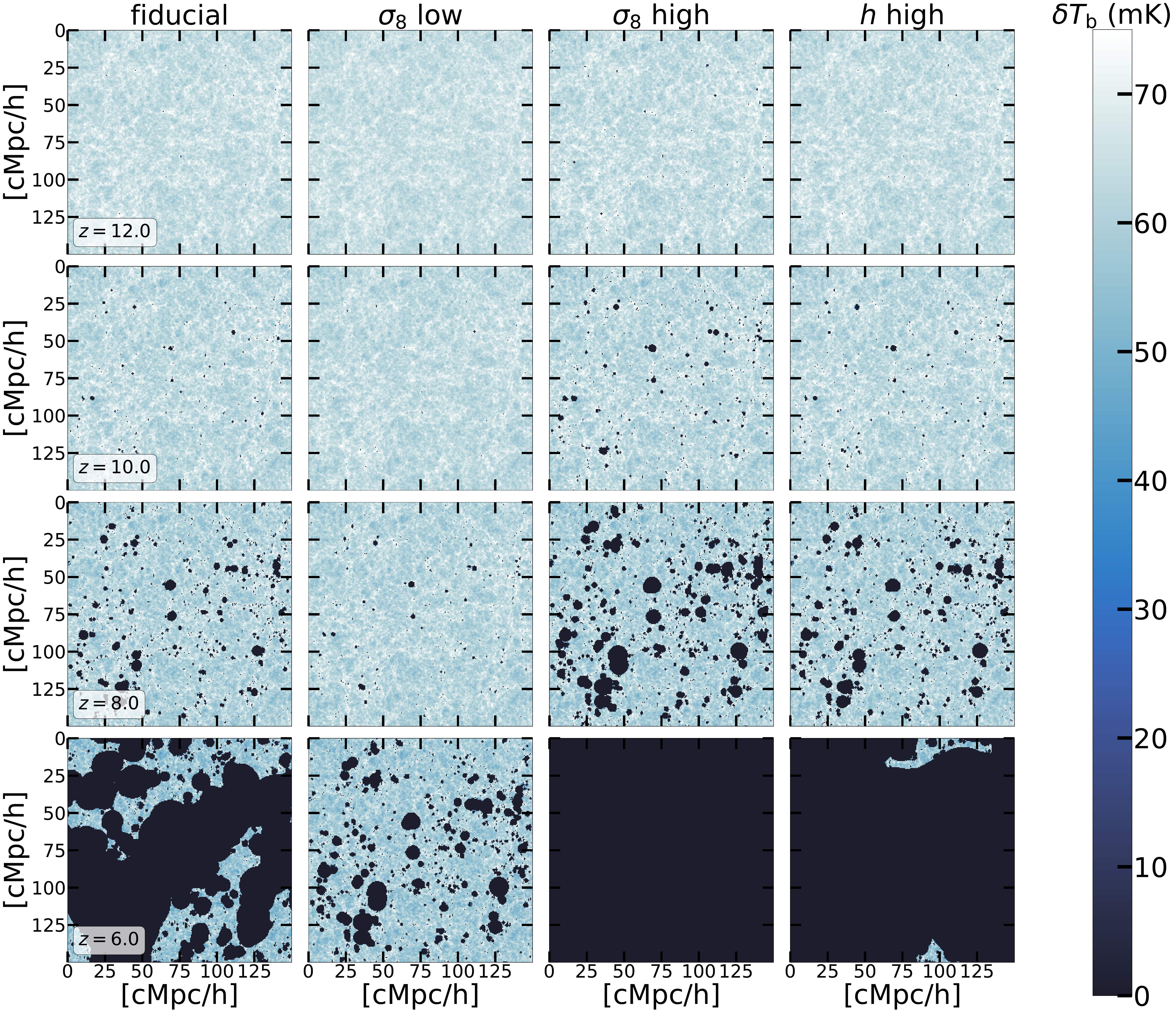}
\captionof{figure}{Maps of $\delta T_{\rm b}$ of the A-series middle slices of single cell thickness (i.e., $\approx 600~h^{-1}~\rm ckpc$) for the four cosmological models in the unconstrained case at $z = 12, 10, 8,~\rm and~6$ (from top to bottom). Note that here, due to the higher matter clustering driving up star formation, the $\sigma_8$ high model is already fully ionized by $z = 6$, and the $h$ high model is approaching complete reionization as well. On the other hand, the $\sigma_8$ low case still has large neutral regions.
}
\label{fig_midslice_funconst}
\end{minipage}

\clearpage
\noindent\begin{minipage}{\textwidth}
\centering
\includegraphics[width=\columnwidth,keepaspectratio]{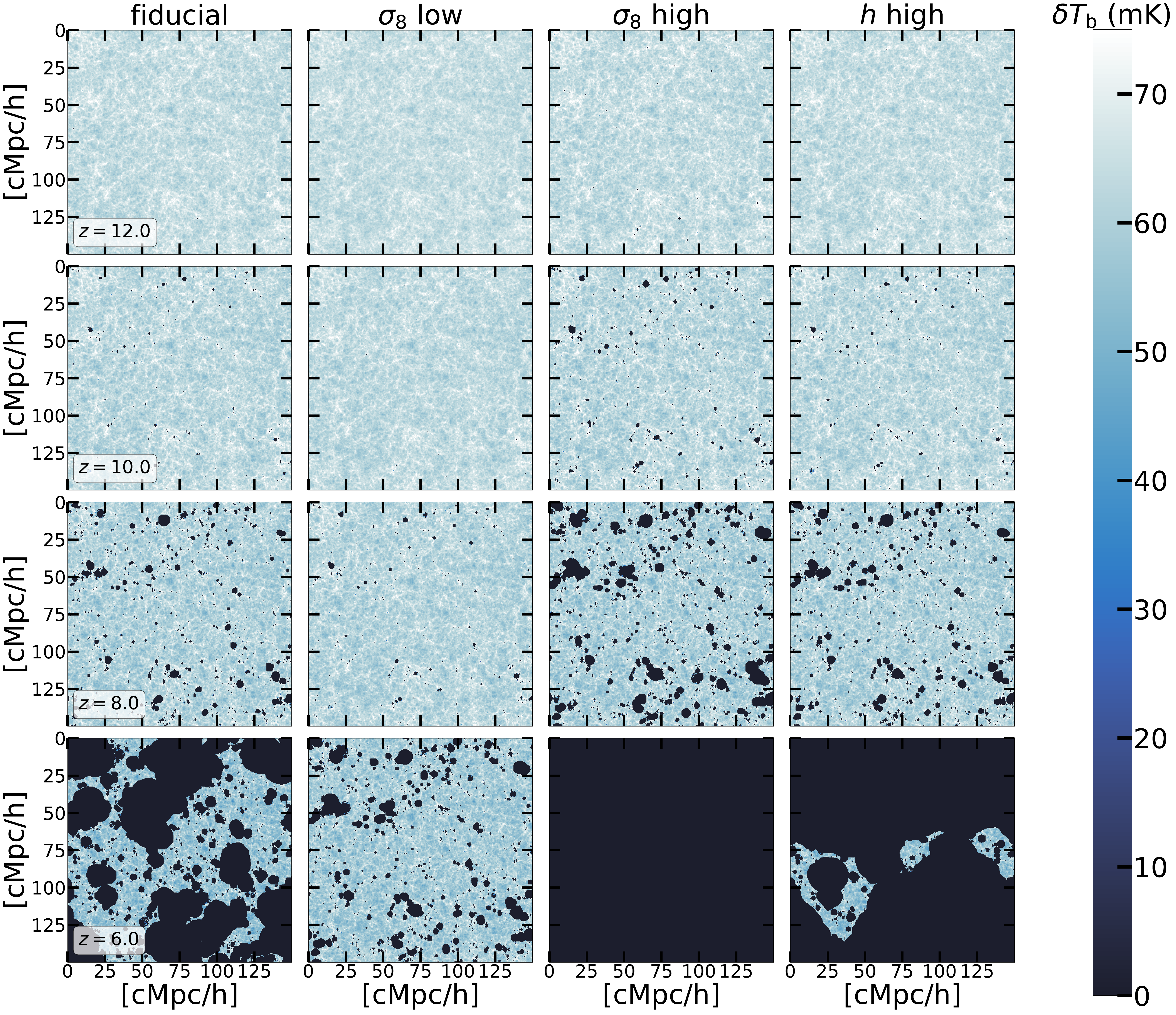}
\captionof{figure}{Maps of $\delta T_{\rm b}$ of the B-series middle slices of single cell thickness (i.e., $\approx 600~h^{-1}~\rm ckpc$) for the four cosmological models in the unconstrained case at $z = 12, 10, 8,~\rm and~6$ (from top to bottom).
}
\label{fig_midslice_punconst}
\end{minipage}


\bsp	
\label{lastpage}
\end{document}